\documentclass[a4paper, 11pt, twoside]{report}

\usepackage{fancyhdr}

\usepackage{tipa}

\newcommand{\clearemptydoublepage}{\newpage{\pagestyle{empty}\cleardoublepage}}

\usepackage{amsmath}
\usepackage{amsfonts}
\usepackage{bbm}
\usepackage{graphicx}
\usepackage{color}
\usepackage{longtable}

\newcommand{\dd}{\mathrm{d}}
\newcommand{\Diag}{\mathrm{Diag}}
\newcommand{\Tr}{\mathrm{Tr}}

\newcommand{\bea}{\begin{eqnarray*}}
\newcommand{\eea}{\end{eqnarray*}}
\newcommand{\Ntot}{N_{\textrm{tot}}}

\newcommand{\beq}{\begin{equation}}
\newcommand{\eeq}{\end{equation}}

\newcommand{\rk}{\mathrm{k}}

\newcommand{\bx}{\mathbf{x}}
\newcommand{\bq}{\mathbf{q}}
\newcommand{\bk}{\mathbf{k}}

\newcommand{\bI}{\mathbbm{1}}
\newcommand{\bC}{\mathbf{C}}
\newcommand{\bP}{\mathbf{P}}
\newcommand{\bX}{\mathbf{X}}
\newcommand{\bY}{\mathbf{Y}}
\newcommand{\bA}{\mathbf{A}}

\newcommand{\cT}{\mathcal{T}}
\newcommand{\cM}{\mathcal{M}}
\newcommand{\cO}{\mathcal{O}}
\newcommand{\cA}{\mathcal{A}}
\newcommand{\cB}{\mathcal{B}}

\newcommand{\dT}{\mathbb{T}}
\newcommand{\dR}{\mathbb{R}}

\newcommand{\bi}{\begin{itemize}}
\newcommand{\ei}{\end{itemize}}
\newcommand{\bv}{\begin{verbatim}}
\newcommand{\ev}{\end{verbatim}}
\newcommand{\av}[1]{$\langle #1 \rangle$}

\definecolor{gray}{rgb}{0.9,0.9,0.9}

\newcommand{\R}{\mathbb{R}}

\newcommand{\point}[1]{\textbf{#1}}

\graphicspath{{Plot/}}

\title{Causal Dynamical Triangulations in Four Dimensions}
\author{Atg}

\begin{document}

\thispagestyle{empty}
\begin{center}
\vspace*{-20mm}
{\bf \Large Jagiellonian University}\\[5mm]
{\bf \large Faculty of Physics, Astronomy and Applied Computer Science}\\[8mm]
\includegraphics[width=30mm]{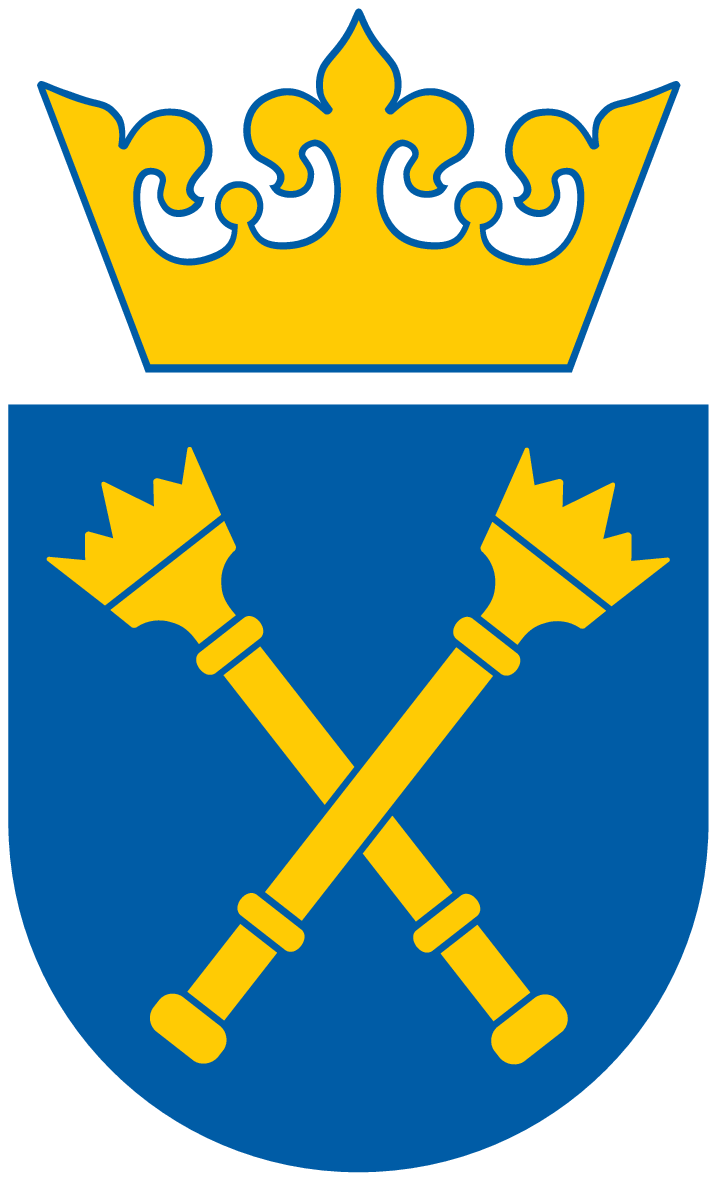}\\[25mm]
{\bf \Huge Causal Dynamical Triangulations\\[1ex] in Four Dimensions}\\[7mm]
{by}\\[7mm]
{\bf \LARGE Andrzej G\"orlich}\\[30mm]
{
\bf 
\large
Thesis written under the supervision of Prof. Jerzy Jurkiewicz,\\
presented to the Jagiellonian University\\
for the PhD degree in Physics
}

\vfill
{\large \bf Krakow 2010}
\end{center}

\newpage

\thispagestyle{empty}
\mbox{}

\newpage

\tableofcontents

\chapter*{Preface}
  \addcontentsline{toc}{chapter}{Preface}
\markboth{Preface}{Preface}

To reconcile classical theory of gravity with quantum mechanics
is one of the most challenging
problems in theoretical physics.
Einstein's \emph{General Theory of Relativity},
which supersedes the  Newton's law of universal gravitation, known since 17$^{\textrm{th}}$ century,
is a geometric theory perfectly describing gravitational interactions 
observed in the macroscopic world.
On the other hand \emph{Quantum Mechanics} 
is indispensable for a description of microscopic physics.
At the energy scales corresponding to the 
Planck length $\ell_{Pl}$, due to the \emph{uncertainty principle}, 
both quantum effects and gravitational interactions must be taken into account
enforcing the necessity of quantizing also the gravitational field\footnote{The Planck 
length $\ell_{Pl} = \sqrt{\frac{\hbar G}{c^3}} \approx 1.6 \times 10^{-35}\ m$ where
$c$ is the speed of light in vacuum, $G$ is the gravitational constant, and $\hbar$ is the reduced Planck constant.}.

The efforts to quantize directly the theory of gravitation
using perturbative expansion meet serious theoretical problems, if such a theory is
to be viewed as a fundamental theory and not only an effective theory.
General relativity without matter field is perturbatively non-renormalizable at two loops
\cite{tHooft, Goroff}.
Newton's constant $G$,
which plays a role of the coupling constant of gravity,
has a dimension $[G] = -2$ in mass units and thus, as argued by Heisenberg,  
makes the theory power-counting non-renormalizable.
Modification of the theory by adding higher derivative terms to the action
can make it renormalizable but spoils the unitarity or the spacetime diffeomorphism invariance.

A way to deal with this problem
is to go beyond a conventional quantum field theory.
\emph{String Theory} attempts to unify all known interactions,
the strong, weak, electromagnetic and also gravitation,
treating general relativity as an effective low-energy limit of some fundamental theory \cite{Polczynski}.

Another set of theories postulate that quantum theory of gravitation
has to be treated non-perturbatively.
\emph{Loop Quantum Gravity} is an approach to quantum gravity, 
which introduces new ways of treating gravity at the
Planck scale by implementing the Dirac's procedure
of canonical quantization to general relativity \cite{Ashtekar,Lewandowski,Rovelli}.
\emph{Spin Foams} \cite{Perez} may be viewed as path integral formulation of 
Loop Quantum Gravity thus gaining some similarities with 
the model of \emph{Causal Dynamical Triangulations} (CDT), which is the 
subject of this thesis.

From Loop Quantum Gravity originates a  reduced 
model called \emph{Loop Quantum Cosmology} \cite{Bojowald}.
In the simplest case of isotropic geometries 
the model describes for example evolution of the scale factor
using the effective Hamiltonian.
Although in Causal Dynamical Triangulations we look
at the same observable, there is a significant difference between the two approaches.
In CDT we analyze the full quantum geometry and at the end we integrate out 
all degrees of freedom except the scale factor rather than introducing a reduction.

The class of lattice approaches to quantum gravity may be divided into two groups.
The first is based on fixed triangulations, containing  the
Regge-Calculus with edge-lengths as the dynamical variables \cite{HamberWilliams1, HamberWilliams2, Wil1}.
The second is using random triangulations,
and includes Causal Dynamical Triangulations with fixed link lengths.
The approach of Dynamical Triangulations started with two-dimensional Dynamical Triangulations,
a model of discrete random surfaces which appeared in various aspects.
Let us mention only a small number of them.
Its history begins with a combinatorial approach studied by Tutte\cite{Tutte}.
By 't Hooft it was put in the context of two-dimensional \emph{Random Matrix Models} in the large $N$-limit and \emph{Quantum Chromodynamics} \cite{tHooft2}.
This allowed to make a connection with a two-dimensional Euclidean quantum gravity \cite{David, Kazakov},
also coupled to matter fields \cite{Diseases, Kazakov2}.
Although analytical tools proved to be very powerful,
there was still a problem with the definition of time which appeared to have a different scaling than the space.
This inspired Ambj\o{}rn and Loll to impose the causality condition which led to a formulation of the two-dimensional Causal Dynamical Triangulations 
\cite{CDT1}.
The continuum limit of this model predicted the same two-loop amplitude as 
in the two-dimensional pure gravity theory obtained in
the proper-time gauge in the continuum formulation \cite{Nakayama}.
However, even in two dimensions incorporation of matter fields was possible only by the use of numerical methods.
The three-dimensional generalization was also widely studied \cite{CDT3D1, CDT3D2},
such models reveal a phase diagram with two phases similar to 
phases $A$ and $C$ of the four-dimensional model.

In this dissertation we discuss the four-dimensional model of Causal Dynamical Triangulations
without matter fields. 
It is a mundane lattice approach to quantum gravity using only standard
quantum field theory and piecewise linear manifolds as a regularization,
refraining from invoking exotic ingredients or excessive fine-tuning.
There are premises 
based on the \emph{Renormalization Group} (RG) approach
that the \emph{asymptotic safety} scenario,
first justified by Weinberg \cite{Weinberg}, postulating the existence of a 
non-Gaussian ultra-violet fixed point is realized \cite{Kawai, Reuter, Codello, HamberRG}.
The presented model attempts to define a non-perturbative
quantum field theory which has as its infrared limit
general relativity and on the other hand a nontrivial ultraviolet limit,
which is in the spirit of the renormalization group approach,
although the tools used by CDT are distinct from RG techniques.
This indicates that lattice gravity theory may play the same role for quantum
gravity as lattice field theory plays for quantum field theory.

{\it Subject of the thesis.}
The purpose of this thesis is to present recent results obtained in
the framework of four-dimensional model of Causal Dynamical Triangulations.
In particular we give answers to questions like~: \emph{how} does a background geometry emerge dynamically,
\emph{what} does it correspond to and \emph{how} to describe quantum fluctuations around the average geometry. 
\newpage
{\it Structure of the thesis. }
The thesis is organized as follows.

In Chapter \ref{Chap:Intro} we introduce methods of four-dimensional Causal Dynamical Triangulations
and derive the foundations needed for further numerical and analytical computations.
At the end of the \emph{Introduction},
the author presents a list of results, where he participated during
the preparation of this thesis.

The logic of  issues presented in the next four Chapters
corresponds to a structure from \emph{top} to \emph{bottom}:
we begin with a discussion of global macroscopic four-dimensional properties at long distances,
and end with short-range fractal structure of slices.

After reviewing in Chapter \ref{Chap:PhaseDiagram} the phase diagram
of the theory we demonstrate the relation with Ho\v{r}ava-Lifshitz gravity 
and the physical importance of the de Sitter space.

In Chapter \ref{Chap:Macroscopic} we prove that a background geometry 
which emerges dynamically corresponds to the maximally symmetric solution
of the \emph{minisuperspace} model, namely the Euclidean de Sitter space.
We study in detail the emerged geometry and show 
that it is genuinely four-dimensional and
that in terms of lattice spacing it resembles an elongated spheroid.

The work presented in Chapter \ref{Chap:Quantum} aims to reconstruct
the effective action describing quantum fluctuations of the scale factor around the semiclassical average.
The resulted action is shown to agree with the discretization of the minisuperspace action.

Chapter \ref{Chap:Slice} contains results of measurements of
Hausdorff dimension and spectral dimension
limited to  hypersurfaces of constant time.
Here we give a direct evidence for a fractal geometry of spatial slices.

In the last Chapter \ref{Chap:Implementation} we describe in detail 
the Monte Carlo algorithm, 
used to obtain  results presented in this work,
and its implementation.

Finally, in Conclusions we briefly discuss the main features of the four-dimensional model of Causal Dynamical Triangulations.

\clearemptydoublepage

\chapter{Introduction to Causal Dynamical Triangulations}
\label{Chap:Intro}

The method of \emph{Causal Dynamical Triangulations} (CDT) 
is a non-perturbative and background independent approach to quantum theory of gravity.
This model was proposed some years ago by J. Ambj\o rn, J. Jurkiewicz and R. Loll
with the aim of defining a lattice formulation of quantum gravity from first principles \cite{Ajld4, Reco, Dyna, Art}.
The foundation 
of the model reviewed in this dissertation
is the formalism of path-integrals applied to quantize a theory of gravitation.
The standard way to define the quantum mechanical Feynman's path integral 
consist in introducing a discreteness of the time coordinate 
$t_i = \varepsilon \cdot i$ (for $i = 1, \dots, N, \ \varepsilon = \frac{T}{N}$)
along non-classical trajectories of a particle
which allows to express the path integral by means of
$N$ ordinary integrals over particle positions
and finally taking the continuum limit $N \to \infty$ \cite{FeynmanHibbsPI, KleinertPI}.

The Causal Dynamical Triangulations method is a natural 
generalization of this discretization procedure to higher dimensions.
In the path integral formulation of quantum gravity,
the role of a particle trajectory plays 
the geometry of four-dimensional spacetime.
Causal Dynamical Triangulations
provide an explicit recipe of calculating the path integral
and specify the class of virtual geometries
which should be superimposed in the path integral.
K. Wilson highlighted the significance of
lattice field theory as an underlying non-perturbative definition of continuum 
quantum field theory \cite{Wilson}.
Following this route, 
we hope that the lattice technique
using causal dynamical triangulations as a regularization
has the potential to play the same role in quantum gravity.
Let us emphasize that no ad hoc discreetness of spacetime is assumed from the outset,
and the discretization appears only as a regularization,
which is intended to be removed in the continuum limit.
The presented approach, has the virtue
that it allows quantum gravity to be
relatively easily represented and studied by computer simulations.

{\bf The action.}
Classical theory of gravitation, General Relativity,
in contrast with other known interactions describes the 
dynamics of spacetime geometry where the 
considered degree of freedom is the metric field $g_{\mu \nu}(x)$.
The nonvanishing curvature of the underlying spacetime geometry
is interpreted as a gravitational field.
The starting point for construction 
of the quantum theory of gravitation is the classical Einstein-Hilbert action
($\{-,+,+,+\}$ signature and sign convention as in \cite{Hawking, Gibbons})
\begin{equation}
 S_{EH}[g_{\mu \nu}] = \frac{1}{16 \pi G} \int_\cM \dd^4 x \sqrt{- \det g} \left(R - 2 \Lambda\right),
 \label{Eq:SEH}
\end{equation}
where $G$ and $\Lambda$ are respectively the Newton's gravitational constant and the cosmological constant,
$\cM$ is the spacetime manifold equipped with a pseudo-Riemannian metric $g_{\mu \nu}$
with Minkowskian signature $\{-, +, +, +\}$
and $R$ denotes the associated Ricci scalar curvature \cite{Misner, Wald}. 
Throughout this thesis we will use the \emph{natural Planck units} $c = \hbar = 1$,
whereas the dependence on Newton's gravitational constant $G$ will be kept explicit.
We shall consider only closed manifolds,
or more specifically, assume that the topology of $\cM$ is $S^1 \times S^3$,
and thus we pass over the Gibbons-–Hawking-–York boundary term.

{\bf Partition function.}
Path-integrals are one of the most important tools used for 
the quantization of classical field theories.
The path integral or partition function of quantum gravity is defined 
as a formal integral over all spacetime geometries, also called histories,
\begin{equation}
 Z = \int \mathcal{D_M}[g]\, e^{i S_{EH}[g]}. 
\label{Eq:ZCont}
\end{equation}
In this expression we should integrate over geometries $[g]$, 
i.e. equivalence classes of spacetime metrics $g$ 
with respect to the diffeomorphism group $Diff_\cM$ on $\cM$.
When integrating over metrics, 
one needs to divide out the volume of $Diff_\mathcal{M}$,
\beq
\mathcal{D_M}[g] = \frac{\mathcal{D_M} g}{Vol[Diff_\mathcal{M}]} .
\label{Eq:MeasureCont}
\eeq
The partition function (\ref{Eq:ZCont}) has a hidden dependence on 
coupling constants $G$ and  $\lambda$.
Knowledge of the generating function $Z$ of a quantum field theory,
after inclusion of the source terms,
allows to calculate all correlation functions,
i.e. vacuum expectation values of products of field operators,
and provides the  complete information about the theory.

\begin{figure}[t!]
\begin{center}
\includegraphics[width=0.5\textwidth]{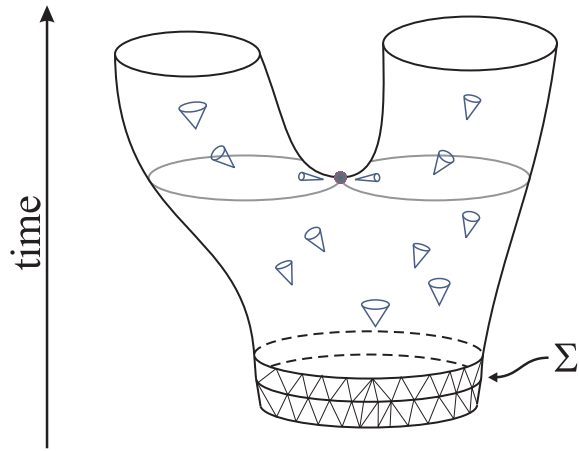}
\end{center}
\caption{
A visualization of a two-dimensional triangulation
with a light-cone structure and a branching point marked.
In Causal Dynamical Triangulations spatial slices are not allowed to split,
which prevents singularities of the time arrow.
}
\label{Fig:Branching}
\end{figure}

{\bf Causality.}
The underlying assumption of CDT is causality condition,
which as we shall see,
will have a significant impact on
desirable properties of the theory.
In \cite{Tei1, Tei2} Teitelboim
advocated the requirement of causality by demanding that  
\emph{only those histories for which the final three-geometry
lies wholly in the future of the initial one} should contribute
to the path integral (\ref{Eq:ZCont}).
The approach of Causal Dynamical Triangulations originates
from this doctrine.
In a gravitational path integral with the correct, Lorentzian
signature of spacetime one should sum over causal geometries only.
As an implication of causality,
we will consider only globally hyperbolic 
pseudo-Riemannian manifolds,
which allow introducing a global proper-time foliation.
The leaves of the foliation are spatial three-dimensional Cauchy surfaces $\Sigma$.
Because topology changes of the spatial slices are often associated with causality violation,
we forbid the topology of the leaves to alter in time.
Fig \ref{Fig:Branching} illustrates a triangulation with imposed foliation
which violates the causality condition. 
The spacetime topology may be written as a product $\cM = I \times \Sigma$,
where $I$ denotes an interval. 
For simplicity, we chose 
the spatial slices to have a fixed topology $\Sigma = S^3$, that of a three-sphere.
Moreover, we establish periodic boundary 
conditions in the time direction.
Therefore, we assume spacetime topology to be $\cM = S^1 \times S^3$,
where $S^1$ corresponds to time and $S^3$ to space.
Such selection of $\cM$ obviates the discussion of boundary conditions for the Universe.

\section{Causal triangulations}
The functional integration in (\ref{Eq:ZCont}) is somewhat formal.
To make sense of the gravitational path integral,
the Causal Dynamical Triangulations model uses a standard method of regularization,
and replaces the path integral over geometries by 
a sum over a discrete set $\dT$ of all causal triangulations $\cT$.
In other words Causal Dynamical Triangulations serve as a regularization 
of smooth spacetime histories present in the formal
path integral (\ref{Eq:ZCont})
with piecewise linear manifolds.
Let us now define in greater detail, what we understand 
as a \emph{causal triangulation}.

\begin{figure}[t!]
\begin{center}
\includegraphics[width=0.99\textwidth]{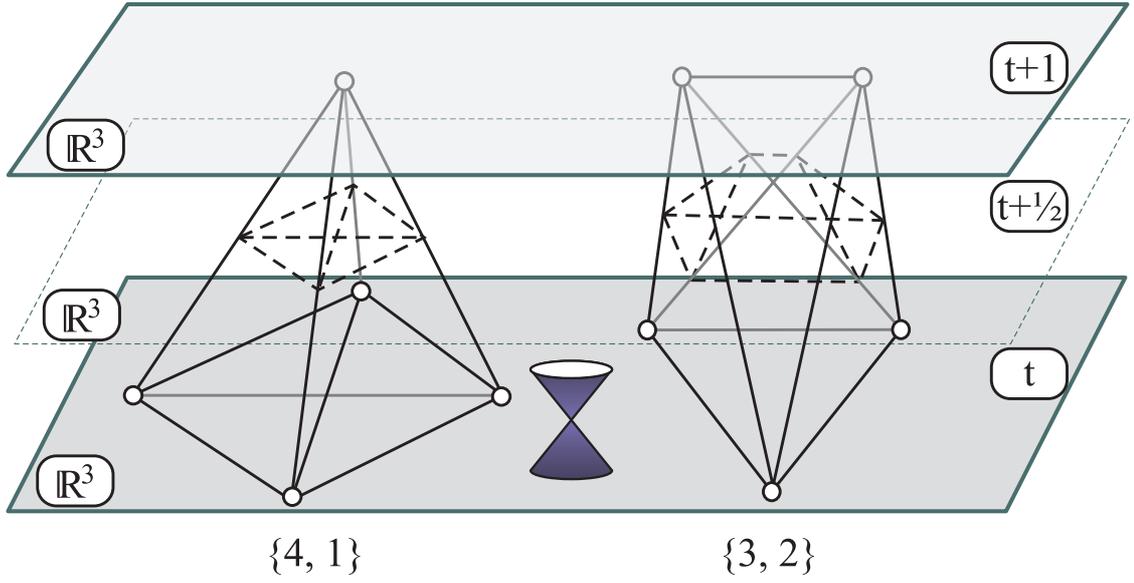}
\end{center}
\caption{
A visualization of fundamental building blocks of four-dimensional
Causal Dynamical Triangulations - four-simplices (solid line).
The simplices join two successive slices $t$ and $t+1$,
and are divided into two types:
$\{4, 1\}$ and $\{3, 2\}$.
The simplices are equipped with the flat Minkowski metric
imposing the light-cone structure (blue drawing).
The dotted line illustrates the cross section of a simplex
by a three-dimensional hyperplane placed between the slices,
at \emph{time} $t+1/2$.
}
\label{Fig:Sympleksy}
\end{figure}

The building blocks of four dimensional Causal Dynamical Triangulations are four-simplices.
A simplex is a generalization of a triangle, 
which itself is a two-dimensional simplex, 
to higher dimensions.
An $n$-simplex is an $n$-dimensional polytope  with $n + 1$ vertices.
Concerning a $n$-simplex, sub-simplices of dimension $n-1$ are called \emph{faces},
while $n-2$ dimensional sub-simplices are called \emph{hinges} or \emph{bones}.
Each four-dimensional simplex is composed of five vertices (0-dimensional simplices), 
all of them connected to each other.
It consists of 10 links (1-dimensional simplices) and of 10 triangles.
The boundary of a simplex is built of five tetrahedral faces (3-dimensional simplices).
Each four-simplex is taken to be a subset of a four-dimensional Minkowski 
spacetime together with its inherent light-cone structure
thus the metric inside every simplex is flat.
Fig. \ref{Fig:Sympleksy} presents a visualization
of four-simplices together with a  light-cone sketch.
As will be explained later, there are two types of simplices.

A $n$-dimensional simplicial manifold, with a given topology, 
is obtained by properly gluing pairwise $n$-simplices along common faces.
The neighborhood of each vertex (i.e. set of simplices sharing this vertex)
should be homeomorphic to a $n$-dimensional ball.
In this thesis we only consider four-dimensional simplicial manifolds topologically isomorphic
with $\cM = S^1 \times S^3$.
A simplicial manifold takes over a metric from simplices
of which it is built and which are equipped with a flat Minkowski metric.
A simplicial manifold armed with a metric is called a piecewise linear space.
In general such $n$-dimensional complex 
can not be embedded in $\dR^n$ which signifies a nonvanishing curvature.
The curvature is singular and localized only at hinges,
which in the four dimensional case correspond to triangles.

As a consequence of the \emph{causality} requirement
we consider only globally hyperbolic manifolds
with a proper-time foliation structure.
In the Causal Dynamical Triangulations approach 
the spatial leaves of the foliation are 
called \emph{slices} and are enumerated by a discrete \emph{time} coordinate $i$. 
At each integer proper-time step $i$,
a spatial slice itself forms a closed three-dimensional piecewise linear manifold $\cT^{(3)}(i)$ with a fixed topology of a three-sphere $S^3$
and induced metric which has a Euclidean signature.
It is represented by a triangulation of $S^3$,
made up of equilateral tetrahedra with a side length $a_s > 0$.

Two successive slices, e.g. at time steps $i = t$ and $i = t + 1$,
given respectively by triangulations $\cT^{(3)}(t)$ and $\cT^{(3)}(t + 1)$,
are connected with four-simplices.
The simplices are joined to create 
a four-dimensional piecewise linear geometry.
Such object takes a form of a four-dimensional \emph{slab}
with a topology of $[0,1] \times S^3$ and
has $\cT^{(3)}(t)$ and $\cT^{(3)}(t + 1)$ as the three-dimensional boundaries. 
A set of \emph{slabs} glued one after another builds the whole simplicial complex.
Such connection of two consecutive slices,
by interpolating the \emph{"space"} between them with properly glued four-simplices,
does not spoil the \emph{causal structure}.
Null rays originating from tetrahedra in a slice $t$ 
and propagating through successive flat simplices with Minkowski metric,
always finally reach a slice $t + 1$ directly (without crossing other slices).  
This is true even for very twisted \emph{slabs}.
The triangulation of the later slice wholly lies in the future of the earlier one.

Each vertex of a four-dimensional triangulation is assigned a discrete time coordinate $i$
corresponding to the slice it belongs to.
Because each simplex connects two consecutive spatial slices
and contains vertices lying in both of them,
there are four kinds of simplices: 
$\bullet$ type $\{4, 1\}$ with four vertices lying in a spatial slice $\cT^{(3)}(t)$
and one in the next slice $\cT^{(3)}(t + 1)$,
$\bullet$ type $\{1, 4\}$ obtained by interchanging the role of $\cT^{(3)}(t)$ and $\cT^{(3)}(t + 1)$,
i.e. with one vertex in the earlier slice and four in the next slice.
$\bullet$ type $\{3, 2\}$ with three vertices lying in a spatial slice $\cT^{(3)}(t)$ and two in the next slice $\cT^{(3)}(t + 1)$,
$\bullet$ type $\{2, 3\}$, analogically, defined by interchanging $\cT^{(3)}(t)$ and $\cT^{(3)}(t + 1)$.
The tetrahedra of $\cT^{(3)}(t)$ are bases of four-simplices of type $\{1, 4\}$ and $\{4, 1\}$.
Sometimes we will not distinguish between types $\{1, 4\}$ or $\{4, 1\}$, and treat them as a common type $\{4, 1\}$,
the same concerns $\{2, 3\}$ and $\{3, 2\}$.
Fig. \ref{Fig:Sympleksy} illustrates
four-simplices of type $\{4, 1\}$ and $\{3, 2\}$ connecting slices $t$ and $t + 1$.
The simplices, as a subset of the flat  Minkowski space, are equipped with the light-cone structure.

Similarly, due to the causal structure, we distinguish two types of edges.
The space-like links connect two vertices in the same slice,
they have length $a_s > 0$.
The time-like links connect two vertices in adjacent slices
and have length $a_t$.
There occur no other types of links.
In Causal Dynamical Triangulations, the lengths $a_s$ and $a_t$ 
are constant but not necessarily equal to each other.
Two different triangulations, indeed, correspond to two inequivalent geometries.
By contrast, in the original Regge's model \cite{Regge, Wil1},
the lengths of individual edges may vary,
however, the gauge freedom in the edge lengths may result in an over counting 
of some triangulations.
Let us denote the asymmetry factor between the two lengths by $\alpha$
\beq
a_t^2 = \alpha \cdot a_s^2.
\label{Eq:AtAs}
\eeq
In the Lorentzian case $\alpha < 0$.
The volumes and angles of simplices are functions of $a_s$ and $a_t$
and differ for the two types $\{4, 1\}$ and $\{3, 2\}$.
Because no coordinates are introduced, 
the CDT model is manifestly diffeomorphism-invariant.
Such a formulation involves only geometric invariants like lengths and angles. 
The exact relations are derived in the Appendix A.

Besides the slices with integer time coordinate,
which are built only of equilateral tetrahedra,
we may also introduce slices with half-integer index $i = t + \frac{1}{2}$.
A horizontal section of a $\{4, 1\}$ simplex with a three-dimensional hyperplane
at half of its height is a tetrahedron.
Similarly, by cutting $\{3, 2\}$ with a hyperplane,
we get a triangular prism as a boundary surface.
The dotted lines shown on Fig. \ref{Fig:Sympleksy}
illustrate the horizontal section for both kinds of simplices. 
A half-integer slice constructed of the two types of solids
is also a closed three-dimensional piecewise linear manifolds and
has a topology of three-sphere $S^3$.

\section{The Regge action and the Wick rotation}

The Einstein-Hilbert action (\ref{Eq:SEH}) has a natural realization
on piecewise linear manifolds called the Regge action.
Hereafter, we will denote the number of $k$-dimensional sub-simplices by $N_k$.
Similarly, let $N_{41}$ mean the number of simplices of type $\{4, 1\}$,
and $N_{32}$ the number of simplices of type $\{3, 2\}$.
Because we distinguish only two types of simplices, 
they have to sum up to the total number of simplices,
\[ N_4 = N_{41} + N_{32}.\]
The total physical four-volume of a simplicial manifold $\cT \in \dT$, is given by
\[ V_4 \equiv  \int_{\cT} \dd^4 x \sqrt{|\det g|} = N_{41} \textrm{Vol}^{\{4, 1\}} + N_{32} \textrm{Vol}^{\{3, 2\}}, \]
where $\textrm{Vol}^{\{4, 1\}} $ is a volume of a $\{4, 1\}$-simplex and 
$\textrm{Vol}^{\{3, 2\}}$ is a volume of a $\{3, 2\}$-simplex.
Both quantities are purely geometric and are proportional to $a_s^{\,4}$ and depend on $\alpha$.
Similarly, it can be shown that the global curvature 
\[ \int_{\cT} \dd^4 x \sqrt{|\det g|} R \]
can be expressed using the angle deficits which are localized at triangles,
and is a linear function of total volumes $N_{32}$, $N_{41}$ and the total  number of vertices $N_0$.
Based on the above arguments, the Regge action, 
calculated for a causal triangulation $\cT$,
can be written in a very simple form,
\begin{equation}
S[\cT] \equiv - K_0\ N_0[\mathcal{T}] + K_4\ N_4[\mathcal{T}]  + \Delta\ (N_{41}[\mathcal{T}] - 6 N_0[\mathcal{T}]),	
\label{Eq:SRegge}
\end{equation}
where $K_0$, $K_4$ and $\Delta$ are bare coupling constants,
and naively they are functions of $G, \lambda$ and $a_t, a_s$.
$K_4$ plays a similar role as a cosmological constant,
it controls the total volume.
$K_0$ may be viewed as inverse of the gravitational coupling constant $G$.
$\Delta$ is related to the asymmetry factor $\alpha$ between lengths time-like and spatial-like links.
It is zero when $a_t = a_s$ and does not occur in the Euclidean Dynamical Triangulations.
$\Delta$ will play an important role as it will allow to observe new phases.
The derivation of the Regge action and 
explicit expressions for the coupling constants is comprised in the Appendix A.

{\bf Discrete partition function.}
Causal Dynamical Triangulations provide a 
regularization method of histories
appearing in the formal gravitational path integral (\ref{Eq:ZCont}).
The integral is now discretized by replacing it with
a sum over the set of all causal triangulations $\dT$ weighted with the Regge action (\ref{Eq:SRegge}),
providing a meaningful definition of the partition function,
\begin{equation}
Z \equiv \sum_{\mathcal{T} \in \mathbb{T}} \frac{1}{C_\mathcal{T}} e^{i S[\mathcal{T}]}.
\label{Eq:ZDiscLor}
\end{equation}

{\bf Symmetry factor.}
The fraction $1 / C_\mathcal{T}$ is a symmetry factor,
given by the order $C_\mathcal{T}$ of the automorphism group of a triangulation $\mathcal{T}$.
It might be viewed as the remnant of the division by 
the volume of the diffeomorphism group $Diff_\mathcal{M}$
present in a formal gauge-fixed continuous expression (\ref{Eq:ZCont}) for the partition function $Z$.
The factor $1 / C_\mathcal{T}$ introduces the measure on the set of geometries,
in the same way as does (\ref{Eq:MeasureCont}) in the continuous case.
In fact, we do not even have a mathematical
characterization which geometries should contribute to the path integral.
Therefore, the measure in (\ref{Eq:ZDiscLor}) is an attempt
to define the quantum theory of gravitation.
The next assumption is that piecewise linear geometries 
appearing in CDT are a dense subset in the set of
geometries relevant for the path integral.
There is no straightforward method to calculate $C_\mathcal{T}$ for a general triangulation.
However, we can easily deal with the factor $C_\mathcal{T}$ by replacing 
the sum over \emph{unlabeled triangulations} $\mathcal{T} \in \mathbb{T}$ in (\ref{Eq:ZDiscLor})
with a sum over \emph{labeled triangulations} $\tilde{\mathcal{T}} \in \tilde {\mathbb{T}}$.
To each vertex of a triangulation $\tilde{\mathcal{T}}$ is assigned a unique label
(e.g. an element of a set $\{ v_i \} = \{1, \dots, N_0[\tilde{\mathcal{T}}]$\}).
A triangulation is completely defined by the adjacency relations
of simplices and all sub-simplices in terms of the vertex labels,
preserving topological restrictions.
Two triangulations defined by the same set of vertex labels $\{v_i\}$,
are isomorphic if there exists a bijective map $\phi : \{v_i\} \to \{v_i\}$,
which maps neighbors into neighbors.
The factor $C_\mathcal{T}$ is defined as the number
of such maps - among all $N_0!$ permutations, 
they generate the same set of neighbors,
i.e. the same labeled triangulation.
The number of labeled triangulations isomorphic 
with an unlabeled triangulation $\mathcal{T}$ 
(we may dress it with labels $\{ v_i \}$ in a random way
and apply the above definition of isomorphism class)
is denoted as $\mathcal{N}[\mathcal{T}]$,
in other words it is the number of different labelings of $\mathcal{T}$.
From the definition of $C_\mathcal{T}$ and $\mathcal{N}[\mathcal{T}]$
follows,
\[ C_T \mathcal{N}[\mathcal{T}] = N_0 [\mathcal{T}] ! .\]
The partition function (\ref{Eq:ZDiscLor}) can be written as a sum over labeled causal triangulations,
\begin{equation}
Z = \sum_{\mathcal{T} \in \mathbb{T}} \frac{1}{C_\mathcal{T}} e^{i S[\mathcal{T}]}  = \sum_{\tilde{\mathcal{T}} \in \tilde {\mathbb{T}}} \frac{1}{N_{0}[\tilde{\mathcal{T}}] !} e^{i S[\tilde{\mathcal{T}}]}.
\label{Eq:ZDiscLorLab}
\end{equation}
The division by the number of vertex labels permutations ensures that we effectively sum over inequivalent parametrizations.
Also from the numerical point of view
it is much easier to consider labeled triangulations.
The computer algorithm, described in Chapter \ref{Chap:Implementation},
works on labeled triangulations and 
takes the symmetry factor $C_T$ into account automatically
\cite{Dyna, QuantumGeometry}.

{\bf Wick rotation}
So far, it was impossible to evaluate the four-dimensional partition function (\ref{Eq:ZCont}) or
even its discrete counterpart (\ref{Eq:ZDiscLorLab}) using purely analytical methods.
We may however resort to numerical methods, namely to Monte Carlo techniques which
allow to calculate expectation values of observables defined on piecewise linear manifolds.
The advantage of the Causal Dynamical Triangulations approach, is that 
for a fixed size of triangulations understood as the number of simplices $N_4$,
the number of combinations is finite,
which in general allows to apply numerical calculations.
Nonetheless, this number grows exponentially with the size,
and the critical exponent coincides with the critical value $K_4^{crit}$ of the bare coupling constant $K_4$.
Because of the oscillatory behavior of the integrand (\ref{Eq:ZCont}) or (\ref{Eq:ZDiscLorLab}),
we are still led into problems in defining the path integral, 
also the mentioned numerical techniques are not useful. 
We may evade this problem by applying a trick called Wick rotation,
which, roughly, is based on the analytical continuation of the time coordinate to imaginary values,
and results in the change of the spacetime signature from Lorentzian to Euclidean
and a substitution of the complex amplitudes by real probabilities,
\beq
e^{i S^{Lor}} \to e^{- S^{Euc}}.
\label{Eq:SWick}
\eeq
In Causal Dynamical Triangulations,
due to the global proper-time foliation,  
the Wick rotation is well defined.
It can be simply implemented
by analytical continuation 
of the lengths of all time-like edges, $a_t \rightarrow i a_t$,
\[ a_t^2 = \alpha \cdot a_s^2, \quad \alpha > 0. \]
This procedure is possible, because we have a distinction between time-like and space-like links.
The Regge action rotated to the Euclidean sector, after redefinition applied in (\ref{Eq:SWick}),
$S^{Euc} = - i S^{Lor}$, has exactly the same simple form as its original Lorentzian version (\ref{Eq:SRegge}).
An exact derivation of the Wick rotated Regge action is to be found in the Appendix A.

{\bf Unboundedness of the Euclidean action.}
It is formally easy to perform the Wick rotation of the continuous 
Einstein-Hilbert action (\ref{Eq:SEH}).
However, the corresponding Euclidean action is unbounded from below,
which is a consequence of the \emph{wrong sign} of the kinetic term
for the conformal mode.
There are configurations with arbitrarily large negative value of the action,
thus the Wick rotated contributions (\ref{Eq:SWick})
to the path integral make it ill-defined.
Also in the discretized model some triangulations may have very large 
negative values of the Regge action, but still finite
due to the UV lattice regularization.
The problem of infinities is revived when taking the continuum limit.
Fortunately, in the non-perturbative approaches, like CDT,
the partition function emerges as a subtle interplay of the entropic nature of triangulations,
determined by the measure independent of \emph{bare} coupling constants,
and the \emph{bare} action.
The entropy factor may suppress the unbounded contributions coming from the conformal factor.
This is exactly what happens in CDT.
Together with a convergence of the coupling constants to their critical values,
if such a point exists, the \emph{entropic} and \emph{action} terms should be balanced,
and one hopes to obtain the proper continuum behavior.

Let us note the importance of the causality assumption.
The Causal Dynamical Triangulations approach arose as a modification
of Euclidean Dynamical Triangulations, where no global foliation was imposed.
This resulted in several problems.
As proved by the computer simulations \cite{Observing},
at each point of the spacetime, the Universe could branch off creating a \emph{baby} Universe.
Inclusion of such degenerate geometries to the path integral, 
does not allow for a well-defined continuum theory.
Because the Euclidean simplicial manifolds are bereft of the light-cone structure,
one does not know how to rotate back to the Lorentzian signature and recover causality in a full quantum theory.
Therefore in CDT we ab initio deal with the Lorentzian simplicial spacetimes in four-dimensions
and insist that only causal well-behaved geometries appear in the regularized path integral.

{\bf Expectation values.}
As a consequence of the regularization procedure and Wick rotation to the Euclidean signature,
the partition function (\ref{Eq:ZCont}) is finally written as a real sum
over the set of all causal triangulations $\dT$ (or labeled triangulations $\tilde{\dT}$),
\begin{equation}
Z = \sum_{\mathcal{T} \in \mathbb{T}} \frac{1}{C_\mathcal{T}} e^{- S[\mathcal{T}]}  = \sum_{\tilde{\mathcal{T}} \in \tilde {\mathbb{T}}} \frac{1}{N_{0}[\tilde{\mathcal{T}}] !} e^{- S[\tilde{\mathcal{T}}]}.
\label{Eq:ZDisc}
\end{equation}
We should keep in mind, that the Euclidean Regge action $S[\cT]$, as well as the partition function $Z$ 
depend on bare coupling constants $K_0$, $K_4$ and $\Delta$.
With the partition function (\ref{Eq:ZDisc}) is
associated a probability distribution on the space of triangulations $P[\cT]$
which defines the quantum expectation value
\beq
 \langle \cO \rangle \equiv \sum_{\mathcal{T} \in \mathbb{T}} \cO[\cT] P[\cT],
\quad P[\cT] \equiv \frac{1}{Z}  \frac{1}{C_\mathcal{T}} e^{- S[\mathcal{T}]},
\label{Eq:ExpVal}
\eeq
where $\cO[\tilde{\dT}] = \cO[\dT]$ denotes some observable.
The above partition function defines a statistical mechanical problem  which is free of oscillations
and may be tackled in an approximate manner using Monte Carlo methods.
Equation (\ref{Eq:ZDisc}) is a starting point for computer simulations,
which further allow to measure expectation values defined by (\ref{Eq:ExpVal})
and to obtain physically relevant information.

\section{The author's contribution to the field}

First, let us briefly summarize earlier results obtained within the CDT framework in four-dimensions.
In the publication \cite{Reco}, the phase structure of the model was examined
and three phases were found and characterized, namely phases $A$, $B$ and $C$.
They are also reported in the Chapter \ref{Chap:PhaseDiagram}.
Phase $C$ became especially interesting from physical point of view.
It was indirectly shown that a semiclassical background is generated dynamically. 
This background geometry corresponds to a four-dimensional de Sitter spacetime \cite{Reco, SemiUni}.
For  that  reason phase $C$ is called also a de Sitter phase.
Among others the spectral dimension, Hausdorff dimension and scaling properties were measured \cite{Spectral}.
The repeated measurements are described in the Section \ref{Sec:FourDimensional}.

Further, the author would like to depict his contribution
to the development of four-dimensional Causal Dynamical Triangulations.
The research described in this dissertation may be divided into two parts.
The first part consisted of developing a computer software package allowing 
to perform numerical calculations within the CDT framework.
The author has created a set of programming tools
allowing to carry out simulations and data analysis.
The main component of the package is 
the code generating statistically independent spacetime configurations according to a probability distribution
(\ref{Eq:ExpVal}) using Monte Carlo techniques.
In general, simulations enable to measure the expectation values defined by (\ref{Eq:ExpVal}).
The algorithm is based on a program written in \emph{FORTRAN} by Prof. Jerzy Jurkiewicz 
in collaboration with Prof. Jan Ambj\o rn and Prof. Renate Loll \cite{Dyna}.
It was critically examined and fully rewritten in \emph{C} by the author.
As a part of the implementation of the Monte Carlo method the author
has introduced and customized efficient algorithms and data structures,
which are the domain of computer science.
The used solutions, improvements and code optimization led to important acceleration of the code
(\emph{ca.} 7 times).
Details of the implementation of the Monte Carlo algorithm and of the simulation process
will be given in the Chapter \ref{Chap:Implementation}.
All numerical results presented in this work
have been obtained by the author 
using computer programs written by himself.
Here the author would like to pay tribute to the tools he used to complete the research.
A vast majority of simulations presented in this work was carried out 
on a cluster at the Institute of Physics of Jagiellonian University.
Parallel computations were performed on up to 64 cores.
Some measurements required in total many CPU years of simulations.
The author used the \emph{GNU Compiler Collection} (gcc) and the
\emph{Intel C++ Compiler} (icc) to compile his source code.
\emph{Wolfram Mathematica} and \emph{Gnutplot} allowed him to perform
data analysis and create plots printed in this thesis.
Some of the pictures were obtained with the help of \emph{Open Graphics Library} (OpenGL)
and \emph{CorelDRAW}.

Before performing measurements, 
one has to propose procedures determining \emph{what} and \emph{how} to measure.
A construction of suitable observables which give physically relevant information
is a highly non-trivial task.
For instance, in quantum gravity one cannot speak about the absolute position of a point on the superposition of spacetimes.
Moreover, they required specific numerical tools necessary to their measurement.

The second part consisted of analysis and physical interpretation of results.
The key element introduced by the author was a procedure
eliminating the time-translational freedom,
which allows to superimpose configurations in an unambiguous way.
The method is based on 
a unique designation of the triangulation center
and is described in detail in Section \ref{Sec:Spatial}.
Breaking of the time-translational symmetry opened the door
to further, more detailed analysis of the de Sitter phase.
It allowed to introduce a new observable, 
namely the spatial volume at specific absolute time coordinate.
Previously one could deal only with relative time positions.
The crucial achievement was to directly show that 
the background four-dimensional de Sitter spacetime geometry emerges dynamically.
In Section \ref{Sec:Spatial} the author proves this statement.
Examining the distribution of spatial volume
for a specific slice, the author determined
which region of the configurations is dominated by discrete effects and which can be treated as continuous.
As shown in Section \ref{Sec:Minisuperspace}, the background geometry corresponds
to the classical solution of a reduced \emph{minisuperspace} model.
The implementation of the Monte Carlo algorithm written by the author,
allowed to repeat the measurements, e.g. of the spectral dimension, 
on larger configurations with higher accuracy.
In Section \ref{Sec:FourDimensional} the author presents the measured values 
of Hausdorff and spectral dimension,
and based on the scaling properties shows that 
indeed the background spacetime is a well-defined four-dimensional Universe.
Furthermore the author modified the method of measuring the spectral dimension
consisting in performing the \emph{diffusion process}
to determine the propagator of scalar fields defined on a triangulation.
This method is more efficient and accurate than the previously used \emph{heat-bath} method.
The results of these studies will be presented in a separate publication.
In addition, improved numerical tools allowed to
investigate phase transitions with increased accuracy and to note the 
correspondence between the phase diagram of CDT and \emph{Lifshitz} models,
which form the basis for so called \emph{Ho\v{r}ava-Lifshitz} gravity.
Details of the correspondence will be given in the Chapter \ref{Chap:PhaseDiagram}.

In Section \ref{Sec:Geometry4D} the author
examined in greater detail the global geometry of spacetime in de Sitter phase.
It was shown, that background geometry resembles a four-dimensional 
spheroid, prolate in time direction, 
whose elongation in terms of lattice spacings depends on values of coupling constants.
The author tested that it is  possible to reach a spherical shape.
Even for an individual configuration the appropriate volume functions
did not deviate substantially from the average,
suggesting finite volume fluctuations around background geometry.
Also almost no fractal structure was observed with respect to the four-dimensional definition of distance.

Another consequence of the elimination of the translational mode 
and the existence of background geometry is the possibility to perform a semiclassical
expansion of the spatial volume around the mean value.
This allowed for the determination of the effective action describing fluctuations of the three-volume.
It should be noted here that we did not make any reduction of degrees of freedom,
as is the case of cosmological models and minisuperspace model,
but we integrated out all degrees of freedom except the scale factor.
Moreover, the presented semiclassical expansion is truly non-perturbative
and takes into account both the \emph{entropy factor} and the \emph{bare action}.
However, it turned out that the resulting effective action shows a remarkable consistency with minisuperspace model.
The author describes these results in Chapter \ref{Chap:Quantum}.

In Chapter \ref{Chap:Slice} the author conducted an in-depth analysis of three-dimensional slices
- the surfaces of constant time with respect to the imposed foliation.
He repeated the measurements of Hausdorff dimension and spectral dimension.
In particular, he directly demonstrated the fractal nature of these slices.

\clearemptydoublepage

\chapter{Phase diagram}
\label{Chap:PhaseDiagram}

{\noindent \it This chapter is based on the following publications:
(i) J. Ambj\o rn, A. G\"orlich, S. Jordan, J. Jurkiewicz and R.~Loll,
  \emph{''CDT meets Ho\v{r}ava--Lifshitz gravity''},
  Phys. Lett. B {\bf 690}, 413 (2010);
(ii)   J. Ambj\o rn, A. G\"orlich, J. Jurkiewicz and R.~Loll,
  \emph{''CDT - an Entropic Theory of Quantum Gravity'',}
	[arXiv:1007.2560].
}

\vspace{4ex}

\begin{figure}[t]
\begin{center}
\includegraphics[width=0.99\textwidth]{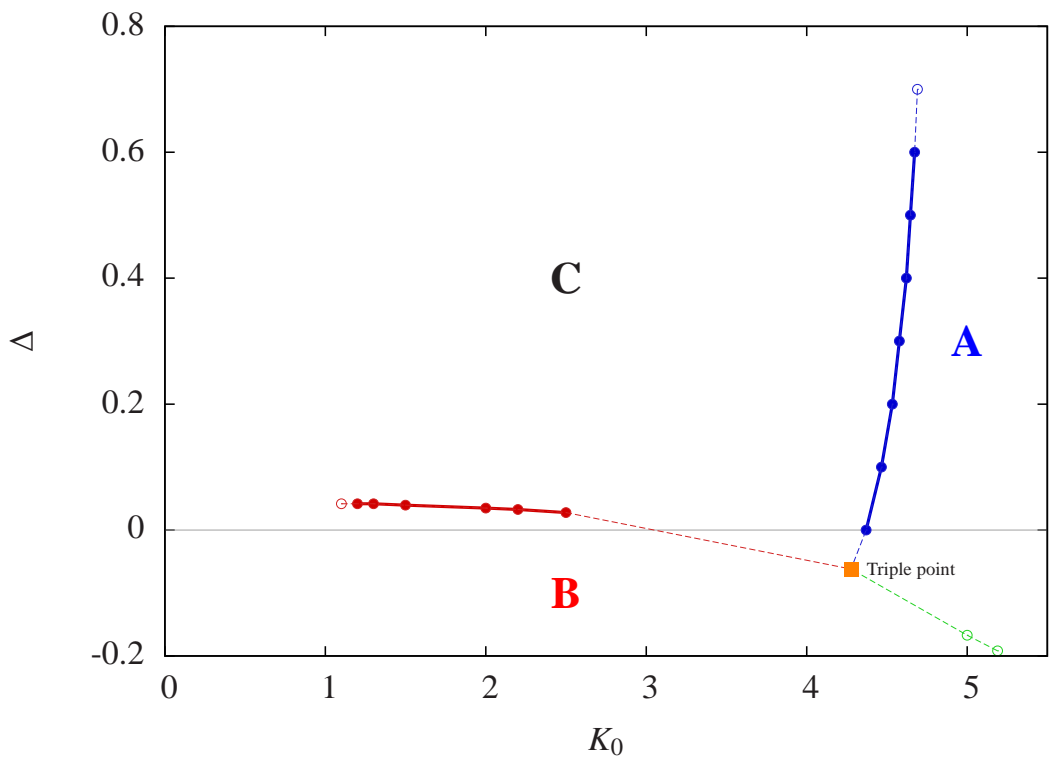}
\end{center}
\caption{
A sketch of the phase diagram of the four-dimensional Causal Dynamical Triangulations.
The phases correspond to regions on the bare coupling constant $K_0-\Delta$ plane. 
We observe three phases:
a \emph{crumpled} phase $A$, a \emph{branched polymer} phase $B$ and the most interesting a genuinely four-dimensional de Sitter phase $C$. 
}
\label{Fig:PhaseDiagram}
\end{figure}

The Causal Dynamical Triangulations model is described by the Regge action (\ref{Eq:SRegge})
which depends on a set of three bare coupling constants $K_0$, $\Delta$ and $K_4$.
All three coupling constants are nonlinear functions of parameters appearing in the
continuous Einstein-Hilbert action, namely $G$ and $\Lambda$, and the asymmetry factor
$\alpha = \frac{a_t^2}{a_s^2}$ which is a regularization parameter.
Explicit relations are derived in the Appendix A.
Parameters have following interpretations:
$K_0$ is proportional to the inverse bare gravitational coupling constant $G$.
$\Delta$ is related to the asymmetry factor between lengths of time-like links $a_t$ and space-like links $a_s$.
It is zero when $a_t = a_s$ and thus is not present in Euclidean Dynamical Triangulations (EDT).
Here, $\Delta$ plays an important role as it allows to observe new phases.
Finally, $K_4$ acts as a cosmological constant $\Lambda$,
it controls the total volume.
For simulation technical reasons it is preferable to keep the total four-volume
fluctuating around some finite prescribed value during Monte Carlo simulations. 
Thus  $K_4$ needs to be tuned to its critical value,
and effectively does not appear as a coupling constant.
The two remaining bare coupling constants $K_0$ and $\Delta$ can be freely adjusted and 
depending on their values we observe three qualitatively different behaviors of a typical configuration.
The phase structure was first qualitatively described in a comprehensive publication \cite{Reco}
where three phases were labeled $A$, $B$ and $C$.
The first real phase diagram obtained due to large-scale computer simulations was described in \cite{CDTHorava}.
The phase diagram, based on Monte Carlo measurements,
is presented in Fig. \ref{Fig:PhaseDiagram}.
The solid lines denote observed phase transition points 
for configurations of size $80000$ simplices,
while the dotted lines represent an interpolation.

In the remainder of this Section we describe the properties of the phases,
and discuss the phase transitions.

\begin{center}
\includegraphics[width=0.53\textwidth]{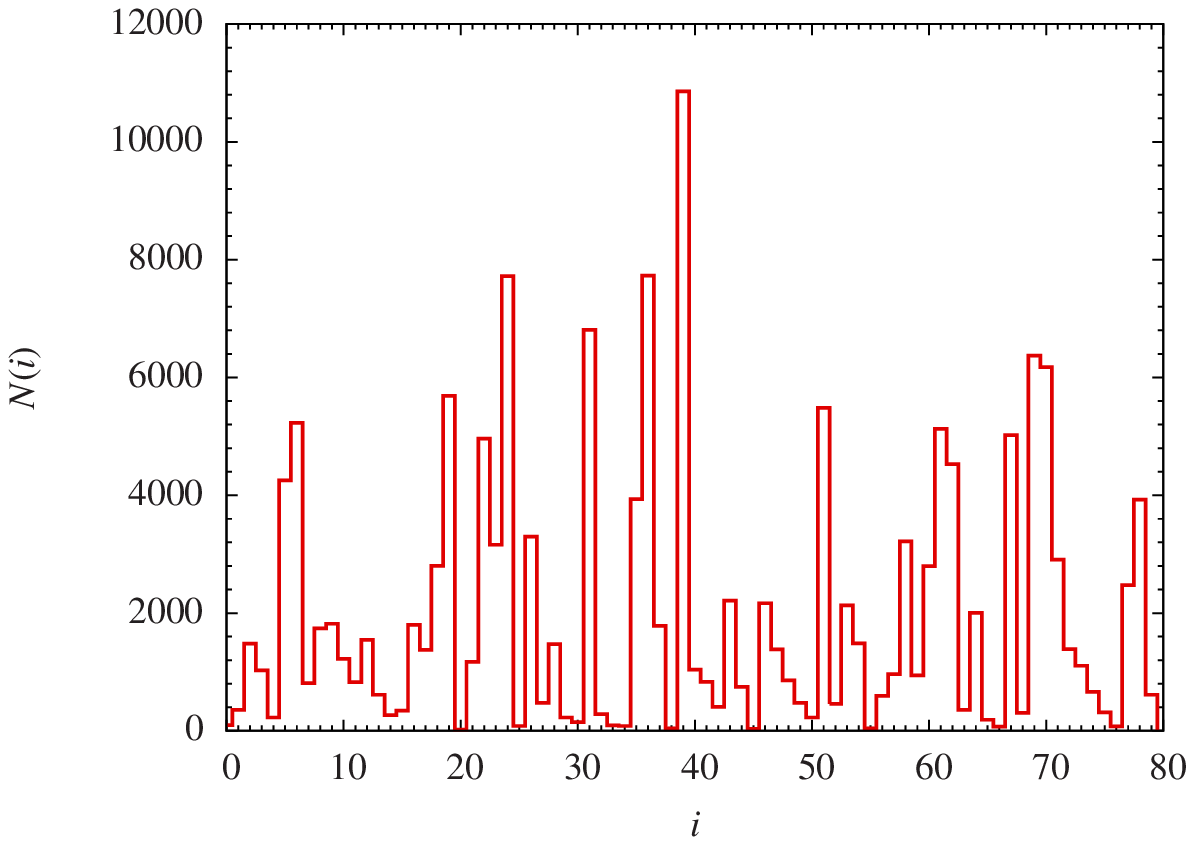}
\raisebox{7mm}{\includegraphics[width=0.46\textwidth]{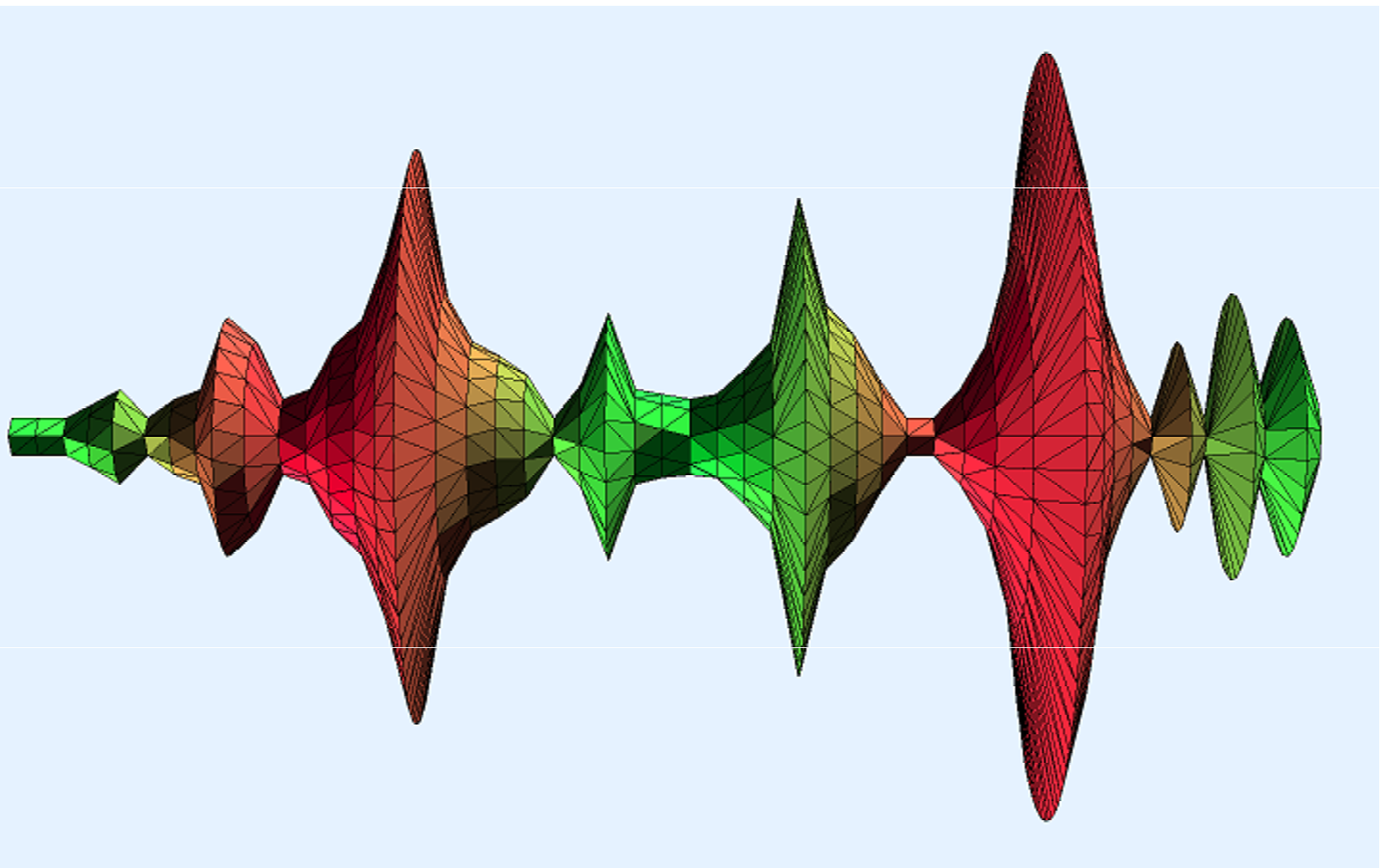}}\\
{\bf Phase A.} 
On the left: Snapshot of spatial volume $N(i)$ for a typical configuration of phase $A$.
The picture on the right presents a two-dimensional visualization of a corresponding triangulation.
The time axis is set horizontally.
\end{center}
\newpage
{\noindent\bf Phase A.}
For large values of $K_0$ (cf. Fig. \ref{Fig:PhaseDiagram}) 
the universe disintegrates into uncorrelated irregular sequences of maxima and minima 
with time extent of few steps.
As an example of a configuration in this phase, the spatial volume distribution $N(i)$,
defined as the number of tetrahedra in a spatial slice labeled by a discrete time index $i$,
is shown in the above figure (on the left).
The picture on the right illustrates a two-dimensional causal triangulation,
embedded in $\dR^3$, with the time axis being horizontal.
In this visualization, slices consist of cyclically connected edges and their circumferences
correspond to spatial volumes.
When looking along the time direction, we observe a number of small universes.
The geometry appears to be oscillating in the time direction.
They can merge and split with the passing of the Monte Carlo time.
These universes are connected by \emph{necks}, corresponding to minima,
i.e. $S^3$ spatial slices not much larger than the smallest possible,
which consists of five tetrahedra glued together.
In the computer algorithm we do not allow these \emph{necks} to \emph{vanish} such that the configuration becomes disconnected. 
This phase is related to so-called \emph{branched polymers phase} present in Euclidean Dynamical Triangulations (EDT)\cite{Thre}.
No spatially nor time extended universe,
like the Universe we see in reality, is observed and the phase $A$ is regarded as non-physical.

\begin{center}
\includegraphics[width=0.53\textwidth]{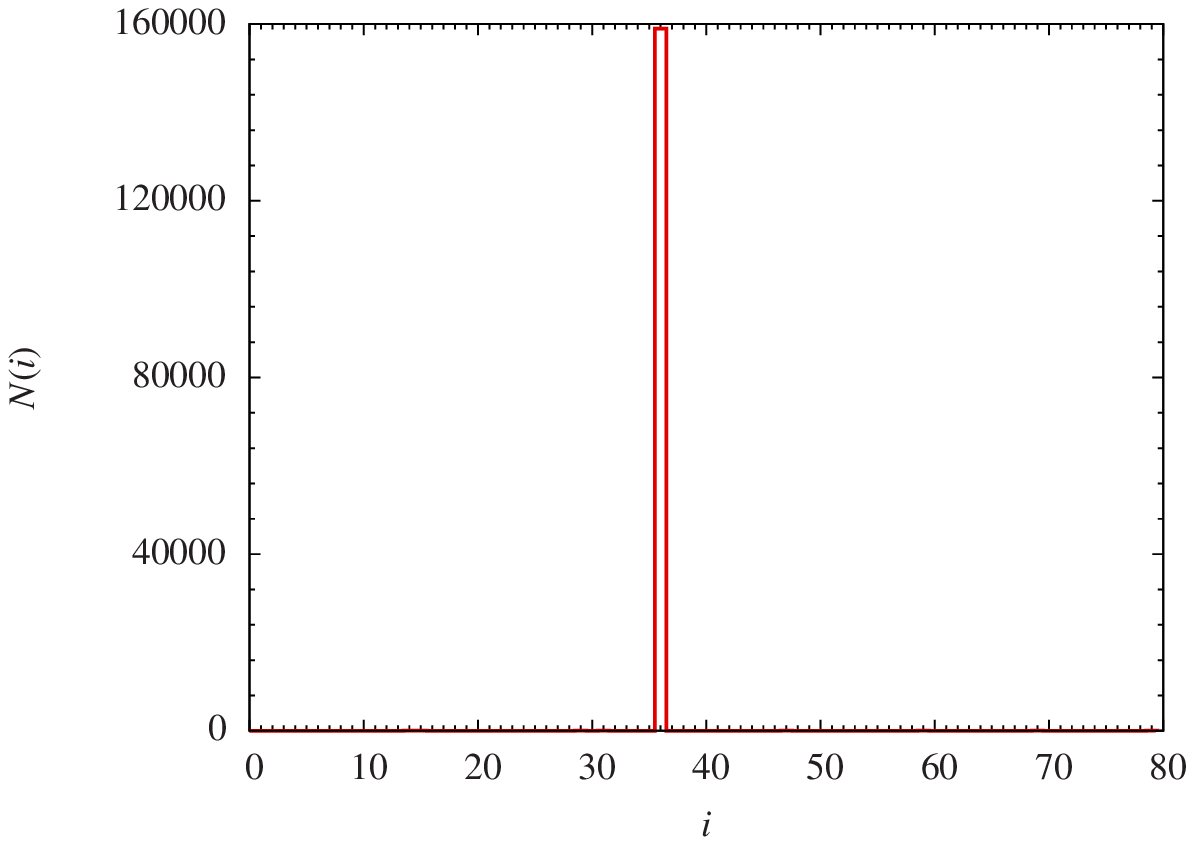}
\raisebox{7mm}{\includegraphics[width=0.46\textwidth]{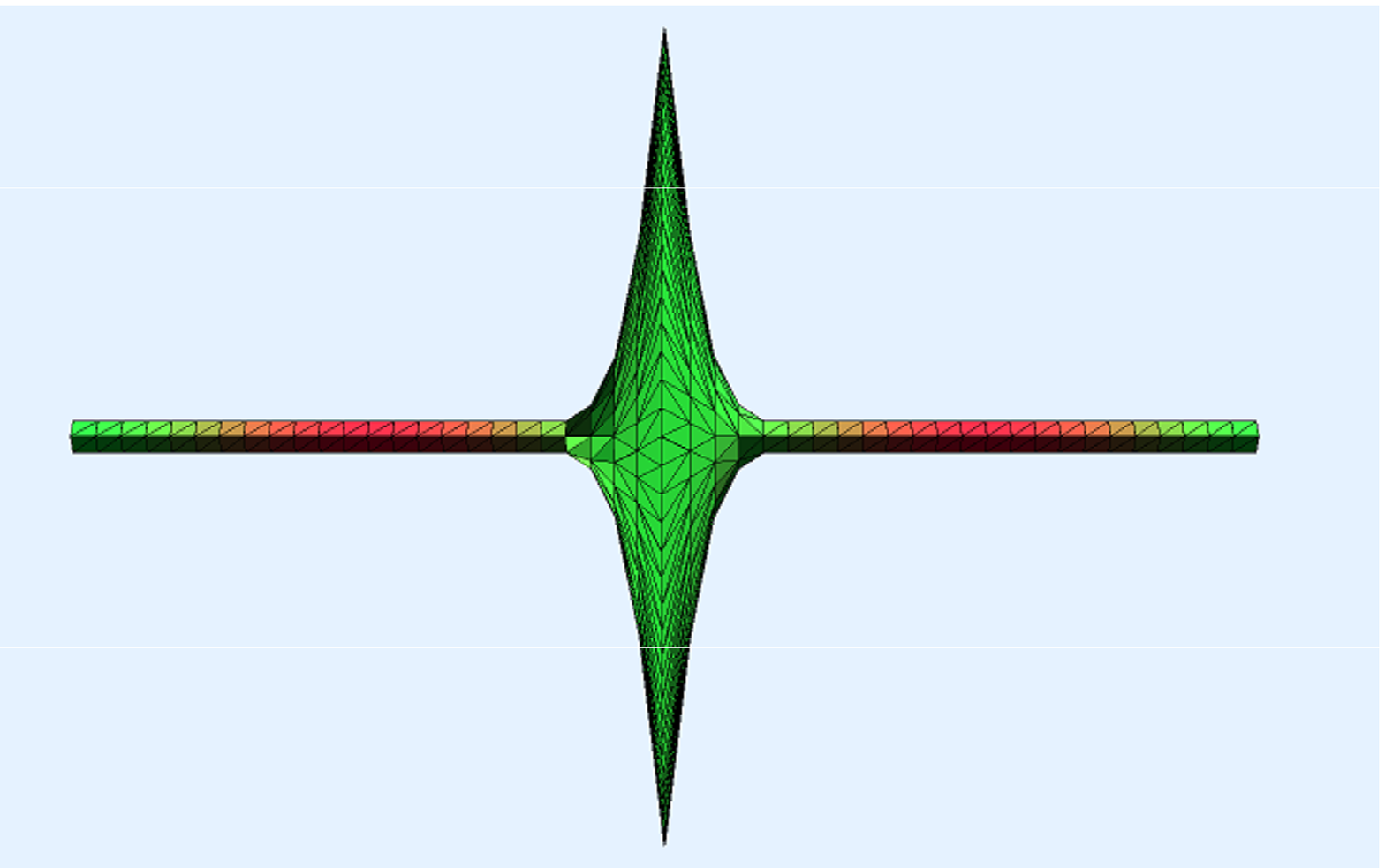}}\\
{\bf Phase B.} 
Snapshot of a spatial volume $N(i)$ for an individual configuration in phase $B$ and
a two-dimensional visualization of a corresponding triangulation.
\end{center}
{\noindent\bf Phase B.}
For small values of $\Delta$ nearly all simplices are localized on one spatial slice.
Although we have a large three-volume collected at one spatial hypersurface of a topology
of a three-sphere $S^3$, the corresponding slice has almost no spatial extent. 
In very few steps it is possible to get from any tetrahedron to any other tetrahedron 
- along a path joining centers of neighboring tetrahedra.
This is possible because many vertices, belonging to the largest slice, have extremely high coordination number,
counted as the number of tetrahedra sharing a given vertex.
The Hausdorff dimension is very high, if not infinite. 
In the case of infinite Hausdorff dimension the universe has neither time extent nor spatial extent, 
there is no geometry in a traditional sense.   
This phase corresponds to so-called \emph{crumpled phase} also known from EDT.
Phase $B$ is also regarded as non-physical.

\begin{center}
\includegraphics[width=0.53\textwidth]{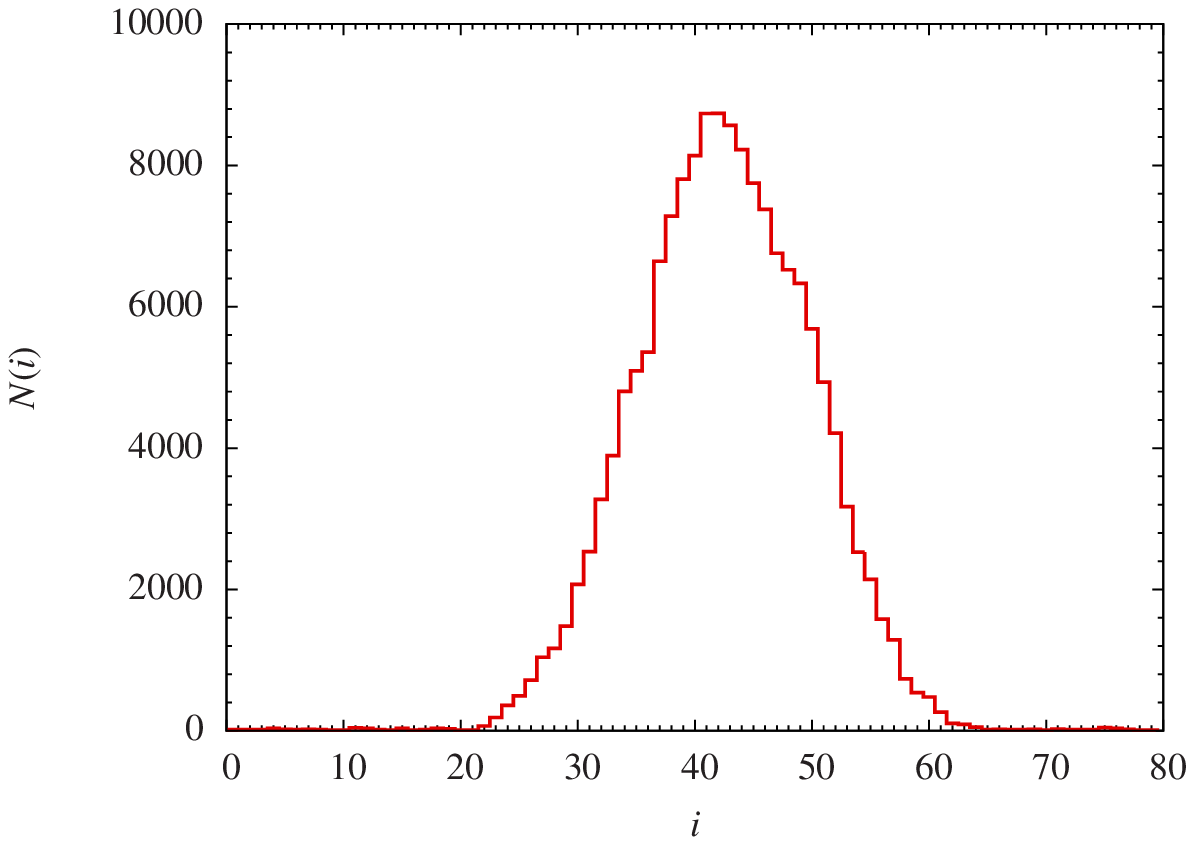}
\raisebox{7mm}{\includegraphics[width=0.46\textwidth]{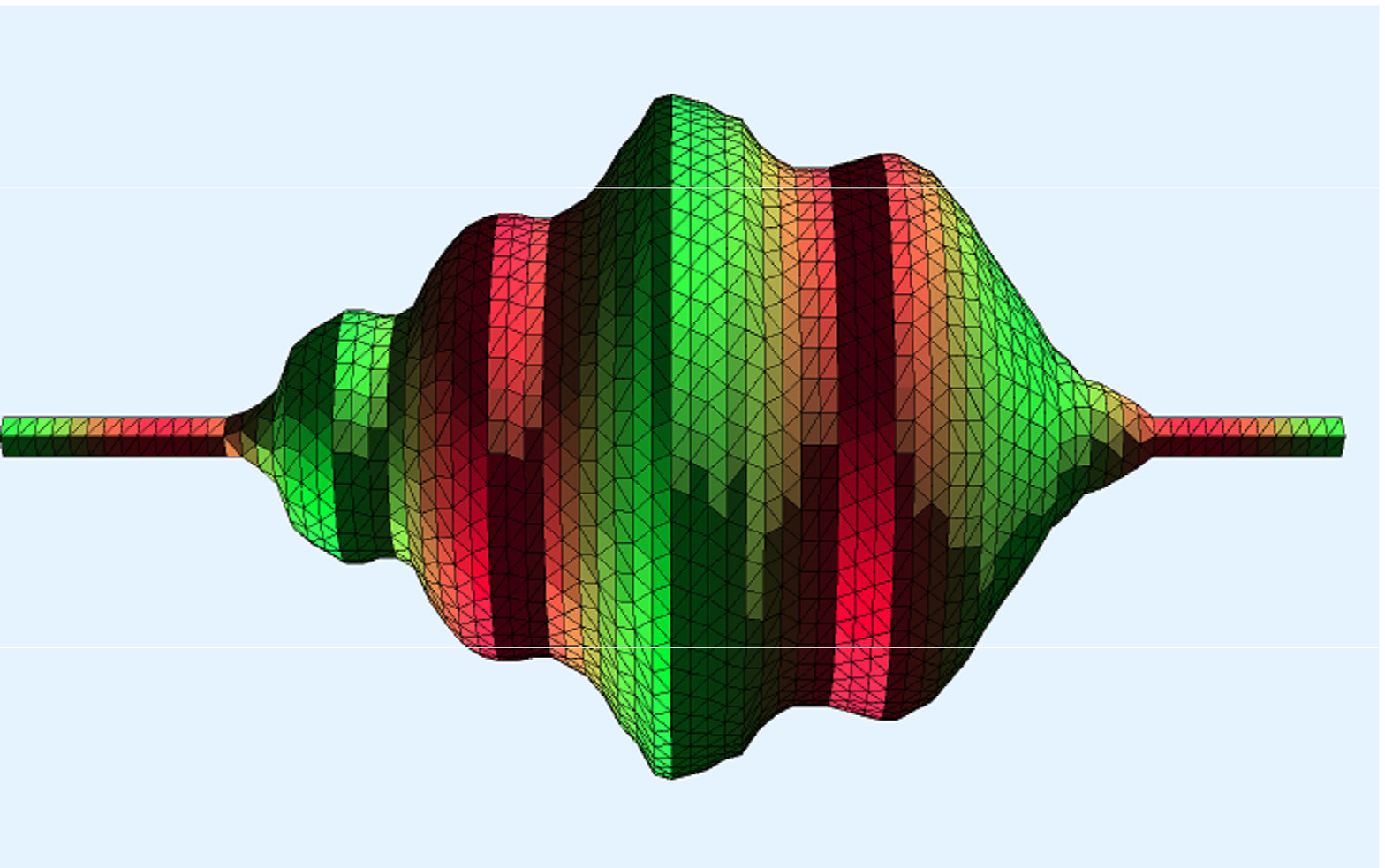}}\\
{\bf Phase C.} 
Snapshot of a spatial volume $N(i)$ for an individual configuration in phase $C$ and
a two-dimensional visualization of a corresponding triangulation.
A typical configuration is bell-shaped with well-defined spatial and time extent.
\end{center}
{\noindent\bf Phase C.}
For larger values of $\Delta$ we observe the third, physically most interesting, phase.
In this range of bare coupling constants,
a typical configuration is bell-shaped and behaves like a well defined four-dimensional manifold (cf. above figure).
The measurements of the Hausdorff dimensions confirm that at large scales the universe is genuinely four-dimensional \cite{Reco}.
Most results presented in this paper were obtained for a total volume of $N_4 = 362000$ simplices,
and for $K_0 = 2.2, \Delta = 0.6, K_4^{\textit{crit}} = 0.9$. 
This point is firmly placed in the third phase $C$ (cf. Fig. \ref{Fig:PhaseDiagram}).
A typical configuration has a finite time extent and spatial extent
which scales as expected for a four-dimensional object.
The averaged distribution of a spatial volume coincides with the distribution of 
Euclidean de Sitter space $S^4$ and thus this phase is also called a de Sitter phase.

\section{Phase transitions}

In this Section we will try to determine the order of the phase transitions.
So far, there is a strong numerical evidence that the 
transition between phases $A$ and $C$ is of first order,
while between phases $B$ and $C$ may be either 
a first-order or a strong second-order.
Since we have at our disposal only results obtained by computer simulations,
we can not verify the transition order with absolute certainty.

Numerical determination of the position and order of the phase transition line
demands very large resources of computation time.
This is caused by several factors. 
To precisely determine the point of transition
one has to perform dense sampling in the space spanned by the coupling constants.
Often the observation of the phenomenon of phase coexistence,
characteristic for the first-order transition,
requires a very accurate tuning of bare coupling constants.
In addition, we have to deal with the critical slowdown.
Monte Carlo algorithms, which work well deep inside the phases,
cease to be effective when approaching a transition point.
The acceptance of Monte Carlo moves drastically decreases and the same moves tend to be repeated.
This significantly extends the autocorrelation time.
It is therefore important to have efficient algorithms,
which often must be customized for a specific transition.
Also finite-size effects affect the position of the transition.
This factor is particularly important for results presented in this Section.
One can still expect shifts in the location of the line as $N_4 \to \infty$.
A characteristic feature of first-order phase transitions is
suddenness of the transition and the occurrence of the hysteresis
with the change of coupling constants values in simulations.
The latter however makes it difficult to find a precise location of a transition point. 
It should be noted that in numerical simulations,
which out of necessity work on finite configurations,
one never observes a true phase transition
which is related to singularities of derivatives of free energy .

\begin{figure}[t]
\begin{center}
\includegraphics[height=95mm]{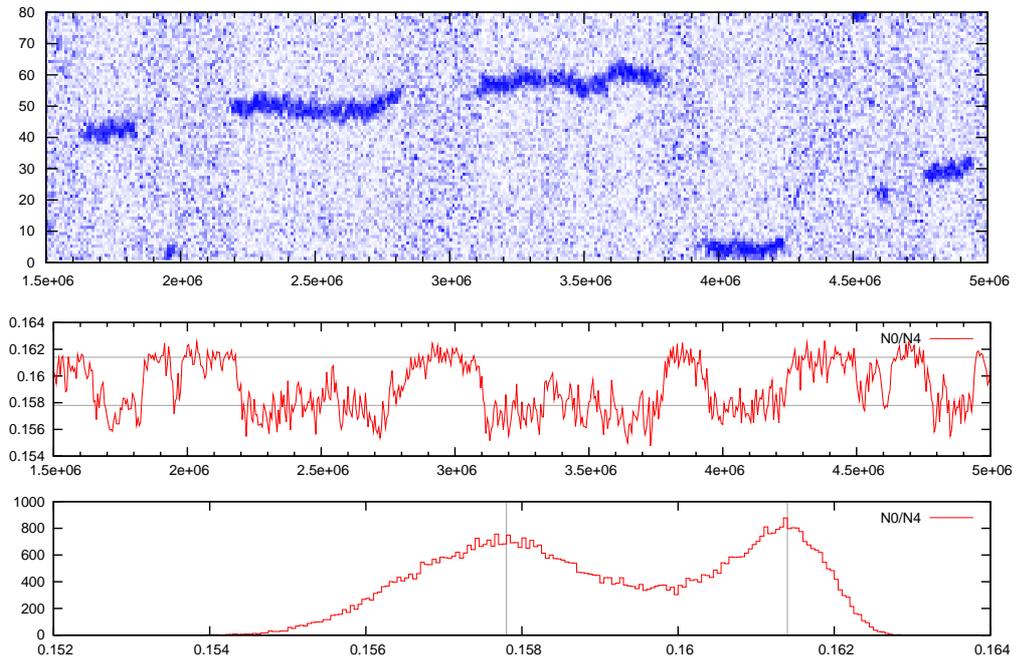} %
\end{center}
\caption{
Transition between phases $C$ (smaller $K_0$) and $A$ (larger $K_0$) 
at $K_0 = 4.711$ and $\Delta = 0.6$ for $N_4 = 120\rk$.
We observe that configurations \emph{jump} between two regimes,
which is a strong evidence of first order transition.
Top: Density plot of the spatial volume as a function 
of Monte Carlo simulation time (horizontal axis) and slicing time $i$ (vertical axis).
Darker colors mean larger spatial volumes $N(i)$.
Middle: Order parameter $N_0 / N_4$, conjugate to $K_0$, as a function of Monte Carlo time.
Bottom: Distribution of the values taken by the order parameter $N_0 / N_4$. Double-peak structure is present.
}
\label{Fig:TransitionCA}
\end{figure}

Let us first consider the $A$-$C$ phase transition
\footnote{The author would like to thank Samo Jordan for carrying out the measurements and providing the results.}.
As can be seen in Fig. \ref{Fig:PhaseDiagram},
this transition line can be approached by changing $K_0$,
and keeping $\Delta$, and for numerical reasons also $N_4$, fixed.
A natural candidate for the order parameter is the variable conjugate
to $K_0$, namely the ratio $N_0/N_4$.
The graph in the middle of Fig. \ref{Fig:TransitionCA} shows
the behavior of the order parameter $N_0/N_4$ as a function
of the Monte Carlo simulation time at the transition line.
As one can see, the order parameter jumps between two values
which characterize different geometries in both phases.
A density plot of the geometry shape, namely the spatial volume $N(i)$,
as a function of Monte Carlo time is shown at the top of Figure \ref{Fig:TransitionCA}.
The quantity $N(i)$ is plotted in the following way.
On the horizontal axis is the simulation time $\tau$ measured in multiples of Monte Carlo moves.
On the vertical axis is the discrete proper-time $i$ corresponding to the slice number.
The value of $N^{(\tau)}(i)$ is determined by the color of a point with coordinates $(\tau, i)$.
White color means a minimal three-volume, 
dark blue means the maximal volume 
and the intermediate values are properly graded.
We can distinguish two types of behavior,
either the three-volume gathers around a particular slice (blue marks) or 
it disintegrates over a whole time axis.
There is a clear correlation between the type of geometry and the value of the order parameter.
Large values of the ratio $N_0/N_4$ correspond to a non-localized geometry,
randomly distributed across the time axis.
This is typical for configurations of phase $A$. 
Small values of the order parameter correspond to the situation 
where the geometry has a finite time extent and is localized around a single maximum.
This is the expected behavior for the configuration of phase $C$.
Clearly one can see that the nature of geometry changes abruptly, 
at fixed values of the coupling constants corresponding to the transition point.
A centered configuration typical for the phase $C$, 
suddenly transforms into a disintegrated configuration typical for the phase $A$,
after a while to reappear in another place.
The graph at the bottom of Fig. \ref{Fig:TransitionCA} shows the distribution 
of values taken by the order parameter $N_0/N_4$.
One sees two peaks, which correspond to different types of the geometry.
This confirms the earlier hypothesis 
that configurations behave as if they were either in phase $C$ or phase $A$ 
and jump between them.
Admittedly, the peaks are slightly blurred,
but this is a consequence of finite-size corrections.
With the increase of the total volume $N_4 \to \infty$,
peaks become sharper \cite{CDTHorava}.
This suggests that, the $A$-$C$ transition is of first-order.

\begin{figure}[t]
\begin{center}
\includegraphics[height=95mm]{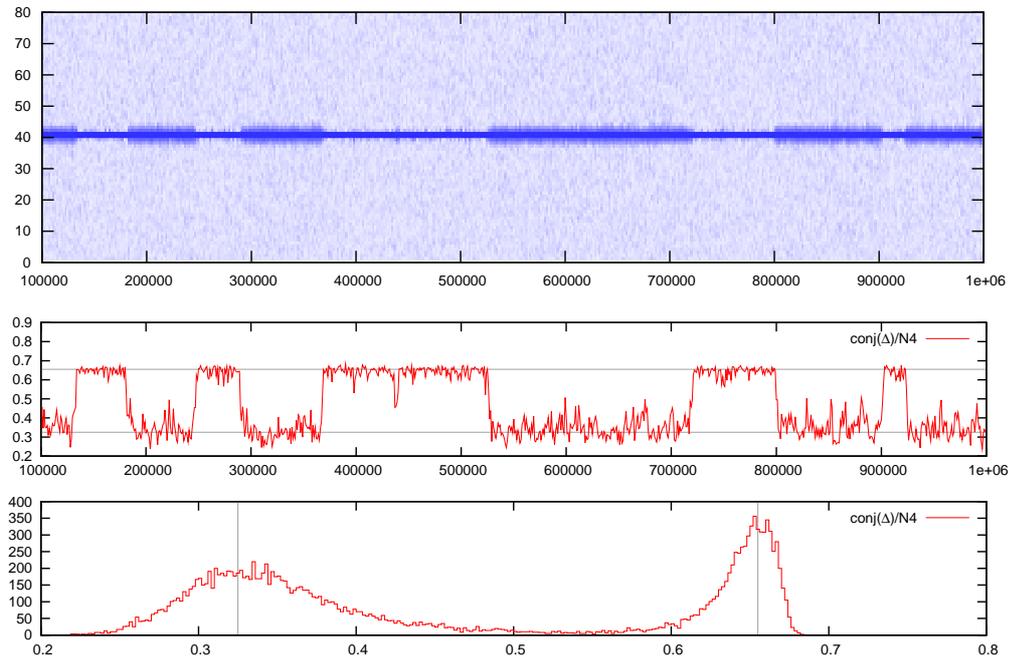} %
\end{center}
\caption{
Transition between phases $C$ (larger $\Delta$) and $B$ (smaller $\Delta$) 
at $K_0 = 2.2$ and $\Delta = 0.022$ for $N_4 = 40k$.
Although the configurations \emph{jump} between two regimes,
but the effect gets weaker with increasing total volume $N_4$.
Top: Density plot of the spatial volume as a function 
of Monte Carlo simulation time (horizontal axis) and slicing time $i$ (vertical axis).
Darker colors mean larger spatial volumes $N(i)$.
Middle: Order parameter $(N_{41} - 6 N_0) / N_4$ ,conjugate to $\Delta$, as a function of Monte Carlo time.
Bottom: Distribution of the values taken by the order parameter. Double-peak structure is present.
}
\label{Fig:TransitionCB}
\end{figure}

In a similar way we depict the measurements obtained for the $B$-$C$ phase transition.
Starting from phase $C$, we can approach phase $B$ by varying a value 
of the coupling constant $\Delta$ (cf. Fig. \ref{Fig:PhaseDiagram}).
Thus the variable conjugate to $\Delta$ (cf. Regge action (\ref{Eq:SRegge})),
namely $(N_{41} - 6 N_0) / N_4$, is used as the order parameter.
Indeed, such choice of the order parameter, 
allows to observe similar effects as in the previous case.
The graph in the middle of Fig. \ref{Fig:TransitionCB} shows the
order parameter as a function of the Monte Carlo time.
The same jumping behavior is observed as for the $A$-$C$ transition.
The graph at the top of Fig. \ref{Fig:TransitionCB} illustrates
the geometry of configurations, by plotting the spatial volume $N^{(\tau)}(i)$
as a function of simulation time $\tau$.
Again there is a strict correlation between the order parameter and the typical shape of the geometry.
Large values of the parameter indicate that the system is in phase $B$.
As can be seen from the top plot, in this case the universe has no time extent 
and the volume is wholly located at one slice.
Small values of the parameter correspond to a system typical of phase $C$,
which has a non-trivial time extent.
The graph at the bottom of Fig. \ref{Fig:TransitionCB} plots the distribution
of values taken by the order parameter.
As can be seen, the peaks  of the distribution are even sharper
and jumps are more evident.
However, the strength of the first-order transition signal is suppressed
with the increasing total volume $N_4$.
Peaks become blurred and start to merge.
Therefore, this result might be an artifact of too small configuration sizes.
Moreover, the position of the transition depends on the total volume.
If the system is near the $B$-$C$ phase transition line,
it happens that for small universe sizes the system is still in phase $C$,
with time extended configurations.
But, when increasing $N_4$, while keeping the coupling constants $K_0$ and $\Delta$ fixed,
the configuration width is decreasing, opposite to the expected scaling,
and the configurations change their character and collapse 
to degenerated shapes of phase $B$.
As already mentioned, due to the critical slowdown,
we are not able to perform reliable measurements for larger volumes.
So far, simulations did not answer the question 
whether the $B$-$C$ phase transition is of first order,
or if we observe a strong second-order transition.

Because phases $A$ and $B$ are not regarded as physically relevant,
we shall not study the phase transition between them.

\section{Relation to Ho\v{r}ava-Lifshitz gravity}

Recently Petr Ho\v{r}ava proposed field-theoretical
approach to quantum gravity inspired by the theory of a Lifshitz scalar.
The modification of General Relativity 
introduces anisotropy between space and time,
which makes the quantum theory power-counting renormalizable and potentially ultra-violet 
complete.
The basic features of the theory, named Ho\v{r}ava-Lifshitz gravity,
are only briefly mentioned in this Section,
for details we refer the reader to the original works 
\cite{HoravaSpectral, HoravaMembranes, HoravaQuantum}.

In the publication \cite{HoravaSpectral}
Ho\v{r}ava points out some similarities between
Ho\v{r}ava-Lifshitz gravity and Causal Dynamical Triangulations.
Namely, the spectral dimension manifests the same
short-range as well as long-range behavior in both theories.
In the CDT framework, the spectral dimension is defined and calculated in the Section \ref{Sec:FourDimensional}.
It occurs, that at short distances it is equal to $2$,
while at large distances it is $4$ as expected for the classical four-dimensional Universe.
Identical scale dependence was obtain by Ho\v{r}ava in his theory.
Another formal similarity of the two theories is the 
assumption of global time foliation, 
which refrains from equal treatment of time and space.
In the de Sitter phase of CDT 
the anisotropy withers
and the symmetry is regained.

\begin{figure}[t!]
\begin{center}
\includegraphics[width=0.9\textwidth]{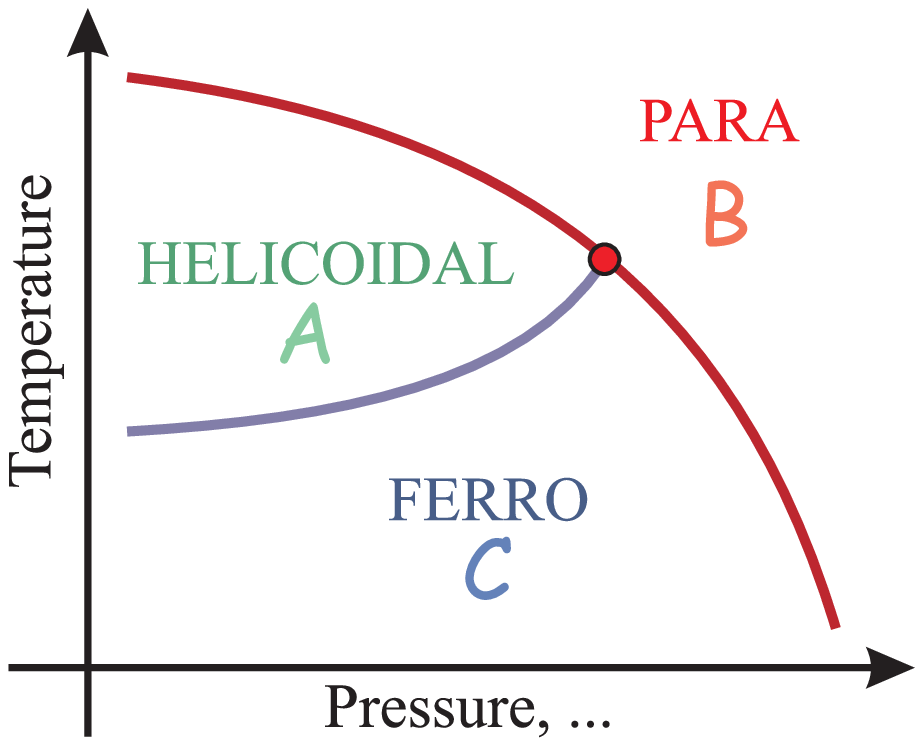}\\
\includegraphics[width=0.3\textwidth]{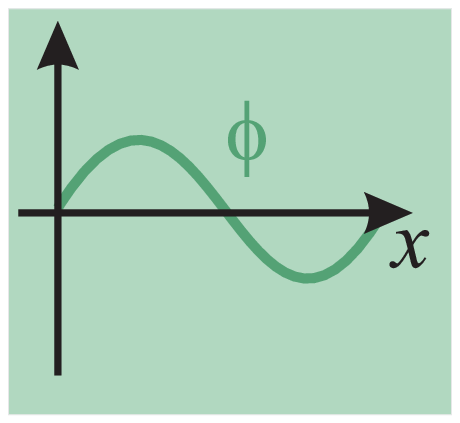}
\includegraphics[width=0.3\textwidth]{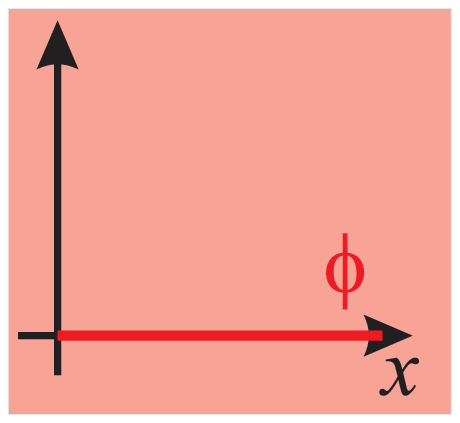}
\includegraphics[width=0.3\textwidth]{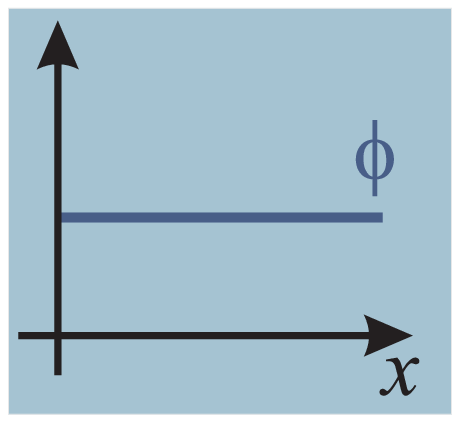}
\end{center}
\caption{
On the top: Sketch of Lifshitz scalar phase diagram with three phases
\emph{helicoidal}, \emph{paramagnetic} and \emph{ferromagnetic}
and the triple point. 
From left to right:
the plot of the order parameter $\phi(x)$ respectively for 
\emph{helicoidal}, \emph{paramagnetic} and \emph{ferromagnetic} phase.
}
\label{Fig:Lifshitz}
\end{figure}

There are however further analogies between the two models.
The Ho\v{r}ava-Lifshitz theory
might be viewed as a generalization of the Lifshitz scalar theory to gravity.
The theory of a Lifshitz scalar was first proposed to describe the tricritical phenomena
and has a phase diagram which consists of three phases meeting at the Lifshitz triple point.
There is a striking resemblance
between the Lifshitz scalar phase diagram and the four-dimensional CDT phase structure.
The Landau free-energy-density functional for a $d$-dimensional effective Lifshitz
theory is given by 
\[ 
F[\phi(x)] = a_2 \phi^2 + a_4 \phi^4 + \dots +  c_2 (\partial_\alpha \phi)^2 + d_2 (\partial_\beta \phi)^2 + e_2 (\partial_\beta^2 \phi)^2 + \dots,
\]
where $\phi(x)$ is the order parameter and 
$\beta = 1 \dots m$, $\alpha = m + 1 \dots d$
\cite{Hornreich, Goldenfeld}.
Distinction between directions $\alpha$ and $\beta$ 
allows for an anisotropic behavior and spatially modulated phases.
A sketch of the phase diagram for the Lifshitz scalar is shown in Fig. \ref{Fig:Lifshitz}.
It is possible to make a one-to-one correspondence between the phases of CDT model
and Lifshitz model if one assumes a following identification.
We can qualitatively characterize  geometries appearing in the three phases of CDT,
and relate them to the Lifshitz mean-field order parameter $\phi(x)$.
Let us introduce a qualitative notion describing the \emph{average geometry},
and denote it as $\langle geometry \rangle$.
In phase $C$, we observe a genuine four-dimensional background geometry (cf. Chapter \ref{Chap:Macroscopic}).
The emerged Universe agrees with the classical notion of four-dimensional geometry,
such that $\langle geometry \rangle > 0$. 
While it might be considered as a four-sphere $S^4$, we can even argue that $\langle geometry \rangle = const$.
As stated at the beginning of this Chapter,
configurations present in phase $B$ have neither spatial nor time extent,
thereby they have no geometry in the traditional sense, and we can write $\langle geometry \rangle = 0$.
Finally, in phase $A$, the configurations appear to be \emph{oscillating} in the time direction,
and $|\partial_t \langle geometry \rangle| > 0$.
So far, we do not know how to define a more quantitative measure of the \emph{average geometry}
which could be identified with the order parameter.

For $m > 0$, anisotropy is present and the system reveals three phases
\footnote{Here we use the \emph{magnetic} analogy to name the Lifshitz phases,
but the phase diagram is applicable to variety of other systems.}:
\begin{description}
\itemsep=0pt
\item[Helicoidal.] For $d_2 < 0$ we have a modulated phase
called \emph{helicoidal} phase. 
The order parameter is oscillating in $\beta$ directions.
For $m = 1$, we can identify this direction with the time present in CDT.  
In this phase we have $|\partial_t \phi(x)| > 0$. 
Therefore, the \emph{helicoidal} phase can be identified with phase $A$ of four-dimensional CDT.
The order parameter is illustrated at the bottom left picture of Fig. \ref{Fig:Lifshitz}.
\item[Paramagnetic.] For $d_2 > 0,\ a_2 > 0$ we observe a \emph{paramagnetic} phase
with a vanishing order parameter, $\phi(x) = 0$.
The order parameter is depicted at the bottom middle plot of Fig. \ref{Fig:Lifshitz}.
The \emph{paramagnetic} phase can be identified with phase $B$ present in CDT.
\item[Ferromagnetic.] For $d_2 > 0,\ a_2 < 0$, we have a \emph{ferromagnetic} phase.
The order parameter is constant but nonzero, $|\phi(x)| > 0$,
and is shown at the right of Fig. \ref{Fig:Lifshitz}.
Hence, the \emph{ferromagnetic} phase can be identified with phase $C$ of four-dimensional CDT.
\end{description}

To complete the picture, 
let us consider the nature of phase transition lines.
Most often transition between the \emph{ferromagnetic} 
and \emph{paramagnetic} phases in the Lifshitz theory
is of second order. 
In CDT it corresponds to the  $B$-$C$ phase transition.
So far, Monte Carlo simulations did not settle whether 
it is a first-order or a strong second-order transition.
For Lifshitz scalar with $m = 1$ it might happen that the transition between
the \emph{ferromagnetic} and \emph{helicoidal} phases is of first order,
which would agree with the $A$-$C$ phase transition in CDT.

We can summarize, 
that the phase diagram of the Lifshitz scalar
and the measured phase diagram of four-dimensional Causal Dynamical Triangulations model
are strikingly similar.

\clearemptydoublepage

\chapter{The macroscopic de Sitter Universe}
\label{Chap:Macroscopic}

{\noindent \it This chapter is based on the following publications:
(i)   A.~G\"orlich,
  \emph{"Background Geometry in 4D Causal Dynamical Triangulations"},
	Acta Phys. Pol. B \textbf{39}, 3343 (2008);
(ii)  J.~Ambj\o rn, A.~G\"orlich, J.~Jurkiewicz and R.~Loll,
  \emph{"Planckian Birth of the Quantum de Sitter Universe",}
  Phys. Rev. Lett. \textbf{100},  091304 (2008);
(iii)
  J.~Ambj\o rn, A.~G\"orlich, J.~Jurkiewicz and R.~Loll,
  \emph{"Geometry of the quantum universe"},
  Phys. Lett. B {\bf 690}, 420 (2010).
}\\[4ex]

In this Chapter we pass over local degrees of freedom of the quantum geometry,
and reduce the considerations to volumes of spatial slices.
In Causal Dynamical Triangulations, the causality condition 
is ensured by imposing on configurations
a global proper-time foliation and keeping the topology of the leaves fixed. 
Due to the discrete structure, the successive spatial slices,
i.e. hypersurfaces of constant time, are labeled by a discrete \emph{time} parameter $i$.
The index $i$ ranges from $1$ to $T$, and because of time-periodic boundary conditions
the time slice $i = T + 1$ is cyclically identified with the time slice $i = 1$.
To each vertex of a triangulation is assigned an integer \emph{time} coordinate.
All slices are built of equilateral spatial tetrahedra,
each being a base of one simplex of the type $\{1, 4\}$ and one of the type $\{4, 1\}$.
By construction they are \emph{glued} in the way to form
a simplicial manifold of a topology of a three-sphere $S^3$.
The topology of spatial slices is not allowed to change in time. 

\section{Spatial volume}
\label{Sec:Spatial}

The spatial volume $N(i)$, also called the three-volume, is defined as the number of 
tetrahedra building a spatial slice $i = 1 \dots T$.
Because each spatial tetrahedron is a face shared by two simplices of type $\{1, 4\}$ and $\{4, 1\}$,
the three-volume $N(i)$ sums up to the total volume $\Ntot$ equal
\beq 
\Ntot \equiv \sum_{i = 1}^{T} N(i) = \frac{1}{2} N_{41} .
\label{Eq:SumNi}
\eeq
Spatial volume $N(i)$ is an example of the simplest observable
providing information about the large scale \emph{shape} of the universe
appearing in CDT path integral.
An individual spacetime history contributing to the partition function 
is not an observable in the same way as a trajectory of a particle in the
quantum-mechanical path integral is not.
However, it is perfectly legitimate to talk about the
\emph{expectation value} $\langle N(i) \rangle$ as well as about the fluctuations around the mean.
The average is defined by equation (\ref{Eq:ExpVal}). 
Due to the lack of analytical tools allowing to deal with the  path integral (\ref{Eq:ZDisc}) of four-dimensional  
Causal Dynamical Triangulations,
we have to resort to numerical computations in order to calculate expectation values.
The Monte Carlo simulations generate a sequence of spacetime geometries,
more precisely causal triangulations $\cT$, according to the probability distribution (\ref{Eq:ExpVal}).
Configurations are then used to calculate the average.
The measurement of the expectation values of observables is explained in detail in Chapter \ref{Chap:Implementation}. 
In this Chapter we show that the \emph{background geometry},
described by the average $\langle N(i) \rangle$, emerges dynamically.
In the next Chapter we study quantum fluctuations of $N(i)$ around its average.
\begin{figure}[t]
\centering
\includegraphics[width=0.9\textwidth]{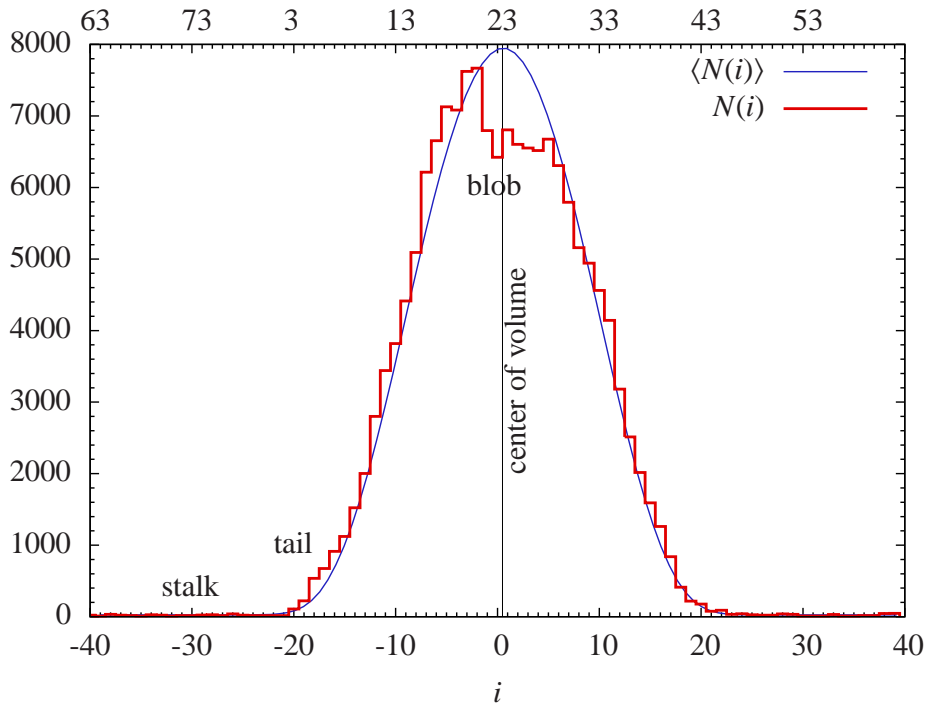}
\caption{
The spatial volume $N(i)$ of a randomly chosen typical configuration 
in phase $C$ ($K_0 = 2.2, \Delta = 0.6$).
The bottom axis corresponds to shifted time index $i$ with fixed position of the \emph{center of volume}.
The original time coordinate is presented on the top axis.
We can distinguish three qualitatively different parts of the configuration:
$\bullet$ the \emph{stalk} with small fluctuations around the minimal slice structure,
$\bullet$ the \emph{tail} which volume is larger than the minimal value, but still very strong discrete effects are present,
and $\bullet$ the \emph{blob} which volume may be considered as a continuous variable.
In the background, the blue curve plots the average spatial volume $\langle N(i) \rangle$
for the above values of coupling constants.
}
\label{Fig:PhaseCNi}
\end{figure}

\begin{figure}
\centering
\includegraphics[width=0.495\textwidth]{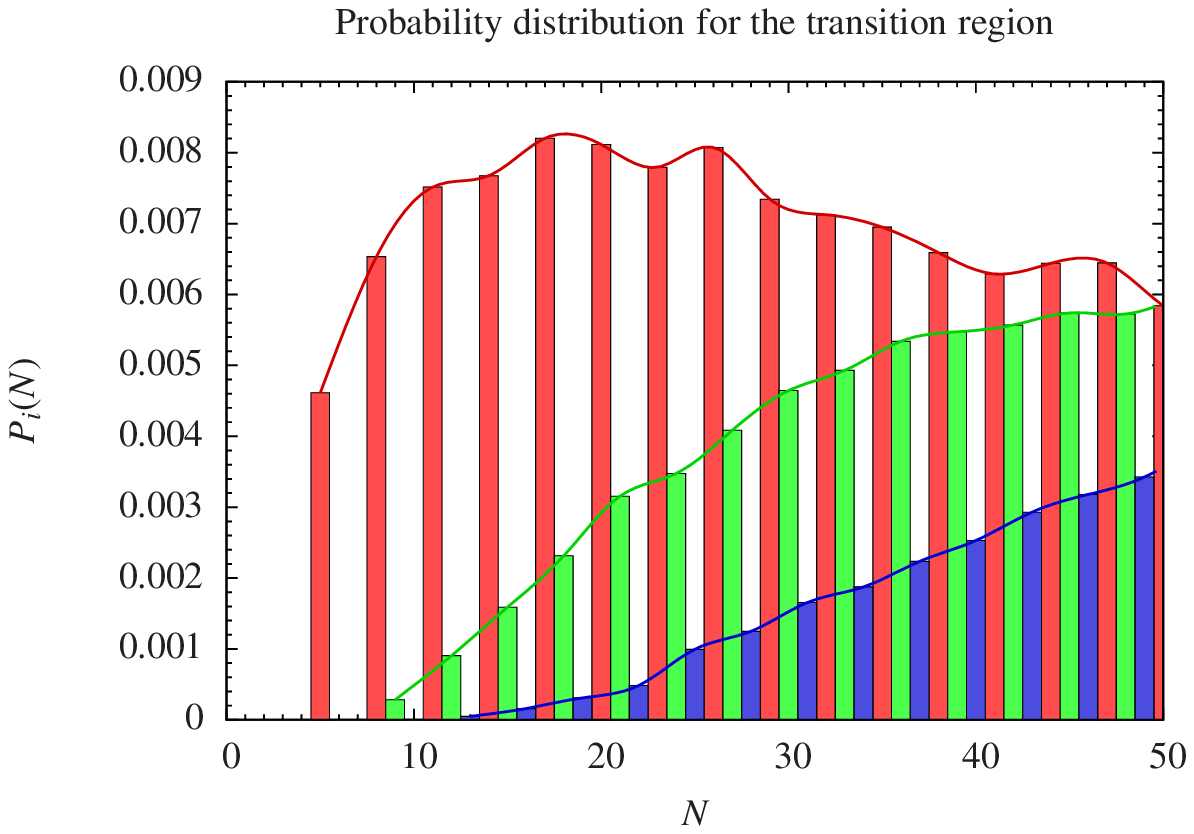}
\includegraphics[width=0.495\textwidth]{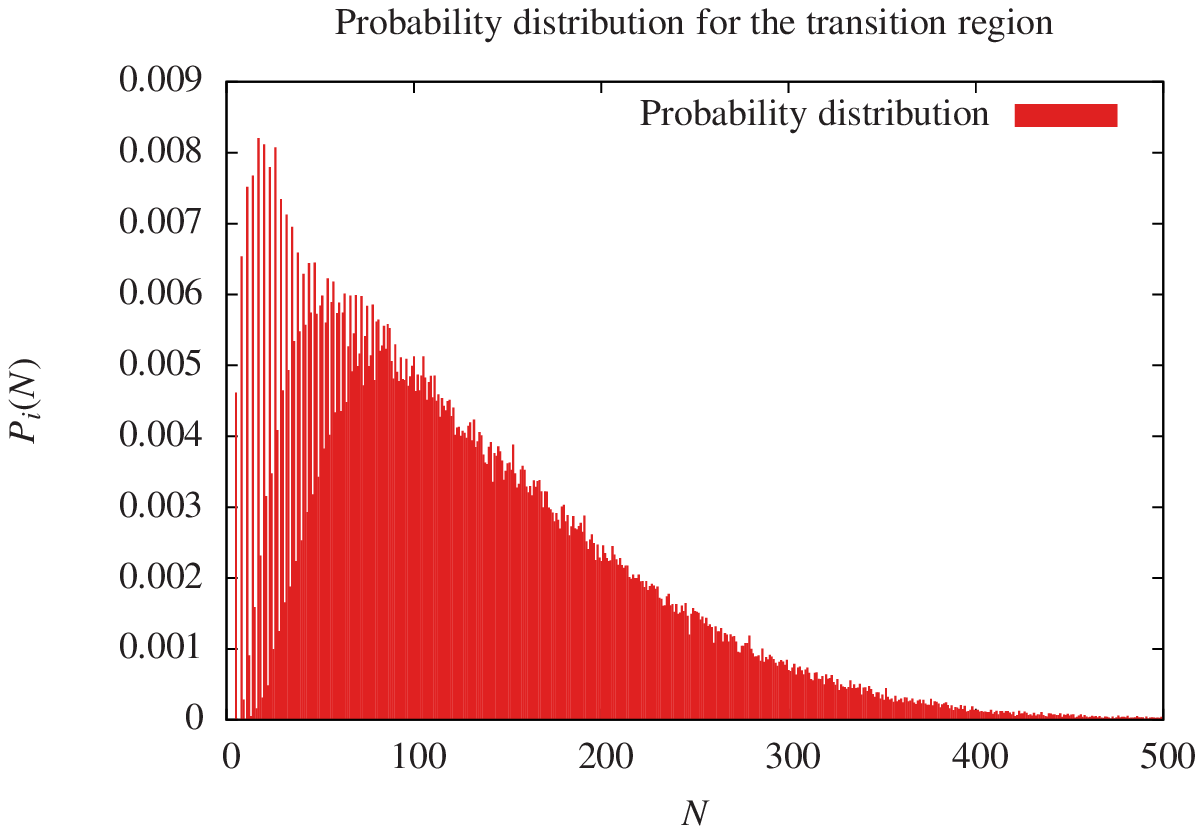}
\caption{A probability distribution in the transition region inside the \emph{tail} (for $i = 17$). For small $N$ the distribution splits into 3 families (left chart). For large N the split disappears but the distribution is highly asymmetric (right chart). }
\label{Fig:Tail}
\end{figure}

All results reported in this Chapter correspond to one particular point of the phase diagram
firmly placed in phase $C$, and
given by the following values of bare coupling constants: $K_0 = 2.2$, $\Delta =0.6$,
volume $N_{41} = 160 000$ and the time-period $T = 80$.
In this phase, the plot of $N(i)$ for an individual configuration
is bell-shaped with a well-outlined \emph{blob}.
Below we describe how a typical triangulation of de Sitter phase is built
and how to properly average $N(i)$.
Fig. \ref{Fig:PhaseCNi} shows the spatial volume $N(i)$ of a typical configuration 
as a function of a discrete time $i$.
For the range of discrete volumes $N_4$ under study,
the Universe \emph{does not} extend over the entire axis,
but rather is localized in a region much shorter than $T = 80$ time slices.
We can distinguish three qualitatively different parts of the configuration
(marked on Fig. \ref{Fig:PhaseCNi}):
\begin{itemize}
\itemsep=0pt
\item {\bf The stalk}.
The volume of a spatial slice is bounded from below by the volume of the minimal structure
consisting of five tetrahedra connected to each other.
Such a structure is the smallest non-degenerate triangulation of $S^3$.
This kinematical constraint ensures that the triangulation remains a simplicial manifold, in which, for example, 
two $4$-simplices are not allowed to share more than one common face, 
and that it forms a connected space.
The stalk is a region which is not much larger than the minimal structure. 
In Fig. \ref{Fig:PhaseCNi} it is visible for $i = -40 \dots -22$ and $i = 22 \dots 39$.
The plot at the left of Fig. \ref{Fig:StalkBlob} presents the probability distribution $P_i(N)$
which describes the probability that the three-volume of spatial slice $i$ is equal to $N(i) = N$.
The distribution is the same along all slices in the stalk.
Thus, the mean volume is constant in this region.
Apparently, there are very large lattice artifacts.
The distribution $P_i(N)$ shown in Fig. \ref{Fig:StalkBlob}  splits into three groups:
red for $N(i) = 5 + 3 k$, green for $N(i) = 5 + 3 k + 1$ and blue for $N(i) = 5 + 3 k + 2$,
where $k$ is a natural number.
Such behavior is an internal feature of the measure imposed on the set of triangulations 
and does not depend on the used Monte Carlo algorithm.
The smallest possible volume is most probable and corresponds to the minimal structure with $5$ tetrahedra,
the slices larger by three tetrahedra are obtained by adding a vertex into one of the tetrahedra.
To conclude, the slices in the stalk are usually at most several times larger than $5$ tetrahedra
and are dominated by discretization artifacts.

\begin{figure}
\centering
\includegraphics[width=0.495\textwidth]{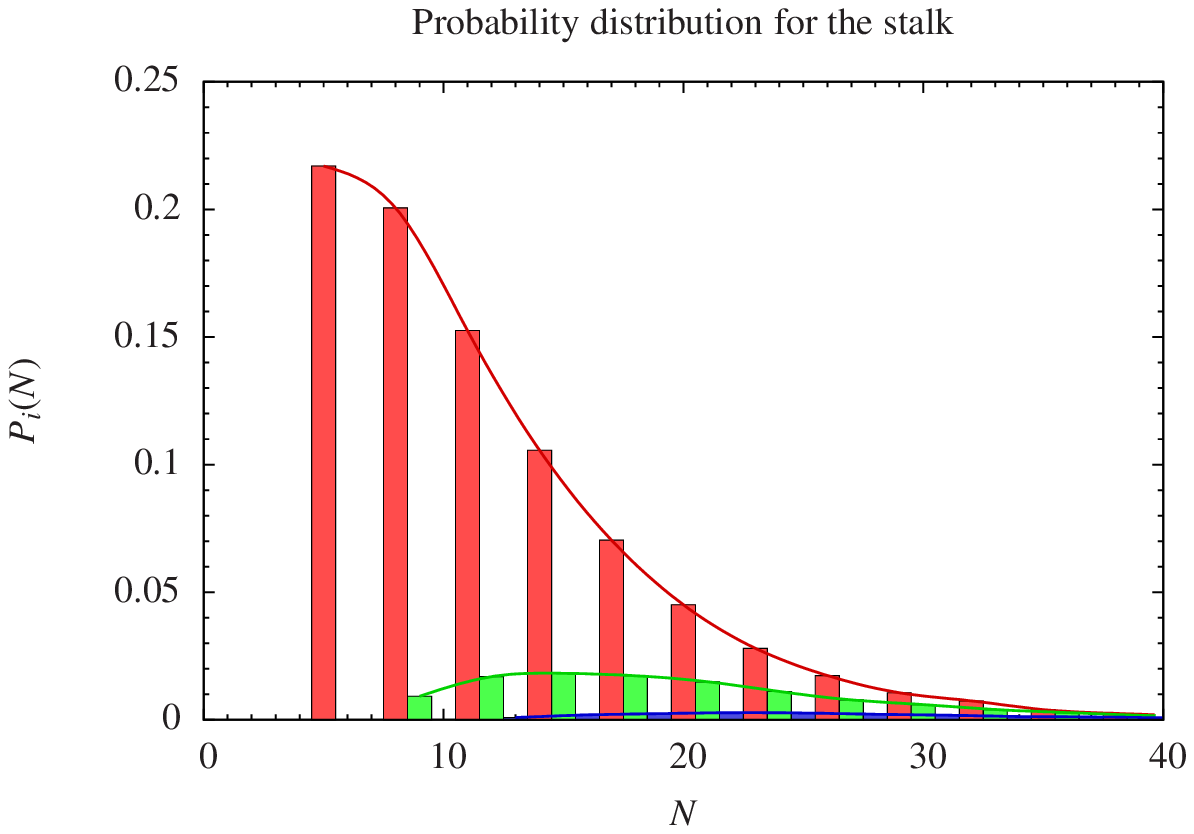}
\includegraphics[width=0.495\textwidth]{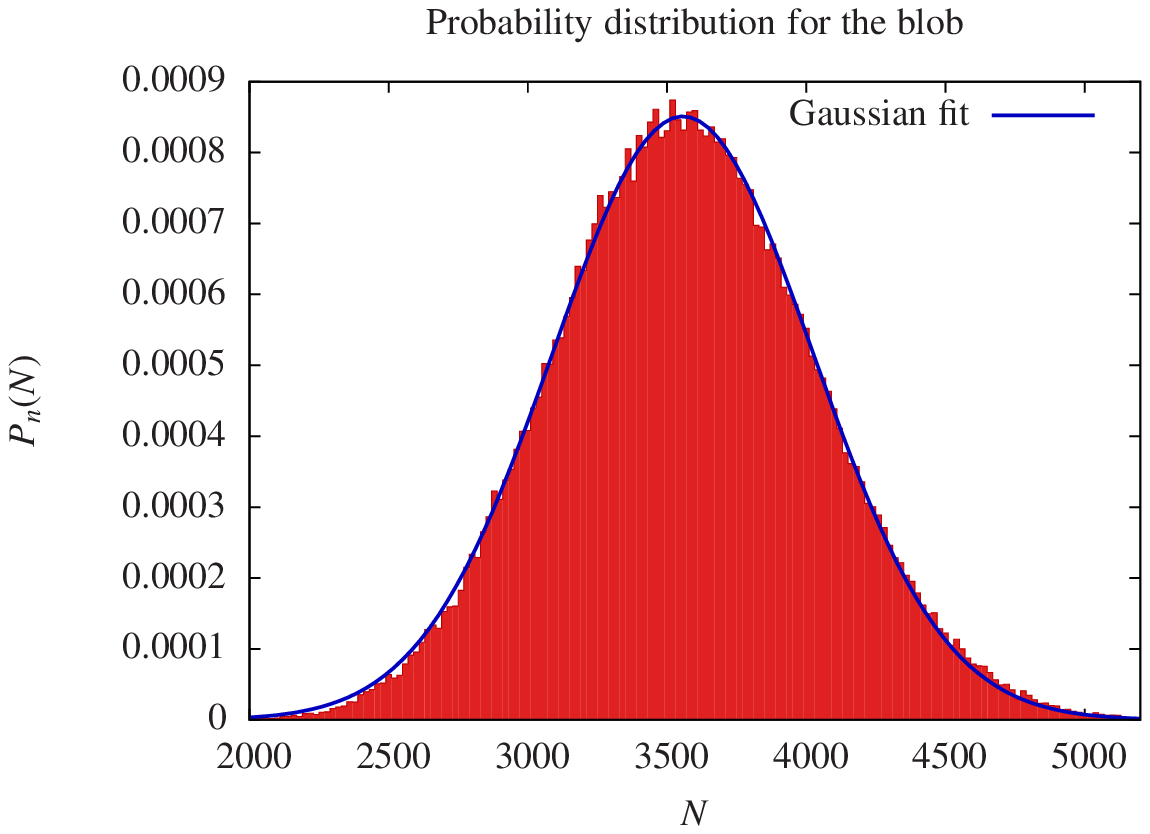}
\caption{On the left: a probability distribution $P_i(N)$ in the \emph{stalk}. On the right: a probability distribution in the blob}
\label{Fig:StalkBlob}
\end{figure}
\item {\bf The blob}. 
For slices with average volume much larger than the cut-off size ($\langle N(i) \rangle \gg 5$),
the probability distribution $P_i(N)$ becomes continuous and is well described by a normal distribution
centered at the expectation value $\langle N(i) \rangle$.
The right plot of Fig. \ref{Fig:StalkBlob} presents  $P_i(N)$ for a central slice.
The \emph{blob} is a range of slices which fulfill above conditions, and it is evident for $i = -14 \dots 14$ in Fig \ref{Fig:PhaseCNi}.
Inside a blob lattice artifacts become irrelevant 
and a semiclassical picture 
of Gaussian volume fluctuations around the mean is very reliable.
\item {\bf The tail}. The intermediate region between the stalk and the blob is called the \emph{tail}. 
Here, the  volume is larger than the minimal value, 
but still very strong discrete effects are visible.
It is present in Fig. \ref{Fig:PhaseCNi} for $|i| = 15 \dots 21$.
Fig. \ref{Fig:Tail} shows the volume distribution $P_i(N)$ for a particular slice in the tail.
As shown in the right plot, the probability distribution is highly asymmetric and depends on the slice number.
The left plot presents the zoomed part of the right picture, for $N = 1 \dots 100$.
As in the case of the stalk, for $N < 100$ the three groups are present, and can be explained by the same mechanism.
For larger volumes, the distribution decays exponentially and the split becomes unclear.
\end{itemize}

The Einstein-Hilbert action (\ref{Eq:SEH})
is invariant under time translations $t \to t+\delta$.
As a consequence, the same holds for the Regge action (\ref{Eq:SRegge})
under discrete time translations.
Because configurations are periodic in time,
a straightforward average $\langle N(i) \rangle$ is meaningless,
as it would give a uniform distribution with a constant value, independent of $i$.
From Fig. \ref{Fig:PhaseCNi} it is clear that in phase $C$ 
the time translation symmetry is \emph{spontaneously broken}.
In order to perform a meaningful average of the spatial volume $\langle N(i) \rangle$,
this should be taken into account and an appropriate time shift is needed.
This is done by the proper centering of configurations,
i.e. by fixing the position of the \emph{center of volume} $i_{CV}$.
Because of the time translation invariance and periodicity, 
the following \emph{centering} procedure is introduced:
\begin{itemize}
\item Given a triangulation with spatial volumes $N(i)$,
we define the \emph{center of volume} position relative to a slice $i$ as
\beq
\mathrm{CV}(i) \equiv \sum_{\delta = -T/2}^{T / 2 -1} (\delta - 0.5)\cdot N(i \oplus \delta),
\label{Def:CV}
\eeq
where we used the \emph{addition modulo} $T$ defined by $i \oplus j \equiv 1 + \textrm{mod} (i + j - 1, T)$
and $\textrm{mod} (i, T)$ is the remainder of the division of $i$ by $T$.
Because of the periodicity in time, the operator $\oplus$ is used instead of $+$. 
\item We find the value of $i \in \{1, 2, \dots, T\}$ which minimizes $\left| CV(i) \right|$,
and denote this by $i_{CV}$.
Due to the periodicity, $\left| CV(i) \right|$ has two minima, namely 
$i_{CV}$ and $i_{CV} \oplus T/2$. 
We choose the value of $i_{CV}$ which is closer to the largest slice $i_{max}$
($i_{max}$ roughly localizes the \emph{blob} center).
\item The Regge action is invariant also under time reflection $i \to T + 1 - i$.
Therefore, we want the real \emph{center of mass} to lie 
as close to $i_{CV} + 0.5$ as possible, i.e. in the interval $\langle i_{CV}, i_{CV}+1)$.
This explains the term $- 0.5$ in (\ref{Def:CV}).
\item Finally, we shift the definition of time, by an integer number of time steps,
and fix the position of the \emph{center of volume} $i_{CV}$ to $0$,
\[ N(i) \leftarrow N(i_{CV} \oplus i), \quad  i \in {-T/2 + 1, \dots, T/2}. \]
\end{itemize}
Fig. \ref{Fig:PhaseCNi} presents the spatial volume $N(i)$
for a particular configuration as a function of the shifted time $i$ (red line).
The data represent the effect of the centering procedure, 
and the shifted time $i$ is labeled at the bottom axis,
while the original time index is shown at the top axis.
The blue curve shows the average volume $\langle N(i) \rangle$.
Evidently, the average position of the \emph{center of volume} is localized at $0.5$.

\begin{figure}
\centering
\includegraphics[width=0.9\textwidth]{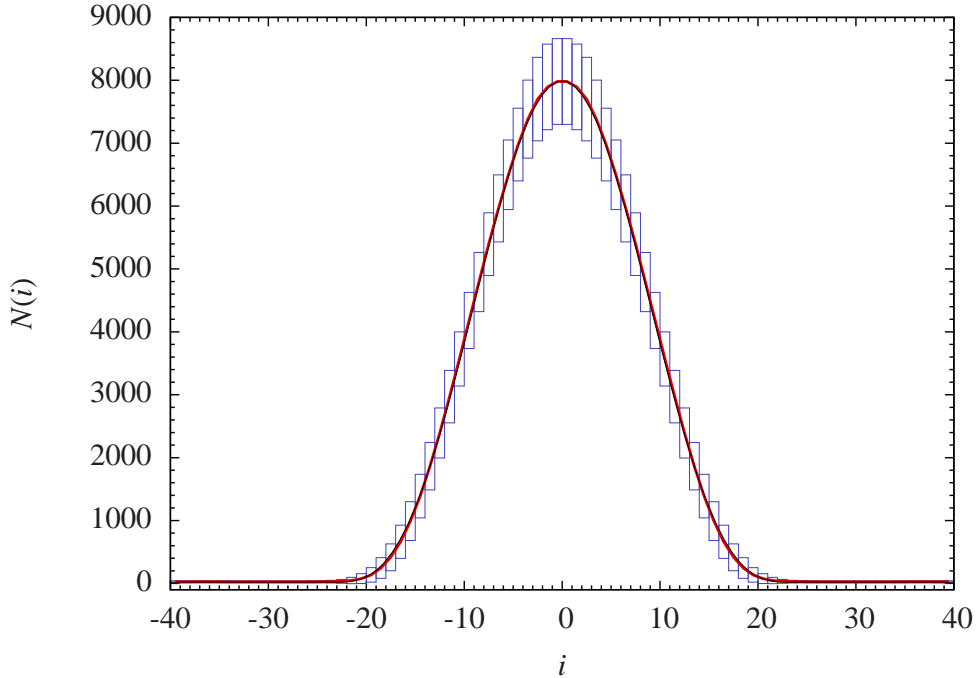}
\caption{
Background geometry $\langle N(i) \rangle$:
Monte Carlo measurements for fixed $N_{41} = 160 000, K_0 = 2.2, \Delta = 0.6$.
The best fit (\ref{Eq:AvNi}) yield indistinguishable curves at given plot resolution.
The bars height indicate the average size of quantum fluctuations.}
\label{Fig:AvNi}
\end{figure}

Having fixed the position of the center of volume along the time-direction
for all contributing triangulations, by applying the above shift procedure,
we can now perform superpositions of configurations and define the expectation value $\langle N(i) \rangle$
as a function of the discrete time $i$.
The average is measured using Monte Carlo techniques,
\begin{equation}
 \langle N(i) \rangle \approx \frac{1}{K} \sum_{k = 1}^{K} N^{(k)}(i),
\label{Eq:MCAvNi}
\end{equation}
where the brackets $\langle \dots \rangle$ mean averaging over the whole ensemble of causal triangulations
weighted with the Regge action (\ref{Eq:SRegge})
and
the expectation value is approximated be a sum over $K$ statistically independent Monte Carlo configurations.
The result for  $k$th realization, is denoted by a superscript ${(.)}^{(k)}$, where $k = 1 \dots K$.
The details of evaluating the sum (\ref{Eq:MCAvNi}) are described in Chapter \ref{Chap:Implementation}.
Fig. \ref{Fig:AvNi} shows the average spatial volume $\langle N(i) \rangle$ (red line)
measured at a point in the phase $C$, $K_0 = 2.2$ and $\Delta = 0.6$.
The height of the blue boxes visible in the plot
indicates the amplitude of spatial volume fluctuations for each $i$
given by $\sigma_i = \sqrt{\langle N(i)^2 \rangle - \langle N(i) \rangle^2}$.
Results obtained by simulations show that the average geometry,
in the \emph{blob} and \emph{tail} region,
is extremely well approximated by a formula
\begin{equation}
\bar{N}(i) \equiv \langle N(i) \rangle = H\cdot\cos^3(i / W),	
\label{Eq:AvNi}
\end{equation}
where $W$ is proportional to the time extent of the \emph{Universe}
and $H$ denotes its maximal spatial volume.
The fit $H\cdot\cos^3(i / W)$ is also plotted in Fig.~\ref{Fig:AvNi} with a thin black line,
but it is indistinguishable from the \emph{empirical} curve.
The \emph{background geometry} given by the solution (\ref{Eq:AvNi})
is consistent with a geometry of a four-sphere $S^4$
and corresponds to the Euclidean de Sitter space,
the maximally symmetric solution of classical Einstein equations 
with a positive cosmological constant \cite{Background,Nonp}.
This is one of the most important results described in this dissertation.
We presented the direct evidence, that the \emph{background geometry} of a four-sphere emerges dynamically.
We did not put by hand any background.
Moreover, neglecting the stalk,
which by construction has a nonzero volume,
we spontaneously end up with the $S^4$ topology,
although we started with $\cM = S^1 \times S^3$.

\section{The minisuperspace model}
\label{Sec:Minisuperspace}

The shape of the three-volume $\bar{N}(i) = H\cdot\cos^3(i / W)$ emerges as a classical solution
of the \emph{minisuperspace} model.
The minisuperspace model appears for example
in quantum cosmological theories developed by Hartle and Hawking 
in their semiclassical evaluation of the wave function of the Universe \cite{HartleHawking}.
This model assumes spatial homogeneity and isotropy, which means that all degrees of freedom except the three-volume (scale factor) are \emph{frozen}.
In CDT model we have the opposite situation, no degrees of freedom are excluded, 
instead we integrate out all of them but the scale factor. 
Nevertheless in both cases results demonstrate high similarity.
Neglecting internal structure of time slices,
let us introduce a spatially homogeneous and isotropic metric on a Euclidean spacetime with $S^1 \times S^3$ topology,
\beq
\dd s^2 = \dd \tau^2 + a^{2}(\tau) \dd \Omega_3^2,
\label{Eq:MiniMetric}
\eeq
where $a(\tau)$ is the \emph{scale factor} depending on the \emph{proper time} $\tau$ 
and $\dd \Omega_3^2$ denotes the line element on $S^3$.
For such metric the \emph{physical volume} of a spatial slice for a given time $\tau$ equals
\[ v(\tau) = \int \dd \Omega_3 \sqrt{\det g|_{S^3}} = 2 \pi^2 a(\tau)^3. \]
and the scalar curvature is given by
$R = \frac{6}{a^2} (1 - \dot{a}^2 - a \ddot{a})$, 
where the dot denotes a derivative w.r.t. $\tau$.
The Euclidean version of the Einstein-Hilbert action (\ref{Eq:SEH}) is given by \cite{Hawking, Gibbons}
\beq
S_{EH}^{Euc} [g_{\mu \nu}] = - \frac{1}{16 \pi G} \int \dd \tau \ \dd \Omega_3 \  \sqrt{\det g} (R - 2 \Lambda).
\label{Eq:SEHEuc}
\eeq
The above action calculated for the metric (\ref{Eq:MiniMetric}) up to boundary terms takes the form
\beq
S[a] = \frac{2 \pi^2}{16 \pi G} \int \dd \tau \left( - 6 a \dot a^2 - 6 a + 2 \Lambda a^3\right),
\label{Eq:MiniAct}
\eeq
and is called the minisuperspace action.

The Euclidean Einstein-Hilbert action (\ref{Eq:SEHEuc}) suffers from the unboundedness
of the conformal mode,
which is caused by the \emph{wrong sign} of the kinetic term,
as is reflected in the standard minisuperspace action (\ref{Eq:MiniAct}).
The same \emph{wrong sign} is present in CDT, the Regge action (\ref{Eq:SRegge})
is also unbounded from below, but the regularization makes the action finite for each individual triangulation.
There is however a strong evidence \cite{Dasgupta, Antoniadis}, that after integrating out all degrees of freedom,
except the scale factor, 
which means taking into account the non-perturbative measure, 
one obtains a \emph{positive} kinetic term in (\ref{Eq:MiniAct}).
As we shall see in the next Chapter, 
this is exactly what happens in four-dimensional CDT.
The non-perturbative path integral over causal triangulations
takes into account both the entropy factor, counting the number of configurations,
and the \emph{bare} action.
As a consequence, 
the \emph{effective} action for $N(i)$
is equal to the minisuperspace action (\ref{Eq:MiniAct})
with an opposite sign and removed unboundedness.

For sufficiently small values of $G$ the curvature term, 
generating \emph{negative} sign of the kinetic term,
will overcome the entropy factor,
and histories with large time oscillations of the scale factor
will dominate the path integral. 
In the Appendix A we derive relations between \emph{bare} coupling constants,
which show that $K_0$ is approximately proportional to $\frac{1}{G}$.
For large values of $K_0$ we expect to observe triangulations
with large oscillations in time direction,
and this is exactly what happens in phase $A$.

Nonetheless, the \emph{overall sign} of the action does not affect 
the classical solution of equations of motion.
For the Lagrangian $L = - a \dot a^2 - a + \frac{1}{3} \Lambda a^3$,
the Euler-–Lagrange equation reads
\beq
\frac{\partial L}{\partial a} - \frac{\dd}{\dd \tau} \frac{\partial L}{\partial \dot a} 
=  \dot a^2 + 2 a \ddot a + \Lambda a^2 - 1 = 0. \nonumber
\label{Eq:MiniMotion}
\eeq
The classical trajectory, solving the above nonlinear differential equation,
is given by
\beq
\bar a(\tau) = R \cos \left(\tau/R\right), \quad R = (\Lambda / 3)^{-1/2}. 
\label{Eq:ATrajectory}
\eeq
Turning back to the spatial volume variable, 
the minisuperspace action (\ref{Eq:MiniAct}) can be rewritten as
\beq
S[v] = - \frac{1}{24 \pi G} \int \dd \tau \left(\frac{\dot v^2}{v} + \beta v^{1/3} - 3 \Lambda v \right).
\quad \quad \beta = 9 (2 \pi^2)^{2/3},
\label{Eq:MiniActV}
\eeq
The classical trajectory (\ref{Eq:ATrajectory}) of physical volume equals 
\beq
\bar{v}(\tau) = 2 \pi^2 R^3 \cos^3 \left(\frac{\tau}{R} \right).
\label{Eq:Trajectory}
\eeq
The physical volume $\bar{v}(\tau)$ describes
the maximally symmetric space for a positive cosmological constant,
namely the Euclidean \emph{de Sitter Universe}
or a geometry of a four-sphere $S^4$ with radius $R$.
This result is in 
agreement with the relation (\ref{Eq:AvNi}) for $\bar N(i)$ found in numerical simulations.
The de Sitter space \emph{emerges dynamically} as a \emph{background geometry} in the CDT model.
In the next Chapter we give a direct evidence,
that quantum fluctuations of the spatial volume $N(i)$
are also governed by the minisuperspace action (\ref{Eq:MiniActV}),
up to the \emph{overall sign}.

\section{The four dimensional spacetime}
\label{Sec:FourDimensional}

In this Section, we prove that
the Universe coming out in the CDT model is genuinely four dimensional.
To support this statement we check the scaling properties and measure the spectral dimension
of the appearing triangulations.

{\bf Scaling dimension.}
Up to now, the measurements were performed for only one value of the total volume $\Ntot$.
Keeping the coupling constants $K_0$ and $\Delta$ fixed,
which na\"ively means that the geometry of simplices is not changed,
we measure the expectation value $\bar{N}(i)$ for different total volumes $\Ntot$.

For the scaling dimension $d_H$ time intervals should scale as $\Ntot^{1/d_H}$
and the volume-independent time coordinate $t$ scales as a function of a discrete time $i$ as
\beq
t \equiv \Delta t \cdot  i, \quad \Delta t \equiv \Ntot^{-\frac{1}{d_H}} .
\label{Def:Scaledt}
\eeq
In the continuum limit $\Ntot \to \infty$, the time step should vanish $\Delta t \to 0$.
Each configuration is composed of a \emph{stalk} and a \emph{bulk}.
The kinematical constraint limits the minimal volume of slices
lying in the thin \emph{stalk}.  
Hence, one should use the effective total volume
and subtract simplices which, out of necessity, lie in the stalk.
Therefore, in the scaling relations (\ref{Def:Scaledt}) and (\ref{Def:nt})
appears the effective volume $\Ntot = \frac{1}{2} N_{41} - T s$,
where $s \approx 10$ is  the average spatial volume in the \emph{stalk}.
This subtraction has a very small effect 
for large configurations and vanishes in the continuum limit,
but might be relatively important for the smaller ones.

\begin{figure}[t]
\centering
\includegraphics[width=0.9\textwidth]{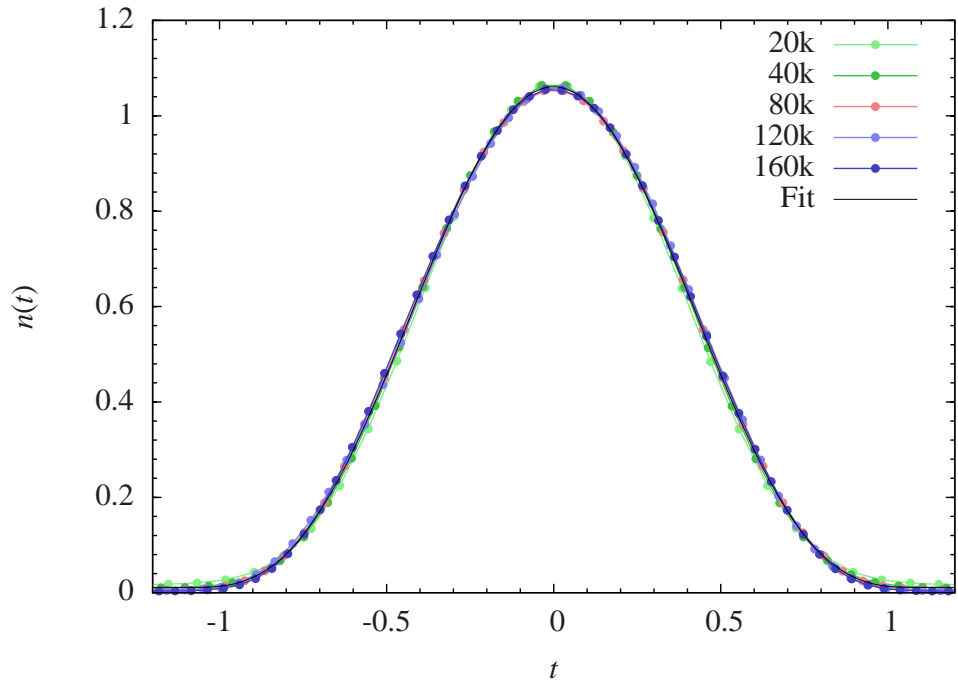}  %
\caption{
Average scaled spatial volume $\bar{n}(t)$ for a variety of total volumes $\Ntot$
calculated for the scaling dimension $d_H = 4$.
Measured in Monte Carlo simulations for $K_0 = 2.2$ and $\Delta = 0.6$.
We omit the error bars not to obscure the picture.
The thin black line plots the fit $\bar{n}(t) = \frac{3}{4 B} \cos^3 \left(t / B\right)$,
where $B = 0.69$. 
}
\label{Fig:ScaledAvNi}
\end{figure}

To compare spatial volume distributions $N(i)$ for geometries with different volumes $\Ntot$,
we introduce the scaled three-volume $n(t)$.
It has to be normalized and compatible with the time scaling (\ref{Def:Scaledt}).
This is satisfied by the following definition,
\beq
n(t) \equiv \Ntot^{-1+\frac{1}{d_H}}  N(i), \quad \bar{n}(t) = \langle n(t) \rangle  .
\label{Def:nt}
\eeq
For very large $\Ntot$ the time interval $\Delta t$ is close to zero
and in the continuum limit the sum over discrete time steps can be replaced by an integral
\beq
\int \dd t \ldots \leftrightarrow \sum_i \Delta t \ldots .
\label{Eq:ContT}
\eeq
Definition (\ref{Def:nt}) of the scaled spatial volume $n(t)$ 
and equation (\ref{Eq:SumNi}) for the total discrete volume $\Ntot$
provide the normalization condition 
\[  \int n(t) \dd t = \frac{1}{\Ntot} \sum_i N(i) = 1. \]

\begin{figure}
\centering
\includegraphics[width=0.9\textwidth]{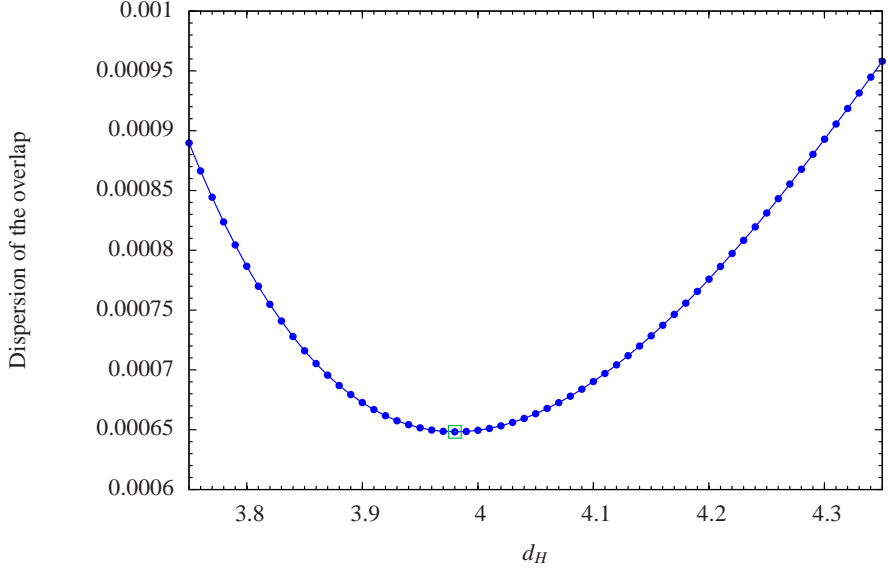} %
\caption{
Error of the overlap of scaled spatial volumes
shown in Fig. \ref{Fig:ScaledAvNi} for various values of the scaling dimension $d_H$.
The point corresponding to the best fit is marked with a rectangle.}
\label{Fig:ErrorDH}
\end{figure}
Now it is possible to directly compare $n(t)$ for various total volumes
and check for which value of the scaling dimension $d_H$ the overlap is the best.
To estimate the optimum  value of $d_H$, we use a similar method as in \cite{Reco}.
For given values of coupling constants $K_0$, $\Delta$ and total volume $\Ntot$,
the average $\bar N(i)$ is measured and scaled in accordance with (\ref{Def:Scaledt}) and (\ref{Def:nt})
for fixed $d_H$,
\[ (i,\ \bar{N}(i) ) \to \left (t = \Ntot^{-\frac{1}{d_H}} i,\ \bar n(t) = \Ntot^{-1+\frac{1}{d_H}} \bar{N}(i) \right) .\]
The new series of points is then interpolated.
Fig. \ref{Fig:ScaledAvNi} shows the scaled three-volumes $n(t)$ using $d_H = 4.0$
for several values of total volumes $\Ntot$.
In order to obtain the best fit of $d_H$, the error function is defined as follows.
For a given value of the scaled time $t$,
there is one value of the interpolated curve $n(t)$ per each total volume $\Ntot$.
The error function is defined as the variance of these values
averaged over a number (of order $T$) equally distributed points $t$ in the bulk region. 
The scaling dimension $d_H$ is chosen as the value corresponding to the minimum of the error function , 
i.e. to the best overlap of scaled spatial volumes $n(t)$.
Fig. \ref{Fig:ErrorDH} shows the dependence of the error function on $d_H$.
The minimum corresponds to
\beq
d_H = 3.98 \pm 0.10.
\label{Eq:DHFit}
\eeq
The error of determination of $d_H$ was estimated using the Jackknife method \cite{Efron, Efron2}.
The expected value $d_H = 4$ is very close to the measured result,
and is well within the margin of error.
The result gives strong evidence, that the \emph{Universe} which emerges in Causal Dynamical Triangulations
is genuinely \emph{four-dimensional}.

Since $\bar{n}(t)$ is normalized and is obtained by the scaling of $\bar{N}(i)$
which is given by equation (\ref{Eq:AvNi}), it is expressed by the formula
\beq
\bar n(t) = \frac{3}{4 B} \cos^3 \left( \frac{t}{B} \right), 
\label{Eq:ntcos}
\eeq
where $B$ depends only on the coupling constants $K_0$ and $\Delta$,
but not on $\Ntot$. 
For $K_0 = 2.2$ and $\Delta = 0.6$,  the measured values is $B \approx 0.69$.
The curve (\ref{Eq:ntcos}) with adjusted $B$ is drawn with a black line in Fig. \ref{Fig:ScaledAvNi},
the fit is remarkably good.

From equations (\ref{Def:nt}) and (\ref{Eq:ntcos})
and the scaling dimension $d_H = 4$ results a following expression 
for the three-volume $\bar{N}(i)$
\beq
\bar{N}(i) =  \frac{3}{4} \frac{\Ntot^{3/4}}{B} \cos^3 \left( \frac{i}{B \Ntot^{1/4}} \right).
\label{Eq:AvNiFull}
\eeq
As expected for a four-dimensional spacetime,
the time extent $T_{univ}$ of the \emph{blob}, measured in units of time steps, 
scales as $T_{univ} \sim \pi B \cdot \Ntot^{1/4}$.
The expression (\ref{Eq:AvNiFull}) specifies (\ref{Eq:AvNi}) 
and is only valid in the extended part of the Universe
where the spatial three-volumes are larger than the minimal cut-off size.

{\bf Physical volume.}
Let us relate the discrete spatial volume $N(i)$ with the physical volume
$v(\tau)$ of hypersurfaces of a constant time.
The classical solution of the physical volume $\bar{v}(\tau)$ is given by the formula (\ref{Eq:Trajectory}),
while the average discrete volume $\bar{N}(i)$ by the formula (\ref{Eq:AvNiFull}).
Up to some factors they are expressed by the same function.
Henceforth, we make a key assumption that the average configuration described by $\bar{N}(i)$
in fact has a geometry of a four-sphere $S^4$ given by $\bar{v}(\tau)$.
The physical total four-volume of a four-sphere with a radius $R$ equals 
\beq
V_4 = \int_{-\frac{\pi}{2} R}^{\frac{\pi}{2} R} \bar{v}(\tau) \dd \tau = \frac{8 \pi^2}{3} R^4.
\label{Eq:V4Cont}
\eeq
On the other hand, the four-volume of a triangulations is given by (\ref{Eq:ReggeVol}),
\beq
V_4 = \textrm{Vol}^{\{4, 1\}} a^4 N_{41}  + \textrm{Vol}^{\{3, 2\}} a^4 N_{32} = \left(\textrm{Vol}^{\{4, 1\}} + \xi \textrm{Vol}^{\{3, 2\}}\right) a^4 N_{41} = C_4 a^4 \Ntot, 
\label{Eq:V4Disc}
\eeq
where $\xi = \frac{N_{32}}{N_{41}}$ and $C_4 = 2 \left(\textrm{Vol}^{\{4, 1\}} + \xi \textrm{Vol}^{\{3, 2\}}\right)$,
which is interpreted as the average four-volume shared by one spatial tetrahedron.
Here, $a=a_s$ is the cut-off length, i.e. the lattice constant.
The continuum time $t$ defined by (\ref{Def:Scaledt}) and the discrete time $i$
are proportional to the proper time $\tau$ (cf. (\ref{Eq:MiniMetric})), 
\beq
 \tau = \sqrt{g_{tt}} \cdot t = \sqrt{g_{tt}} \cdot \Delta t \ i, \quad \Delta t = \Ntot^{-1/4} .
\label{Eq:ContTau}
\eeq
Therefore, a slab between slices $i$ and $i + 1$ has a proper-time extent $\Delta \tau$
and a four-volume
\beq
 v(\tau) \Delta \tau = v(\tau) \sqrt{g_{tt}} \Delta t = C_4 a^4 N(i) = \Ntot^{3/4} C_4 a^4 n(t). 
\label{Eq:VtauNi}
\eeq
The above equation is consistent with formulas (\ref{Eq:V4Cont}) and (\ref{Eq:V4Disc})
which determine the total four-volume of the emerging
de Sitter space with a radius $R$ equal
\beq
R = \left (\frac{3 C_4 \Ntot}{8 \pi^2} \right)^{1/4} a .
\label{Eq:Radius}
\eeq
In the next Chapter, we will use the above relation to derive the renormalized value
of the cut-off $a$.
Comparing formulas (\ref{Eq:Trajectory}) and (\ref{Eq:ntcos})
it follows that the proper-time extent of the de Sitter Universe is 
$\pi R$, while in terms of the time $t$ it is equal $\pi B$, 
hence
\beq
\sqrt{g_{tt}} = \frac{\tau}{t} = \frac{R}{B}. 
\label{Eq:gtt}
\eeq
Assuming such scaling relations between physical and discrete volume (cf. (\ref{Eq:VtauNi})),
and between proper and discrete times (cf. (\ref{Eq:ContTau})),
we ensure that the empirically derived formulas (\ref{Eq:ntcos}) or (\ref{Eq:AvNiFull})
describe a Euclidean de Sitter Universe for all $\Ntot$.

{\bf Spectral dimension.}
Another quantity revealing information about the geometry 
is related to the diffusion phenomena,
namely it is the spectral dimension $d_S$.
On a $d$-dimensional Riemannian manifold with a metric $g_{\mu \nu}(\bx)$,
the probability density $\rho(\bx, \bx_0; \sigma)$
describes the probability of finding a diffusing particle at position $\bx$
after some fictitious diffusion time $\sigma$ with an initial position at $\sigma = 0$ fixed at $\bx_0$.
The evolution of  $\rho(\bx, \bx_0; \sigma)$ is controlled by the diffusion equation
\beq
\partial_\sigma \rho(\bx, \bx_0; \sigma) = \triangle_g \rho(\bx, \bx_0; \sigma),
\label{Eq:Diffusion}
\eeq
with the initial condition 
\beq
\rho(\bx, \bx_0; \sigma = 0) = \frac{1}{\sqrt{\det g(\bx)}} \delta(\bx - \bx_0),
\label{Eq:DiffusionInitial}
\eeq
where $\triangle_g$ is the Laplace operator corresponding to $g_{\mu \nu}(\bx)$.
The \emph{return probability} describes the probability of finding a particle at the initial point
after diffusion time  $\sigma$.
The \emph{average return probability} $P(\sigma)$, supplying a global information about the geometries,
is given by  
\[ P(\sigma) = \left\langle \frac{1}{V_4} \int \dd^d \bx \sqrt{\det g(\bx)} \rho(\bx, \bx; \sigma)\right \rangle \ ,\]
where $V_4 = \int \dd^d \bx \sqrt{\det g(\bx)}$ is the total spacetime volume and the average is also performed over the ensemble of geometries.
For infinite flat manifolds the spectral dimension $d_S$ can be extracted from the return probability
due to its definition,
\beq
d_s \equiv -2 \frac{\dd \log P(\sigma)}{\dd \log \sigma}.
\label{Def:Spectral}
\eeq
For Euclidean flat manifold $\R^d$, the spectral and Hausdorff dimensions
are equal to the topological dimension, $d_S = d_H = d$.
For the four-sphere $S^4$, the spectral dimension $d_S = 4$ for short diffusion times,
while for very large times, because of the finite volume, the zero mode of the Laplacian will dominate
and, with the above definition, $d_S$ will tend to zero.

\begin{figure}[t]
\centering
\includegraphics[width=0.9\textwidth]{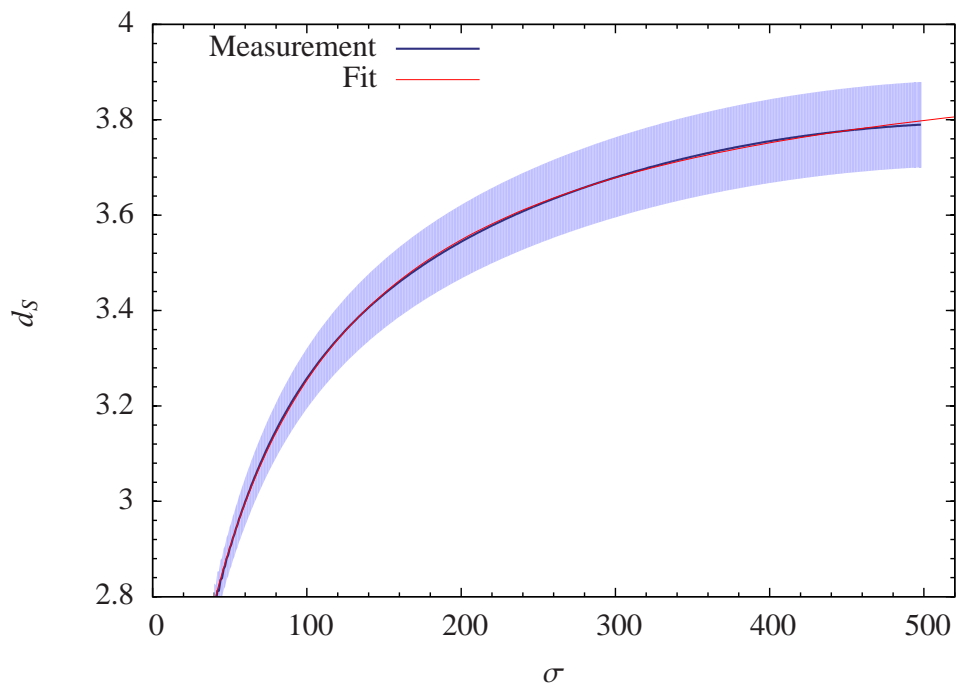}
\caption{
The spectral dimension $d_S$ of the Universe as a function of diffusion time $\sigma$,
measured for $K_0 = 2.2, \Delta = 0.6$ and $N_{4} \approx 368\textrm{k}$.
The blue curve plots the average measured spectral dimension,
while the highlighted area represents the error bars.
The best fit $d_S(\sigma) = 4.02 - \frac{120}{58+\sigma}$ is drawn with red line.
}
\label{Fig:Spectral4D}
\end{figure}
Definition (\ref{Def:Spectral}) is particularly convenient
because it is easy to perform numerical simulations 
which measure the return probability.
In the CDT framework, the spacetime geometry is regularized by piecewise flat manifolds
built of four-simplices.
Let us recall, that after the Wick rotation spacetimes
appearing in the model are Riemannian manifolds equipped with the positive-definite metric tensor.
The diffusion process can be carried out
by implementing the discretized version of the diffusion equation (\ref{Eq:Diffusion})
\[\rho(i, i_0; \sigma + 1) - \rho(i, i_0; \sigma) = 
\Delta \sum_{j \leftrightarrow i} \big(\rho(j, i_0; \sigma) - \rho(i, i_0; \sigma)\big),\]
where $\Delta$ denotes the time step
and the sum is evaluated over all simplices $j$ adjacent to $i$.
Here variables  $i_0, i$ and $j$ denote the labels of simplices
and the diffusion process is running on the dual lattice, 
i.e. the probability flows from a simplex to its neighbors.
Since each simplex has exactly five neighbors,
it is convenient to set $\Delta = \frac{1}{5}$ and the diffusion equation reads
\beq
\rho(i, i_0; \sigma + 1) = \frac{1}{5} \sum_{j \leftrightarrow i} \rho(j, i_0; \sigma).
\label{Eq:DiscreteDiffusion}
\eeq
For a fixed triangulation, consisting of $N_4$ simplices,
the procedure of evaluating $\rho(i, i_0; \sigma)$ is as follows \cite{Reco, Spectral}:
\begin{itemize}
\item We pick an initial four-simplex $i_0$.
Because we are interested in the bulk features,
and the volume distribution as a function of time
is not uniform,
the diffusion always starts from a simplex 
lying in the central slice $i_{CV}$.
We impose the following initial condition (\ref{Eq:DiffusionInitial})
\[ \rho(i, i_0; \sigma = 0) = \delta_{i\, i_0}. \]
\item We iterate the diffusion equation (\ref{Eq:DiscreteDiffusion}) and 
calculate the probability density $\rho(i, i_0; \sigma)$ for consecutive diffusion steps $\sigma$.
\item Finally, we repeat the above operations for a number of random starting points $i_0$ 
adjacent to the central slice (because a simulation of the diffusion process is quite \emph{time-consuming} 
it is repeated $K = 100$ times per configuration)
and calculate the average return probability 
\[ P(\sigma) = \frac{1}{K} \sum_{i_0 = 1}^{K} \rho(i_0, i_0; \sigma). \]
\end{itemize}
In numerical simulations the return probability $P(\sigma)$ is averaged over a number of triangulations
($\sim 1000$) and the spectral dimension $d_S$ is calculated from the definition (\ref{Def:Spectral}).
Fig. \ref{Fig:Spectral4D} shows the spectral dimension $d_S$ as a function of the diffusion time steps $\sigma$
, in the range $40 < \sigma < 500$.
For small values of $\sigma$ ($ < 30$) lattice artifacts are very strong
and the spectral dimension becomes irregular.
Because of the finite volumes of configurations, 
for very large $\sigma (\gg 500)$, the spectral dimension $d_S$ falls down to zero.
In the presented range,
the measured spectral dimension $d_S$ is very well expressed by the formula
\beq
d_S(\sigma) = a - \frac{b}{c + \sigma} = 4.02 - \frac{120}{58 + \sigma},
\label{Eq:dSfit}
\eeq
where variables $a, b$ and $c$ were obtained from the best fit.
As observed, the spectral dimension depends on a diffusion time,
and thus it is \emph{scale dependent}.
Small $\sigma$, means that the diffusion process probes only the nearest vicinity of the initial point.
Extrapolation of results gives the \emph{short-distance} limit of the spectral dimension
\[ d_S(\sigma \to 0) =  1.95 \pm 0.10 .\]
In the \emph{long-distance} limit the spectral dimension tends to
\[ d_S(\sigma \to \infty) =  4.02 \pm 0.05 .\]
The measurements presented here were performed on
configurations twice as large as those used in \cite{Spectral, Reco},
but results are the same.
The short-range value of the spectral dimension $d_S = 2$,
much smaller than the scaling dimension $d_H$,
suggests a fractal nature of geometries
appearing in the path integral at short distances.
This statement is supported by results presented in Chapter \ref{Chap:Slice}
where we give a direct evidence of the fractal nature of spatial slices.
At long distances $d_S = 4$, and configurations resemble a smooth manifold,
as we show in the next Section. 
Amazingly such nontrivial scale dependence of the spectral dimension of the quantum spacetime,
the same infra-red ($d_S = 4$) and ultra-violet ($d_S = 2$) behavior,
is also present in the Ho\v{r}ava-Lifshitz gravity \cite{HoravaSpectral}
and Renormalization Group approach \cite{LauscherFractal}.

\section{Geometry of the Universe}
\label{Sec:Geometry4D}

\begin{figure}
\centering
\includegraphics[width=0.9\textwidth]{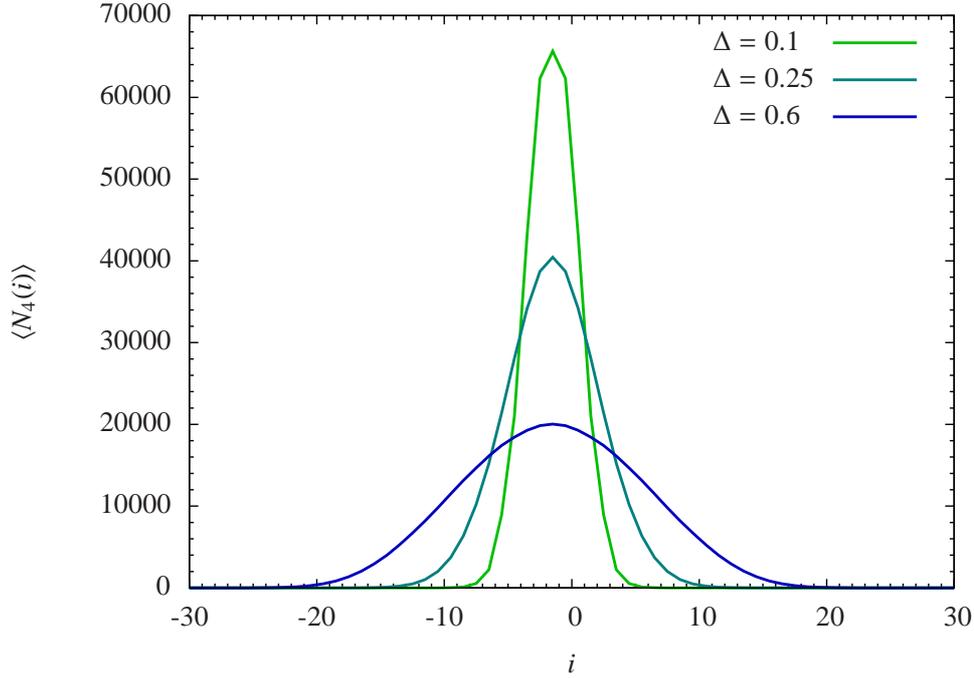}   %
\caption{The average time slab four-volume $\langle N_4(i) \rangle$ shown for $\Delta = 0.1, 0.25, 0.6$ and $K_0 = 2.2$.
$N_4(i)$ denotes the number of all four-simplices contained between slices $i$ and $i + 1$.}
\label{Fig:NiDelta}
\end{figure}

In the previous Sections we identified the background geometry with the geometry of a four-sphere.
The purpose of this Section is to describe the geometry in more detail,
with a particular emphasis on approaching the $B$-$C$ transition line
by varying the coupling constant $\Delta$.
We show that typical configurations in phase $C$ resemble a prolate spheroid stretched in the time direction,
and do not exhibit a fractal nature. 
The three-volume $N(i)$ was defined as a number of 
tetrahedra in the slice $i$, of a centered triangulation.
For convenience, here we will use a slightly different definition of volume.
The spacetime slab between slices $i$ and $i + 1$ has a four-volume denoted by $N_4(i)$, 
which includes simplices of all types $\{4, 1\}$ and $\{3, 2\}$.
The two definitions are approximately equivalent, since $N_4(i) \propto N(i)$.
Fig. \ref{Fig:NiDelta} presents the average slab four-volume $N_4(i)$ 
for three different values of the coupling constant $\Delta$.
In addition to $N_4(i)$, we introduce the radial four-volume $\tilde{N}_4(r)$.
The shell of radius $r$ is defined as a group of simplices
whose \emph{distance} from some initial simplex $i_0$ equals $r$.
This distance is defined as the length of the shortest geodesic connecting the two points.
Although we may determine the exact geodesic on a piecewise linear manifold,
we use a discrete \emph{geodesic distance}
defined as the length of the shortest path between successive centers of neighboring four-simplices.
We expect that our definition of a geodesic will in the continuum limit, 
i.e. on scales sufficiently large compared to the cut-off scale, lead to the same geometric results as 
for exact geodesics.
The number of simplices contained in a shell of radius $r$ is denoted as $\tilde N_4(r)$.
The initial four-simplex is located at $r = 0$, thus $\tilde N_4(r = 0) = 1$.
Each simplex has five neighbors: $\tilde N_4(r = 1) = 5$.
Moving out we find successive shells.
Moreover, we introduce a partition of simplices belonging to a slice $r$
with respect to the time slab $i$ they lie in.
The number of simplices lying simultaneously in shell $r$ and time slab $i$
is denoted as $N_4(i, r)$ and the following relations are satisfied,
\[ N_4(i) = \sum_r N_4(i, r), \quad \tilde N_4(r) = \sum_i N_4(i, r), \]
summing up to the total volume
\[ N_4 = \sum_i N_4(i) = \sum_r \tilde N_4(r). \]

{\bf The fractal structure.}
To examine a fractal nature of a specific configuration,
we define a \emph{diffusion tree}, which measures the
connectedness of radial shells.
Two simplices in a shell of radius $r$
are considered to be connected if one can find a path connecting them
passing only through simplices from the outer shells $r' \geq r$.
A \emph{diffusion tree} is created in the following way:
\begin{enumerate}
\itemsep=0pt
\item For a given configuration we choose the initial simplex $i_0$ from the central slab.
\item We put this simplex to the shell $r = 0$ and remove it from the triangulation.
\item We find the first shell, i.e. all survived simplices which were connected to the original simplex.
We remove all simplices in this shell from the configuration.
\item We find the next shell, i.e. all survived simplices which were connected to at least one simplex from the previous shell.
We remove this shell.
\item We repeat step 4. until no simplices remain. 
After each removal, we check a number of disconnected parts of the remaining triangulation.
\item A set of simplices belonging to the same shell and the same connected component (they may be connected through simplices in further shells,
but not by the removed ones) is represented as a vertex on the graph.
\item A vertex is linked to another vertex in the previous shell, if the corresponding components share neighboring simplices.
Except for the original simplex, each vertex has exactly one \emph{parent}.
\end{enumerate}
\begin{figure}
\centering
\includegraphics[height=9cm]{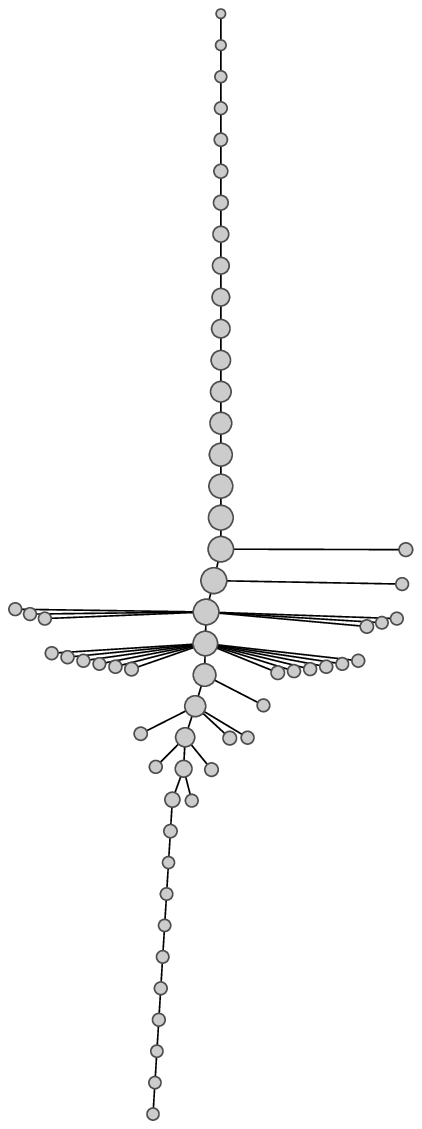}
\includegraphics[height=9cm]{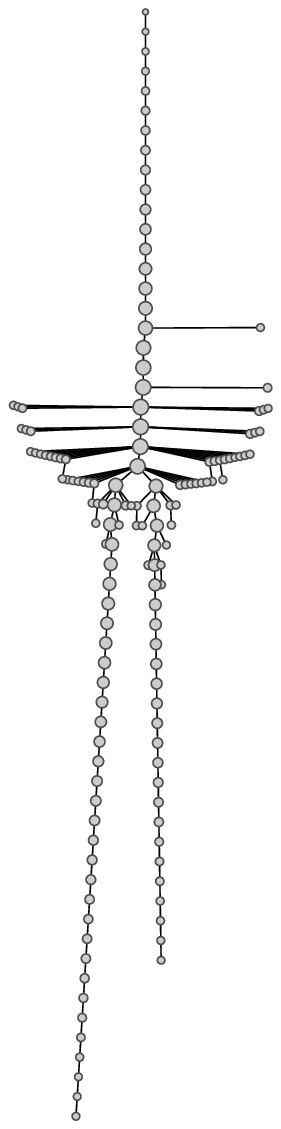}
\includegraphics[height=9cm]{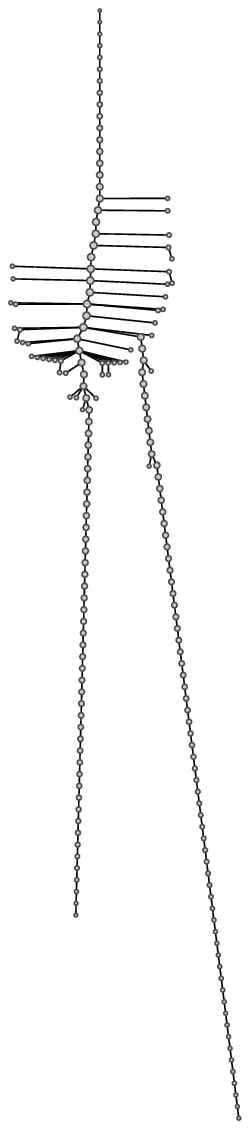}
\caption{
Graphs of the diffusion trees.
From left to right the figures correspond to spacetime configurations measured for $\Delta = 0.10, 0.25$ and $0.60$.
Each vertex represents a connected group of simplices in one shell.
The top node contains only the starting four-simplex.
The distance from the top node equals the shell radius $r$. 
For $\Delta = 0.25$ and $0.60$ the bifurcation of the branch at radius $r_{bif}$ is observed.
}
\label{Fig:DiffTree}
\end{figure}
By construction such a graph forms a tree, 
and the procedure of creating it imitates the diffusion process,
as the shells follow the front of diffusion.
Fig. \ref{Fig:DiffTree} illustrates the \emph{diffusion trees} 
obtained for typical configurations corresponding to three points
with $\Delta = 0.10, 0.25, 0.60$ and $K_0 = 2.2$. 
Only vertices which volume, together with sub-leaves,
is larger than $40$ are visible.
Each node represents a connected component of a shell
and its distance from the top node is equal to the shell radius.
Vertices are drawn as circles whose diameter depends on the size of 
the components they represent.

The average volume $\langle N_4(i) \rangle$ for different values of $\Delta$
is presented in Fig. \ref{Fig:NiDelta}.
Similarly as $\bar{N}(i)$, shown in Fig. \ref{Fig:AvNi},
$N_4(i)$ is expressed by formula $N_4(i) \propto \cos^3 i /W$.
This behavior corresponds to a four-dimensional spheroid,
with three spatial radii of equal length.
Although we know the time extent of the spheroid (cf. Fig. \ref{Fig:NiDelta}),
so far we can not say much about the length of the equator (or equivalently spatial radii) as
we have no information about the 
geometric nature of spatial slices.
An individual configuration usually is given by a deviation of the mean geometry.

\begin{figure}
\centering
\includegraphics[width=0.98\textwidth]{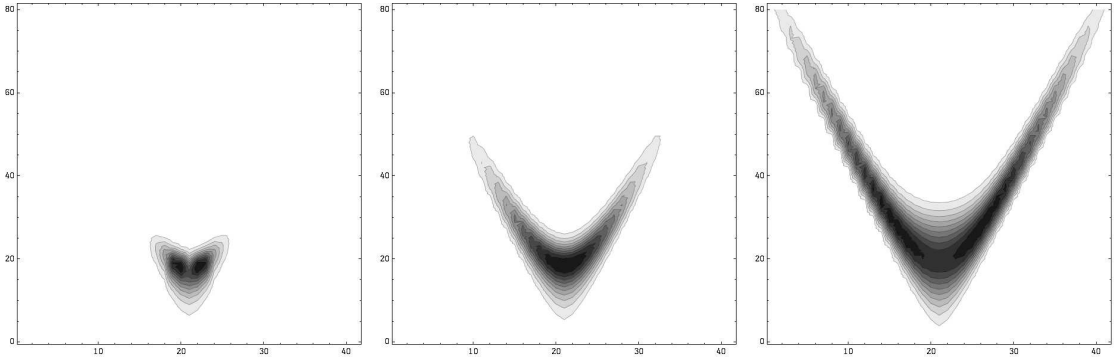}
\caption{The contour plots of $\langle N_4(i,r) \rangle$ for $\Delta = 0.10, 0.25, 0.6$ and $K_0 = 2.2$.
Darker colors mean larger values of $\langle N_4(i,r) \rangle$.
The horizontal axis denotes the discrete time $i$. 
The vertical axis denotes the radius $r$.
}
\label{Fig:Rogi}
\end{figure}

Studying the properties of \emph{diffusion trees}, illustrated in Fig. \ref{Fig:DiffTree},
we indicate that typical configurations, 
corresponding to the two largest values of $\Delta = 0.25, 0.6$,
resemble a prolate spheroid elongated in time direction \cite{Geometry}.
Let us assume that the geometry is indeed a prolate spheroid,
for convenience we can imagine a two-dimensional ellipsoid or a surface of a \emph{cigar}.
For a given triangulation, we start the diffusion 
at the central slice, i.e. on the smallest equator.
At the beginning, the diffusion front propagates in concentric circles,
forming a connected shell, this is illustrated by the main top branch
in Fig. \ref{Fig:DiffTree}. There is only a small number of short outgrows.
However, at some point the diffusion front reaches the antipode
of the starting point and \emph{bifurcates} into two disconnected
fronts moving in opposite directions toward the tips of the spheroid.
This point is denoted by the \emph{bifurcation} radius $r_{bif}$.
The \emph{bifurcation} radius $r_{bif}$
is visible on the graphs of \emph{diffusion trees} in Fig. \ref{Fig:DiffTree}
as the point at which the main branch \emph{bifurcates} into two branches.
Here, $r_{bif}$ is identified with half of the shorter circumference
lying in the central slice $i_{CV}$,
and for $\Delta = 0.25, 0.6$ it is much shorter than the time extent.
While decreasing $\Delta$ the bifurcated branches become less distinct, and eventually vanish,
which suggests that the four-dimensional geometry approaches a spherical shape.
Moreover, the Figures \ref{Fig:DiffTree} indicate that the four-dimensional spacetime geometry
does not reveal a true fractal structure when looking at a shell decomposition of spacetime,
as only short outgrows are present.
It should be noted that the diffusion trees were obtained using a four-dimensional 
definition of the neighborhood.
By contrast, in Chapter \ref{Chap:Slice} we examine
a similar shell decomposition within a spatial three-dimensional hypersurface and show that a fractal structure
is indeed present. In the latter case we use a three-dimensional 
definition of the neighborhood of tetrahedra.

\begin{figure}[t]
\centering
\includegraphics[width=0.85\textwidth]{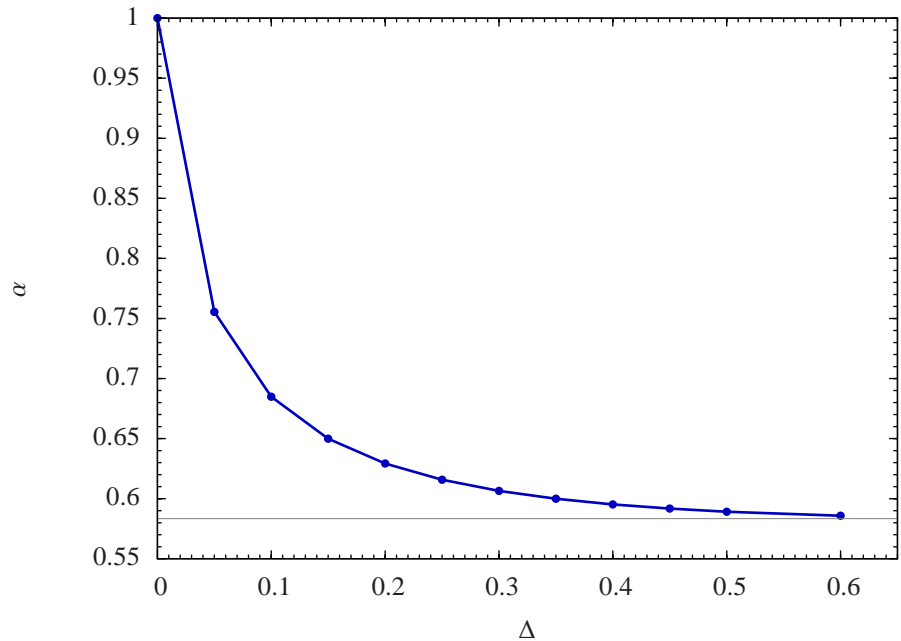}
\caption{
The asymmetry factor $\alpha$ 
between lengths of time-like and space-like edges of the triangulation
as function of $\Delta$ for $K_0 = 2.2$.
The thin horizontal line corresponds to the boundary value $\frac{7}{12}$.
}
\label{Fig:Alfa}
\end{figure}
Analyzing the volume distribution $N_4(i)$ (cf. Fig. \ref{Fig:NiDelta}) and \emph{diffusion trees} (cf. Fig. \ref{Fig:DiffTree})
we argued that geometry of quantum spacetimes appearing in CDT 
resembles a perturbed four-dimensional elongated spheroid.
To support this statement and the dependence on the coupling constant $\Delta$, 
we use a more quantitative observable, namely $\langle N_4(i, r) \rangle$.
In contrary to the diffusion trees, it does not refer to a single configuration,
but is averaged over the ensemble of triangulations and initial points
(for each configuration ca. $100$ starting simplices were chosen from the middle slab $i_{CV}$).
Fig. \ref{Fig:Rogi} presents contour plots of the distribution $\langle N_4(i, r) \rangle$ 
for $\Delta = 0.10, 0.25$ and $0.6$.
For large values of $\Delta > 0.15$ it assumes a $V$-shape in the time-radius plane.
This is an expected behavior for the elongated spheroid.
We start the measurements of $N_4(i, r)$ at $r = 0$ which is a simplex 
in the initial slab $i_{CV}$ which corresponds to a vertical line passing through 
the exact middle of the plot.
Thus, the starting point $N_4(i_{CV}, 0)$ matches to the bottom tips of the Figures \ref{Fig:Rogi}.
When considering further shells, with increasing radius $r$, i.e. going upward in Figures \ref{Fig:Rogi},
the shells spread over successive time slabs.
However, at the \emph{bifurcation} radius,
the shells eventually reach the antipodal point
completely covering the initial slab.
In this case the \emph{bifurcation} radius $r_{bif}$ corresponds to the
largest radius $r$ for which $\langle N_4(i_{CV}, r) \rangle $ is larger than zero.
For $r > r_{bif}$ shells move away from the original slab,
both $r$ and $i$ are growing causing the $V$-shape of the $\langle N_4(i, r) \rangle$ plot
and explaining the elongated shape of  quantum geometries.
As $\Delta$ decreases the V-shape becomes less pronounced (cf. Fig. \ref{Fig:Rogi}),
with an obvious interpretation that the shape of the
spheroid becomes more and more spherical, with approximately equal extent
in spatial and time directions.

So far, the time extent of the Universe
was described in terms of lattice spacing.
However, in four-dimensional CDT the additional coupling constant $\Delta$
is related to the asymmetry factor $\alpha$ of the lengths of time-like and space-like links.
It is defined by equation (\ref{Eq:AtAs}),
and affects the thickness of time slabs.
The factor $\alpha$ is a function of $K_0, \Delta$ and the critical value of $K_4$ 
which can be obtained by inverting equations (\ref{Eq:AppCoupling}).
When $\alpha$ grows, the slabs are getting wider.
As seen in Fig. \ref{Fig:Alfa} this happens when we decrease $\Delta$ and approach the $B$-$C$ phase transition line.
Indeed, together with the decreasing time extent of geometries in terms of a number of slabs,
the slabs themselves are getting wider and the two effects may compensate.

\begin{figure}
\centering
\includegraphics[width=0.85\textwidth]{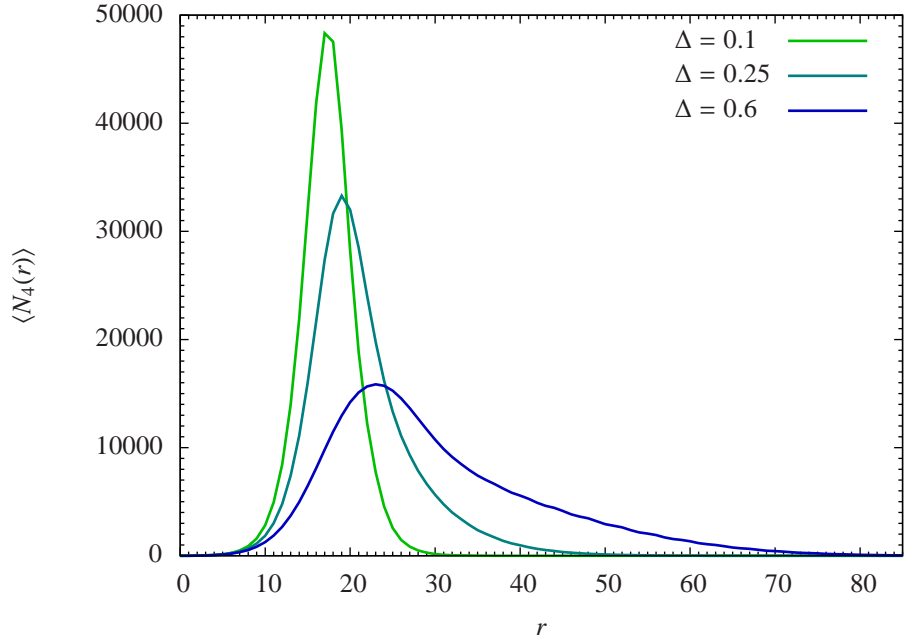}
\caption{The average radial volume $\langle \tilde{N}_4(r) \rangle$ for $\Delta = 0.1, 0.25$ and $0.6$.
$\langle \tilde{N}_4(r) \rangle$ is defined as the number of simplices located exactly at discrete distance $r$,
from an initial simplex lying in the middle slice $i_{CV}$ average both over configurations and starting points. 
}
\label{Fig:NrDelta}
\end{figure}

{\bf The sphericity.}
Let us quantify what we mean by a sphericity of a triangulation.
We introduce the sphericity factor $s$ to measure how spherical is the averaged geometry,
\beq
s \equiv \frac{\sum_{r=0}^{r_{bif}} \langle\tilde{N}_4(r)\rangle}{\sum_{r=0}^{r_{max}} \langle\tilde{N}_4(r)\rangle},
\label{Def:Sphericity}
\eeq
where the bifurcation radius $r_{bif}$ corresponds to the radius $r$ at which
$\langle N_4(i_{CV}, r) \rangle$ falls below some cut-off (here taken to be 4 to suppress the effect of large fluctuations).
$i_{CV}$ is the initial central slice.
Such definition is in agreement with the average bifurcation radius present in diffusion trees.

The distribution $\tilde{N}_4(r)$, as defined at the beginning of this Section,
denotes the number of four-simplices in a shell of radius $r$.
Fig. \ref{Fig:NrDelta} shows $\langle \tilde{N}_4(r) \rangle$ for various values of $\Delta$. 
For the smallest values of $\Delta$ the peak is nicely symmetric and well approximated by $A \cdot \sin^3(r/B)$, 
which corresponds to a sphere.
For larger values of $\Delta$, the elongation of the geometries
is reflected in a large-$r$ tail which cannot be fitted to $A \cdot \sin^3(r/B)$.

\begin{figure}
\centering
\includegraphics[width=0.85\textwidth]{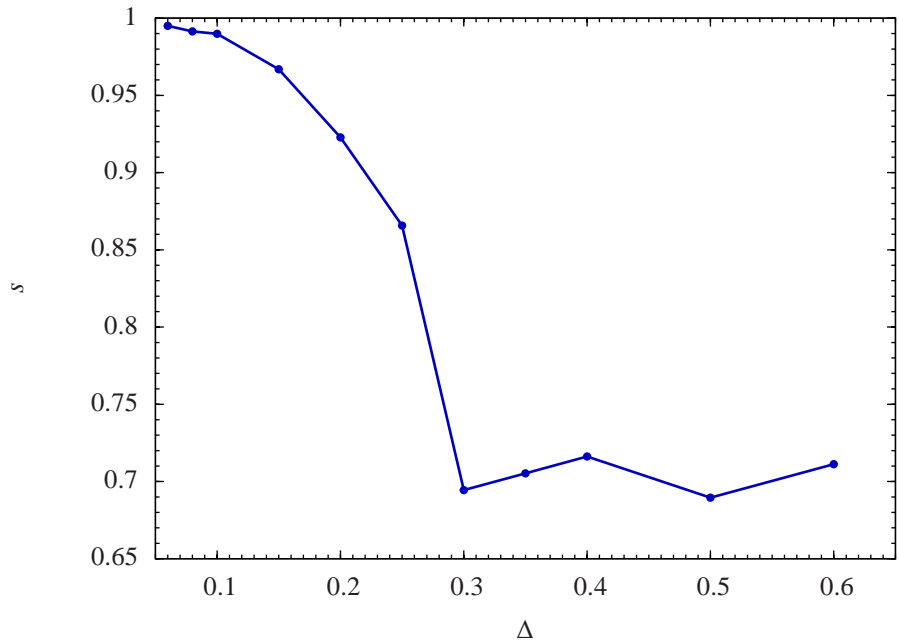}
\caption{Sphericity factor $s$ for $K_0 = 2.2$ and different values of coupling constant $\Delta$.}
\label{Fig:Sphericity}
\end{figure}

The sum in the denominator of ($\ref{Def:Sphericity}$) equals the total volume $N_4$ with the stalk cut away.
The sphericity factor denotes the part of volume which is located up to the bifurcation radius.
For a perfect sphere $S^4$ we have $s = 1$, as whole four-volume is  located between the origin and antipodal point.
For a very elongated object $s$ is close to zero.
Fig. \ref{Fig:Sphericity} shows the sphericity factor $s$, 
for data averaged both over configurations and initial points,
as a function of coupling constant $\Delta$ at fixed $K_0 = 2.2$.
When decreasing $\Delta$ and getting closer to phase $B$,
the sphericity factor grows and approaches to $s = 1$,
i.e. the shape of the \emph{average geometry}
changes from a prolate four-dimensional spheroid, elongated in time direction, and equal radii in spatial directions,
into an almost spherical shape.
This picture is consistent with what we observe for diffusion trees.

To conclude, the dynamically emerged \emph{background geometry} corresponds to the Euclidean de Sitter
solution of the \emph{minisuperspace} model.
It may be very well described by a prolate four-dimensional spheroid
for points deep inside phase $C$.
When approaching phase $B$ the \emph{Universe} is getting narrower 
in terms of lattice spacings, finally reaching a spherical shape.
However, this might be an illusion since one should take into account
the change of slab width.
Lastly, our more detailed investigation shows little or no evidence
of a fractality when looking at a shell decomposition of the spacetime.

\clearemptydoublepage

\chapter{Quantum fluctuations}
\label{Chap:Quantum}

{\noindent \it This chapter is based on the following publications:
(i)  J.~Ambj\o rn, A.~G\"orlich, J.~Jur\-kie\-wicz and R.~Loll,
  \emph{"Planckian Birth of the Quantum de Sitter Universe",}
  Phys. Rev. Lett. \textbf{100},  091304 (2008);
(ii) J.~Ambj\o rn, A.~G\"orlich, J.~Jurkiewicz and R.~Loll,
  \emph{"Nonperturbative quantum de Sitter universe"},
  Phys. Rev. D \textbf{78} (2008) 063544;
(iii)   A.~G\"orlich,
  \emph{"Background Geometry in 4D Causal Dynamical Triangulations"},
	Acta Phys. Pol. B \textbf{39}, 3343 (2008);
(iv) J.~Ambj\o n, A.~G\"orlich, J.~Jurkiewicz, R.~Loll, J.~Studnicki and T.~Trze\'s{}niewski,
  \emph{"The Semiclassical limit of Causal Dynamical Triangulations" in preparation}.
}\\[4ex]

In the previous Chapter we gave the arguments that the dynamically emerging 
background geometry can be obtained as a solution of the minisuperspace model.
Here we will check if quantum fluctuations around the classical trajectory
(\ref{Eq:Trajectory}) are also correctly described by the effective action (\ref{Eq:MiniActV}).
In this Chapter we investigate the properties of the \emph{semiclassical} limit of the 
lattice approach.
Nevertheless, it should be clearly stated that these
considerations are truly non-perturbative,
and take into account both a very important influence of the \emph{entropy factor},
which does not depend on bare coupling constants,
as well as the \emph{bare} action (\ref{Eq:SRegge}).

As before, we integrate out all degrees of freedom except the scale factor,
and restrict our deliberations to three-volume,
thereby reducing the problem to one-dimensional quantum mechanics.
Based on numerical data obtained by computer simulations,
we construct, within the semiclassical approximation, the effective action describing discrete spatial volume $N(i)$
and compare it with the minisuperspace action (\ref{Eq:MiniActV}).
The effective action comes into existence because of a subtle interplay between 
the entropy of configurations, which depends on the path integral measure, and the bare 
action (\ref{Eq:SRegge}).
Let us recall, that the key ingredient which allowed to 
perform a meaningful average of the volume over geometries,
is the removal of the translational zero mode.
The same is true in the case of quantum fluctuations of the spatial volume.

\newpage 
Let us denote the deviation of the three-volume $N(i)$ 
from the expectation value $\bar{N}(i)$ 
by
\[ \eta_i = N(i) - \bar{N}(i) .\]
Imitating the path integral approach to quantum mechanics,
$N(t)$ describes the position of a non-physical particle trajectory, 
giving a contribution to the partition function,
where now $t = i$ is the continuous time coordinate.
Likewise, $\eta(t)$ is a fluctuation from the classical trajectory $\bar{N}(t)$.
In the semiclassical approximation, 
the spatial volume fluctuations $\eta(t)$ are described by a Hermitian Sturm-Liouville operator $P(t)$,
obtained by the quadratic expansion of the effective action around the classical trajectory
\begin{equation}
S[N = \bar{N} + \eta] \approx S[\bar{N}] + \frac{1}{2} \int \eta(t) P(t) \eta(t) \dd t + O(\eta^3).
\label{Eq:ContExpS}
\end{equation}
In the CDT model, the time slicing 
carries with it the discreteness of time coordinate $i$,
thus the Sturm-Liouville operator is substituted by a matrix $\bP$,
and the expansion (\ref{Eq:ContExpS}) is
\begin{equation}
S[N = \bar{N} + \eta] \approx S[\bar{N}] + \frac{1}{2} \sum_{i, j} \eta_i \, \bP_{i j} \, \eta_{j} + O(\eta^3),
\label{Eq:DiscExpS}
\end{equation}
where the sum is performed over time slices $i, j = 1 \dots T$.

\begin{figure}
\begin{center}
\includegraphics[width=0.8\textwidth]{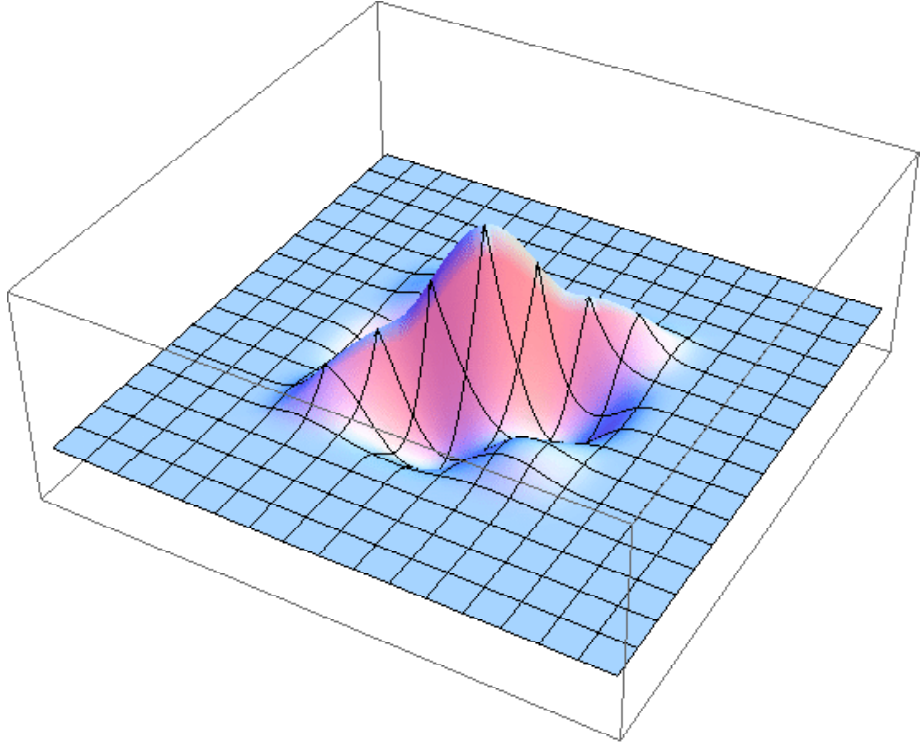}
\end{center}
\caption{The propagator $\bC_{i j} \equiv \langle (N(i) - \bar{N}(i))(N(j) - \bar{N}(j))\rangle$
measured in Monte Carlo simulations for $K_0 = 2.2$, $\Delta = 0.6$ and $\Ntot = 80\rk$.
}
\label{Fig:Cov}
\end{figure}

\begin{figure}
\begin{center}
\includegraphics[width=0.8\textwidth]{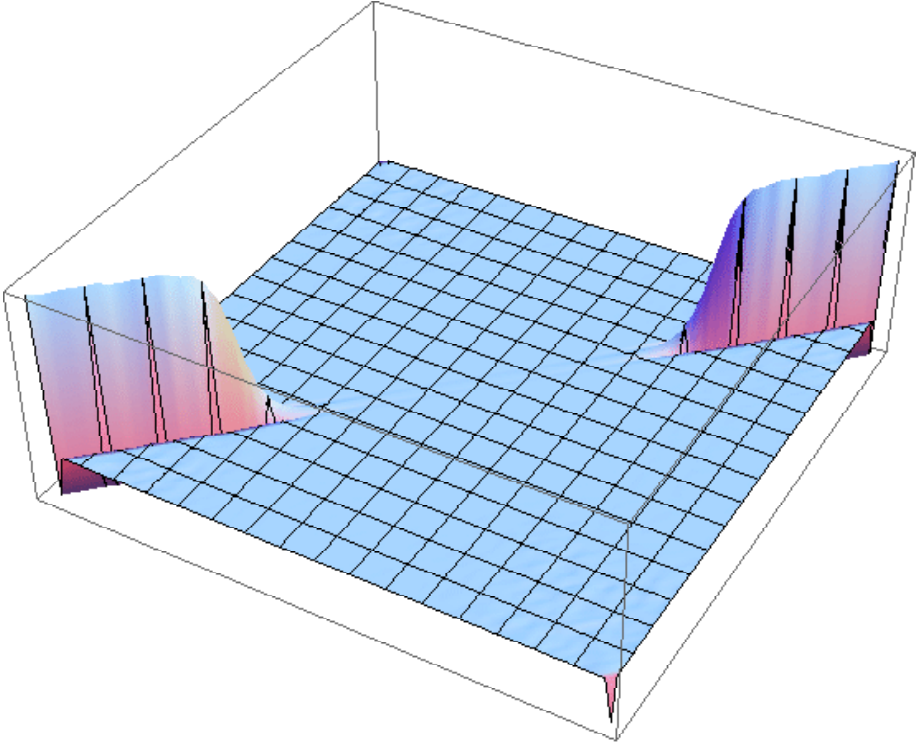}
\end{center}
\caption{The inverse propagator $\bP_{i j}$
extracted from the numerical data for $K_0 = 2.2$, $\Delta = 0.6$ and $\Ntot = 80\rk$.}
\label{Fig:Inv}
\end{figure}

The $\bP$ matrix carries information about quantum fluctuations and 
may be extracted from numerical data obtained by computer simulations.
In analogy to $\langle N(i) \rangle$ (cf. (\ref{Eq:MCAvNi})),
we measure the covariance matrix $\bC$ of volume fluctuations
using Monte Carlo techniques, described in detail in Chapter \ref{Chap:Implementation},
\begin{equation}
\bC_{i j} \equiv \langle \eta_i \eta_{j} \rangle \approx \frac{1}{K} \sum_{k = 1}^{K} \left( N^{(k)}(i) - \bar{N}(i)\right) \left( N^{(k)}(j) - \bar{N}(j)\right).
\label{Def:Prop}
\end{equation}
The covariance matrix $\bC$ is also called the propagator. 
If the quadratic approximation describes properly quantum fluctuations around the average $\bar{N}$,
the following direct relation between the propagator $\bC$ and the matrix $\bP$ is satisfied,
\[ \bC_{i j} = \frac{1}{Z} \int \eta_i \eta_{j} e^{-\frac{1}{2} \sum_{k, k'} \eta_k \bP_{k k'} \eta_{k'}} \prod_s \dd \eta_s = \bP^{-1}_{i j} , \]
where $Z$ denotes the normalization factor.
Fig. \ref{Fig:Cov} and Fig. \ref{Fig:Inv} present the plot of the measured covariance matrix $\bC$ and its inverse $\bP$
for coupling constants $K_0 = 2.2$, $\Delta = 0.6$ and the total volume $N_{tot} = 80\rk$.

For the numerical convenience, measurements were performed only for triangulations
with a fixed total volume $N_{tot} \equiv \sum_{i=1}^{T} N(i)$.
This constraint imposes on the covariance matrix $\bC$ the existence of a zero mode,
preventing from obtaining the $\bP$ matrix by a straightforward inversion of $\bC$.
For each configuration the sum of volume fluctuations vanishes, 
$\sum_{i} \eta_i = \sum_{i} N(i) - \bar{N}(i) = N_{tot} - N_{tot} = 0$.
Hence, the zero mode corresponds to a constant vector $e^{0}$,
\[ \sum_j \bC_{i j} e^{0}_j = 0, \quad e^{0}_j = \frac{1}{\sqrt{T}} .\]
In order to invert the matrix $\bC$
we project it on a subspace orthogonal to the zero mode $e^0$ and then perform the inversion.
This can be achieved in the following way.
First, we add by hand to $\bC$ the dyadic product of $e^0$ and $e^0$.
This shifts the zero eigenvalue, which now equals one,
and allows to invert the redefined matrix. 
Further the zero mode is recovered, by subtracting back the same dyadic product.
The matrix $\bP$, defined as the inverse of $\bC$ on a subspace orthogonal to the zero mode $e^0$,
is given by the following formula (cf. derivation of (\ref{Eq:PCA})),
\begin{equation*}
\bP = (\bC + \mathbf{A})^{-1} - \mathbf{A}, \quad \mathbf{A}_{i j} = e^{0}_i e^{0}_j= \frac{1}{T}.
\end{equation*}

The tentative procedure of dealing with the zero mode
may be avoided by modifying the action by a quadratic term
and allowing the system to perform measurements for an arbitrary total volume.
Such correction removes the artificial zero mode and simplifies the analysis,
but in fact does not affect results presented here.
Together with the extended analysis taking into account also 
half-integer slices built of $\{3, 2\}$ simplices,
the new method will be addressed elsewhere \cite{Semiclassical}.

After calculating the covariance matrix $\bC$, we can get the empirical Sturm-Liouville operator $\bP$
which can be compared with the predictions of the minisuperspace model. 
As may be seen from Fig. \ref{Fig:Inv},
the empirical $\bP$ matrix has to a very good approximation a tridiagonal structure.
Except for the diagonal and both sub-diagonals, the remaining part of the $\bP$ matrix 
becomes indistinguishable from a numerical noise.
Indeed, we will show that such  structure is expected from the minisuperspace action (\ref{Eq:MiniActV}).

The tridiagonal form suggests that the effective action describing fluctuations of $N(i)$ is quasi-local in time,
\beq
S[N(i)] = \sum_{i} f(N(i), N(i + 1)) + \sum _i U(N(i)).
\label{Eq:Sfu}
\eeq
The function $f(x, y)$ describes the \emph{kinetic part} of the effective action.
It couples volumes of successive slices providing the  non-zero subdiagonal elements of $\bP$.
Because both $\bC$ and $\bP$ are symmetric matrices, the function $f$ has to be symmetric in
its arguments, $f(x, y) = f(y, x)$.
The \emph{potential part} is described by  function $U(x)$
which contributes only to the diagonal.
According to the expansion of the effective action (\ref{Eq:DiscExpS})
the elements of the Sturm-Liouville matrix $\bP$ are given by
\begin{align}
\bP_{i j} &= [\bC^{-1}]_{i j} = \left.\frac{\partial^2 S[N]}{\partial N(i) \partial N(j)}\right|_{N=\bar{N}} = \nonumber\\
&= \left\{
\begin{array}{ll}
f^{(1, 1)}(\bar N(i), \bar N(i-1))&, j = i - 1\\
f^{(2, 0)}(\bar N(i), \bar N(i-1)) + f^{(2, 0)}(\bar N(i), \bar N(i+1)) + U''(\bar N(i))&, j = i\\
f^{(1, 1)}(\bar N(i), \bar N(i+1))&, j = i + 1\\
0&, \textrm{otherwise}\\
\end{array}
\right.
\label{Eq:Pfu}
\end{align}
where 
\[ f^{(u, v)} (x, y) \equiv \frac{\partial^{(u+v)} f(x, y)}{\partial^u x \partial^v y}.\]

{\bf Discrete minisuperspace action.}
In the next Sections, 
we determine functions $f(x, y)$ and $U(x)$, and show that the effective action (\ref{Eq:Sfu}) 
corresponds to a discretization of the minisuperspace action (\ref{Eq:MiniActV}) up to an \emph{overall sign}.
Below we derive a discrete version of the minisuperspace action with reversed sign
\beq
S[v] = \int \dd \tau \left( \alpha \frac{\dot v^2}{v} + \beta v^{1/3} - 2 \Lambda v \right),
\label{Eq:MiniCont}
\eeq
which later will be compared to the empirical action.
We have incorporated the factor $\frac{1}{24 \pi G}$ into constants $\alpha, \beta$  and $\Lambda$.
The discretization procedure is not unique,
but up to the order considered here, all 
discretizations are equivalent.
We substitute the physical volume $v(\tau)$ with the discrete volume $N(i)$,
which as stated in Section \ref{Sec:Spatial} may be treated as a continuous variable inside the \emph{blob}.
The \emph{stalk} region is governed by very strong lattice artifacts, 
and therefore is not reliably treated in the semiclassical approximation.
The standard discretization of the time derivative is $\dot{v} \to N(i + 1) - N(i)$,
and the kinetic part is written as
\[ \alpha \frac{\dot{v}^2}{v} \to g_1 \frac{(N(i + 1) - N(i))^2}{N(i + 1) + N(i)} ,\]
while the potential part as
\[ \beta v^{1/3} - 2 \Lambda v \to g_2 N(i)^{1/3} - g_3  N(i). \]
Therefore, a discretized, dimensionless version of (\ref{Eq:MiniCont}) is given by
\beq
S[N] = \sum_i g_1 \frac{(N(i + 1) - N(i))^2}{N(i + 1) + N(i)} + g_2 N(i)^{1/3} - g_3  N(i),
\label{Eq:MiniDisc}
\eeq
which corresponds to
\beq
f(x, y) = g_1 \frac{(x-y)^2}{x+y}, \quad U(x) = g_2 x^{1/3} - g_3 x.
\label{Eq:FU}
\eeq
Further we show that the discrete effective action (\ref{Eq:MiniDisc}) describes 
not only the average $\bar{N}(i)$ (\ref{Eq:AvNi}),
what follows from the classical trajectory of equation (\ref{Eq:MiniCont}),
but indeed also the measured fluctuations $\eta(i)$. 

\section{Decomposition of the Sturm-Liouville matrix}

From the formula (\ref{Eq:Sfu}) follows that
Sturm-Liouville operator decomposes into 
the \emph{kinetic} part $\bP^{kin}$, given by the function $f(x, y)$,
and the \emph{potential} part $\bP^{pot}$, given by the function $U(x)$,
\[ \bP = \bP^{kin} + \bP^{pot}. \]
Only the kinetic part contributes to the sub-diagonal elements
of the tridiagonal matrix $\bP$ (cf. (\ref{Eq:Pfu})).
We make following assumptions about $\bP^{kin}$:\\
$\bullet$ It should be a tridiagonal matrix. The square of the time derivative should couple with the preceding and following time steps.\\
$\bullet$ It should be a symmetric matrix. Covariance matrix is symmetric, and so should be $\bP$.\\
$\bullet$ The sum of elements in each row or column should be zero. 
A constant vector $e^0$ should be a zero mode eigenvector as it is invariant under time translations.\\
Taking into account time-periodicity, this implies a following form of $\bP^{kin}$,
\beq
\bP^{kin} = 
\left(
\begin{array}{cccccc}
k_T + k_1 & -k_1 & 0 & \ldots & 0 & -k_T \\
-k_1 & k_1 + k_2 & -k_2 & 0 & \ldots & 0\\
0 & -k_2 & k_2 + k_3 & -k_3 & 0 & \ldots \\
\vdots & \vdots & \ddots & \ddots & \ddots & \ldots\\
-k_T & 0 & \ldots & 0 & -k_{T-1} & k_{T-1} + k_T \\
\end{array}
\right).
\label{Eq:PKinMat}
\eeq
The matrix $\bP^{kin}$ can be decomposed into parts linearly dependent on $k_i$:
\beq
 \bP^{kin} = \sum_{i=1}^T k_i \bX^{(i)},
\label{Eq:KinDeco}
\eeq
where $\bX^{(i)}$ is a matrix corresponding to the discretization of the second time derivative $\partial^2_t$ at a time $t = i$,
\[ \bX^{(i)}_{j k} = \delta_{i j}\, \delta_{i k} + \delta_{(i+1) j}\, \delta_{(i+1) k} - \delta_{(i+1) j}\, \delta_{i k} - \delta_{i j}\, \delta_{(i+1) k} .\]

In the case of the \emph{potential} term, we have to get rid of the zero mode.
The measured potential $\bP^{pot}$ is a projection of the potential part of the full, 
unconstrained inverse propagator $\tilde \bP^{pot}$ on a subspace orthogonal to $e^0$.
Assuming that the full potential matrix is diagonal
\[ \tilde \bP^{pot} = \Diag(\{u_i\}),  \]
and following equation (\ref{Eq:BP}) derived in the Appendix B,  
we get the empirical propagator inverse
\begin{equation}
\bP^{pot} = (\bI - \mathbf{A}) \, \Diag(\{u_i\}) \, (\bI - \mathbf{A}) = \sum_{i=1}^T u_i \bY^{(i)}, 
\quad \mathbf{A}_{i j} = e^{0}_i e^{0}_j = \frac{1}{T},
\label{Eq:PotDeco}
\end{equation}
which can be decomposed into linear combination of matrices
\[ \bY^{(i)}_{j k} = \delta_{i j} \delta_{i k} - \frac{1}{T}(\delta_{i j} + \delta_{i k}) + \frac{1}{T^2}. \]
Here, $\bI$ denotes the $T \times T$ unit matrix.
The kinetic part remains unaffected by the projection, since $\mathbf{A} \bP^{kin} = 0$.

\begin{figure}[t]
\begin{center}
\includegraphics[width=0.9\textwidth]{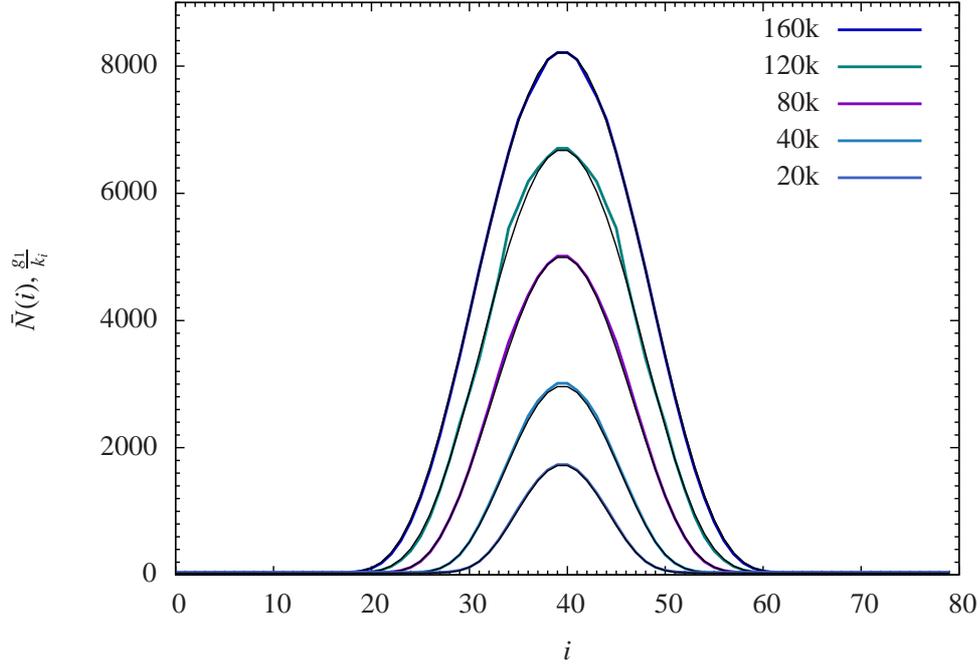} \\
\end{center}
\caption{
Kinetic term: The directly measured expectation values $\bar{N}(i)$ (black line), 
compared to $\frac{g_1}{k_i}$ (thick line) extracted from the measured covariance matrix $\bC$
for $K_0 = 2.2, \Delta = 0.6$ and various total volumes $\Ntot$ ranging from $20000$ to $160000$ simplices.
The theoretical predication 
$\frac{g_1}{k_i} = \frac{1}{2}(\bar{N}(i) + \bar{N}(i + 1))$ is realized with a very high accuracy.
The value of $g_1$ is constant for all volumes $\Ntot$.}
\label{Fig:DecoKin}
\end{figure}

The decomposition of the empirical matrix $\bP$ into a \emph{kinetic} and \emph{potential} part, 
is done using the least square method.
We find such values of $\{k_i\}$ and $\{u_i\}$,
for which the matrix $\bP^{kin} + \bP^{pot}$, given by (\ref{Eq:KinDeco}) and (\ref{Eq:PotDeco}),
is as close as possible to the empirical matrix $\bP$,
i.e. we minimize the residual sum of squares
\beq
\mathrm{RSS}[\{k_i\}, \{u_i\}] \equiv \Tr \left[ \bP - (\bP^{kin} + \bP^{pot}) \right] ^2 .
\label{Eq:PError}
\eeq
We will omit details of the parameter fitting.
In fact, we used the weighted least square method,
giving higher weights to slices with larger volumes, i.e. lying inside the \emph{blob}.
Equation (\ref{Eq:PError}) is quadratic in $\{k_i\}$ and $\{u_i\}$,
and the fitting boils down to calculating traces of products of matrices $\bX^{(i)}$ and $\bY^{(j)}$.

\section{Kinetic term} 

\begin{figure}[t]
\begin{center}
\includegraphics[width=0.9\textwidth]{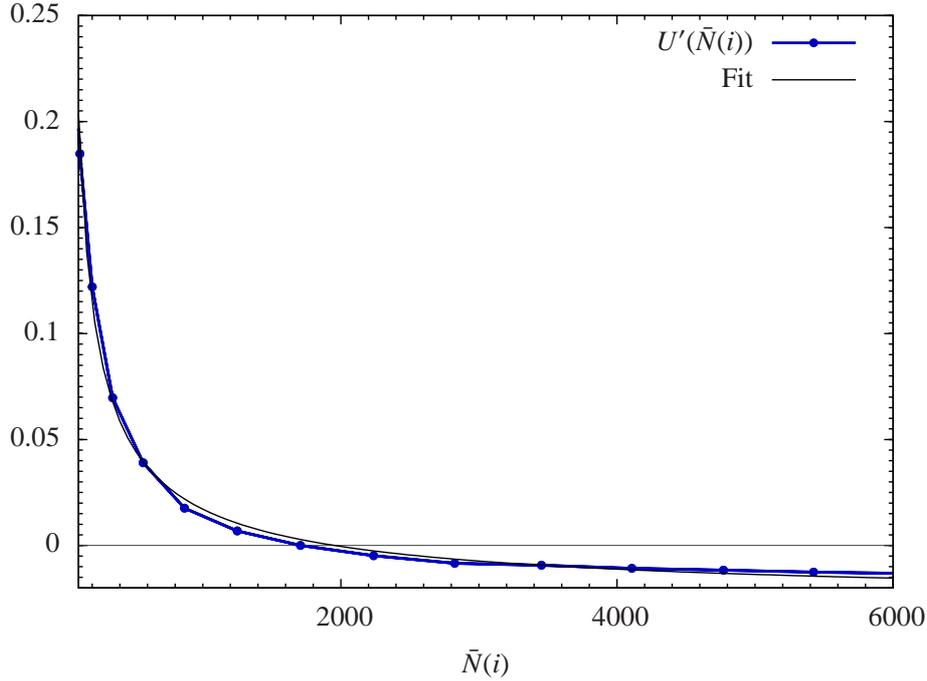}
\end{center}
\caption{The plot of $4\frac{\bar{N}(i+1)}{(\bar{N}(i+1) + \bar{N}(i))^2} + 4\frac{\bar{N}(i-1)}{(\bar{N}(i) + \bar{N}(i - 1))^2} - 2$
as a function of $\bar{N}(i)$ for $K_0 = 2.2, \Delta = 0.6$ and $\Ntot = 80\rk$.
If the minisuperspace action describes correctly fluctuations $\eta(i)$, this function is proportional to the
first derivative of the potential $U'(x)$.
The best fit $a \cdot x^{-c} + b$ plotted with black line corresponds to $c = 0.71 \pm 0.04$.
The minisuperspace model predicts $c = \frac{2}{3}$.
}
\label{Fig:Vprim}
\end{figure}

In this Section we show that the fitted values of the \emph{kinetic} term $\{k_i\}$,
obtained by minimizing residues (\ref{Eq:PError}),
are indeed in agreement with the kinetic part of the 
discrete minisuperspace action (\ref{Eq:MiniDisc}).
For the minisuperspace model,
equations (\ref{Eq:Pfu}) and (\ref{Eq:FU}) give
\beq
  k_i = - \bP_{i i+1} = - f^{(1, 1)}(x, y) = g_1 \frac{8 x y}{(x+y)^3}, \quad x = \bar N(i), \quad y = \bar N(i+1),
\label{Eq:KiMini}
\eeq
and in the zeroth order approximation, $\bar N(i) \approx \bar N(i+1)$, 
we have following behavior of the kinetic term
\begin{equation}
  \frac{g_1}{k_i} = \frac{(\bar{N}(i)+\bar{N}(i + 1))^3}{8\ \bar{N}(i) \cdot \bar{N}(i + 1)} \approx \frac{1}{2}(\bar{N}(i) + \bar{N}(i + 1)).
\label{Eq:ki}
\end{equation}
Fig. \ref{Fig:DecoKin} presents the plot of $g_1/k_i$ 
for the empirical values of $k_i$ and various total volumes $\Ntot$.
The theoretical fit (\ref{Eq:ki}) agrees extremely well with the measured quantities.
Additionally, the effective coupling constant $g_1$ does not depend on $N_{tot}$ in the margin of error.
For $K_0 = 2.2, \Delta = 0.6$, we measured $g_1 = 0.038 \pm 0.002$. 
The kinetic part of quantum fluctuations 
is indeed described by the minisuperspace action (\ref{Eq:MiniCont}).

This success in applying the minisuperspace model to Causal Dynamical Triangulations in four-dimensions,
encourages us to check if we can say something about the potential term making use of the 
knowledge only about the kinetic part and the classical trajectory.
The classical solution minimizes the action,
\[ \forall_i \left. \frac{\partial S[N]}{\partial N(i)}\right|_{\{N(i) = \bar{N}(i)\}} = 0. \]
Assuming that the minisuperspace model is valid, after inserting the function $f(x, y)$ given by (\ref{Eq:FU}) into the action (\ref{Eq:Sfu}) 
we get an equation for the first derivative of the potential $U(x)$
for arguments equal to the average spatial volumes,
\[ U'(\bar{N}(i)) =  4 g_1 \frac{\bar{N}(i+1)}{(\bar{N}(i+1) + \bar{N}(i))^2} + 4 g_1 \frac{\bar{N}(i-1)}{(\bar{N}(i) + \bar{N}(i - 1))^2} - 2 g_1.\]
Fig. \ref{Fig:Vprim} presents the evaluated derivative of the potential, for
empirical values of average spatial volume $\bar{N}(i)$.
Again, the measurements agree within the limit of error with the prediction (\ref{Eq:FU})
\[ U'(x) = \frac{1}{3}g_2 x^{-2/3} - g_3 .\]

\section{Potential term} 

\begin{figure}[t]
\begin{center}
\includegraphics[width=0.9\textwidth]{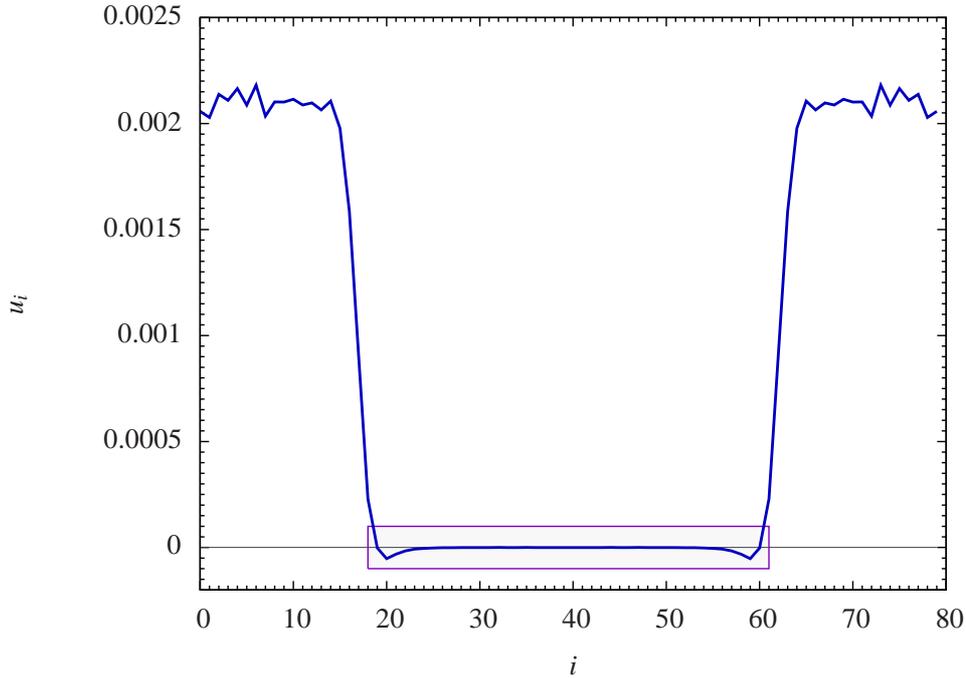}
\end{center}
\caption{
Coefficients $u_i$ of the \emph{potential} part extracted from the empirical matrix $\bP$
obtained in Monte Carlo measurements for $K_0 = 2.2, \Delta = 0.6$ and $\Ntot = 80\rk$.
The coefficients correspond to the second derivative of the potential $u_i = U''(\bar{N}(i))$.
The potential term is dominated by the \emph{stalk} region.
The physically interesting range, corresponding to the \emph{blob}, is highlighted.
}
\label{Fig:DecoPot}
\end{figure}

In this Section we directly show that values of the \emph{potential} term $\{u_i\}$
extracted from the empirical inverse propagator $\bP$
also agree with the minisuperspace model.
Within this framework, following equations (\ref{Eq:Pfu}) and (\ref{Eq:FU}), we expect that
\beq
 u_i = U''(\bar{N}(i)) = - \frac{2}{9}\ g_2\ \bar{N}(i)^{-5/3}.
\label{Eq:UiMini}
\eeq

Fig. \ref{Fig:DecoPot} shows the measured values of coefficients $u_i$ 
extracted from the empirical matrix $\bP^{pot}$.
Because of large statistical errors, it is not an easy task to determine $u_i$.
As can be seen in Fig. \ref{Fig:DecoPot} the plot 
is dominated by the \emph{stalk} which is governed 
by lattice artifacts and can not be reliably treated semiclassically.
The physically interesting region of large volumes
corresponds to relatively small values of $u_i$
as they are expected to falls as $\bar{N}(i)^{-5/3}$.
Due to the projection on the space orthogonal to the zero mode,
the \emph{blob} region is affected by the huge contribution
from the \emph{stalk}.
Moreover, analogically as in the ordinary path-integral approach to quantum mechanics 
when the time step approaches zero, 
in the continuum limit $\Ntot \to \infty$ the potential term 
is sub-dominant w.r.t. the kinetic term for individual spacetime histories in the path integral.

\begin{figure}[t]
\begin{center}
\includegraphics[width=0.86\textwidth]{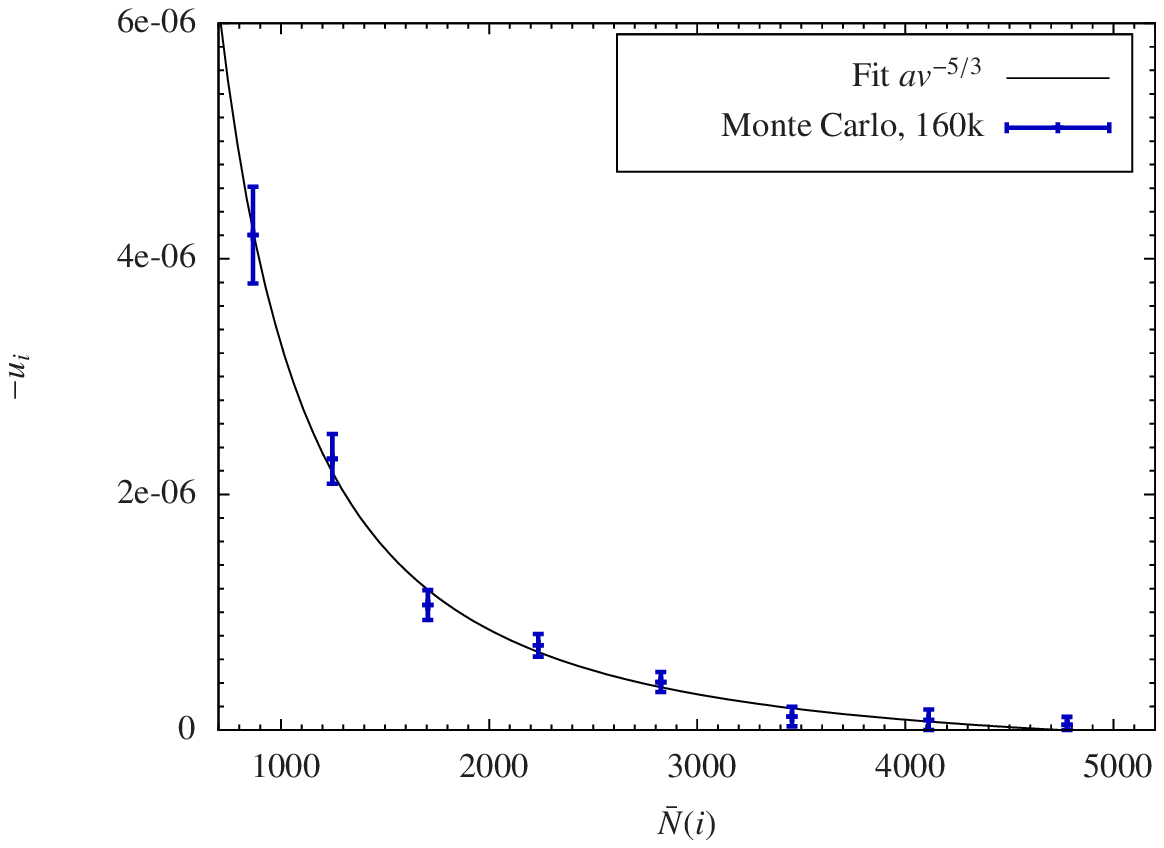}
\end{center}
\caption{
The extracted potential term $u_i$ as a function of average volume $\bar{N}(i)$.
The fit $c_2 \bar{v}^{-5/3}_t$ presents the behavior expected for the minisuperspace model. 
The visible points correspond to the blob region.
}
\label{Fig:Potential}
\end{figure}

Nevertheless, due to sufficiently long Monte Carlo sample, 
the obtained results allowed us to confirm that indeed the 
formula (\ref{Eq:UiMini}) is in agreement with the measurement.
Fig. \ref{Fig:Potential} presents the measured coefficients $-u_i$ 
as a function of the average three-volume $\bar{N}(i)$.
The error bars shown on the plot, were estimated using Jackknife method.
Such a form allows us to directly compare the potential coefficients 
with theoretical predictions $-u_i \propto \bar{N}(i)^{-5/3}$.
The selected range of $\bar{N}(i)$ corresponds to the \emph{bulk},
and is highlighted in Fig. \ref{Fig:DecoPot}.
The best fit of the form $f(x) = a \cdot x^{-c}$ to the empirical values $u_i$
as a function of $\bar{N}(i)$ gives $c = -1.658 \pm 0.096$.
The measured exponent coefficient $c$  is very close to the theoretical value
$c = -5/3$.
The fit $f(x) = a \cdot x^{-5/3}$, corresponding to the potential
(\ref{Eq:FU})
is presented in Fig. \ref{Fig:Potential} with a thin line.
The agreement with the data is good,
the potential part of the effective action is indeed given by
$U(x) = g_2 x^{1/3} - g_3 x$.
There is a very small residual constant term present in this fit,
which presumably is due to the projection onto the space orthogonal to the zero
mode. In view of the fact that its value is quite close
to the noise level with our present statistics, we have simply chosen to ignore it
in the remaining discussion.
Apart from obtaining the correct power $u_i \propto \bar{N}^{-5/3}(i)$,
the coefficient in front of this term is also independent of $\Ntot$.

In summary, we conclude that the measurements of the covariance matrix of quantum volume fluctuations
allowed us to reconstruct the discrete version of the minisuperspace action (\ref{Eq:MiniDisc}) with a high precision.

\section{Flow of the gravitational constant}

In the previous Section we gave a direct numerical evidence
that quantum fluctuations of three-volume are 
very accurately described by the discrete, dimensionless effective action
\beq
S[N(i)] = \sum_i g_1 \frac{(N(i + 1) - N(i))^2}{2 N(i)} + g_2 N^{1/3}(i), 
\label{Eq:SDisc}
\eeq
where we omitted the cosmological constant term since during the measurements
the total volume $\Ntot$ was fixed.
This action comes out as a discretization of the minisuperspace action (\ref{Eq:MiniActV})
with the opposite sign which solves the problem of unboundedness.
Let us note, that it is justified to use the semiclassical approximation
as the distribution of spatial volumes $N(i)$ in the bulk (cf. Fig. \ref{Fig:StalkBlob}) 
is given by Gaussian fluctuations around the mean.

In this Section we establish the relationship between the discrete action (\ref{Eq:SDisc})
and its continuum counterpart (\ref{Eq:MiniActV}),
which will allow us to connect the effective coupling constant $g_1$ with the gravitational constant coupling constant $G$.
At first, let us rewrite the above action in terms of the normalized volume $n(t)$.
Using equations $N(i) = \Ntot^{3/4}\cdot n(t)$ (\ref{Def:nt}) and $\Delta t = \Ntot^{-1/4}$ (\ref{Def:Scaledt}),
where $t = \Delta t \cdot i$, the discrete action can be written as
\[ S[n(t)] = \sqrt{\Ntot} \sum_i \Delta t \ \left[g_1 \frac{1}{2n(t)}  \left(\frac{n(t + \Delta t) - n(t)}{\Delta t} \right)^2 + g_2\ n^{1/3}(t)\right]. \]
and following (\ref{Eq:ContT}), we can represent it in a form of an integral over the continuum time $t$
\beq
S[n(t)] = \sqrt{\Ntot} \int \dd t \ \left[ \frac{1}{2} g_1 \frac{n'(t)^2}{n(t)} + g_2\ n^{1/3}(t)\right]. 
\label{Eq:Snt}
\eeq
As stated in the previous Sections the effective coupling constant $g_1$ does not 
depend on $\Ntot$ when the bare coupling constants are fixed, the same is true for the classical trajectory $\bar{n}(t)$.
Hence, the amplitude of fluctuations of $n(t)$ scales with the total discrete volume $\Ntot$ 
as $\sqrt{\langle (\delta n(t))^2 \rangle} \propto g_1^{-1/2} \Ntot^{-1/4}$
and they vanish in the continuum limit $\Ntot \to \infty$,
while the background $\bar{n}(t)$ stays fixed.
Regarding Fig. \ref{Fig:AvNi}, it means that the height  of the curve will grow as 
$\Ntot^{3/4}$, but the superimposed fluctuations will only grow as $\Ntot^{1/2}$,
and for fixed bare coupling constants the relative size of fluctuations will
go to zero in the infinite-volume limit.
Therefore, 
in order to obtain a proper continuum limit with a finite physical four-volume
$V_4 \propto \Ntot a^4 = const$ and finite fluctuations 
$\sqrt{\langle (\delta v(\tau))^2 \rangle} \propto g_1^{-1/2} \Ntot^{3/4} a^4 = const$
(cf. (\ref{Eq:VtauNi})),
one also has to properly tune the \emph{bare} coupling constants
so that the \emph{effective} coupling constant satisfies $g_1 a^{-2} = const$
while taking $\Ntot \to \infty$ and $a \to 0$.

Further, we rewrite the action (\ref{Eq:Snt}) in terms of the physical volume $v(\tau)$.
Using formulas (\ref{Eq:ContTau}) and (\ref{Eq:VtauNi}), 
we recover a continuum action back from the discrete action (\ref{Eq:SDisc})
\beq
S[v(\tau)] = \frac{g_1 g_{tt}}{2 \sqrt{\Ntot} C_4 a^4} \int \dd \tau \left[ \frac{\dot v(\tau)^2}{v(\tau)} + \tilde{g}_2\ v^{1/3}(\tau)\right]. 
\label{Eq:Svt}
\eeq

\begin{figure}[t]
\begin{center}
\includegraphics[width=0.86\textwidth]{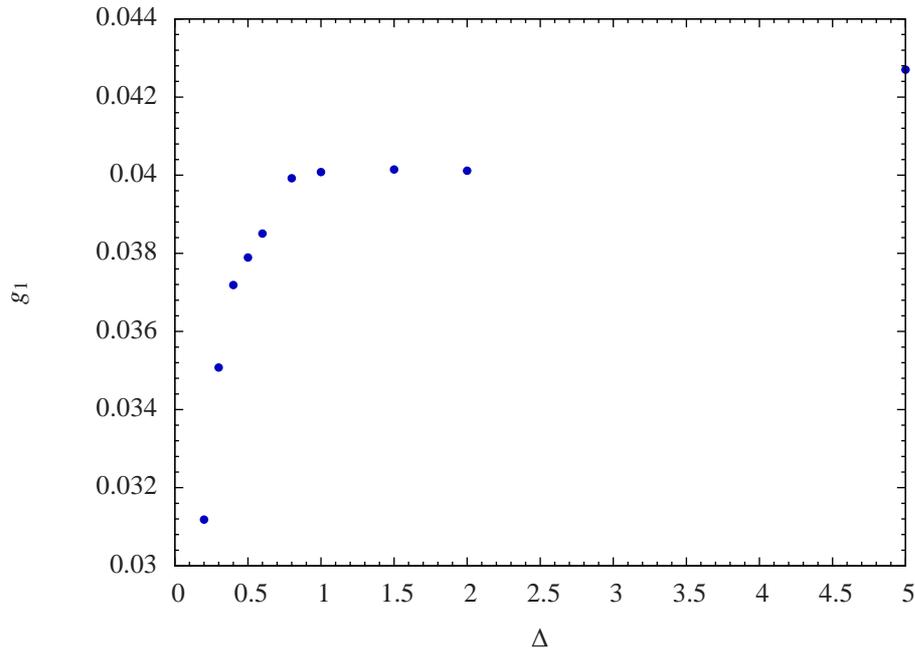}
\end{center}
\caption{
The measured effective coupling constant $g_1$ as a function of bare coupling constant $\Delta$
for $K_0 = 2.2$.
The $B$-$C$ transition point is located c.a. at $\Delta^{crit} = 0.05$.
When approaching phase $B$ from phase $C$ the coupling constant $g_1$ 
diminishes and the fluctuations grow, as expected when reaching phase transition point.
}
\label{Fig:G1Delta}
\end{figure}

This is the second most important result obtained in this dissertation.
Not only the average spatial volume $\bar{v}(\tau)$ obtained in the 
Causal Dynamical Triangulations frameworks is given by the classical trajectory of 
the minisuperspace model,
but also the fluctuations of $v(\tau)$ around the background geometry $\bar{v}(\tau)$
are described by the minisuperspace model up to the overall sign.

It is natural to identify the coupling constant $G$ multiplying the effective action
for the scale factor (\ref{Eq:MiniActV}) with the Newton's gravitational constant $G$.
The effective action describing our computer-generated data (\ref{Eq:Svt})
can be directly compared with the minisuperspace action (\ref{Eq:MiniActV}), 
and using equations (\ref{Eq:Radius}) and (\ref{Eq:gtt}), we get following relations between 
the gravitational constant $G$
and the effective constant $g_1$ \cite{Plan, Nonp},
\beq
G = \frac{2 \sqrt{\Ntot} C_4 a^4}{24 \pi g_1 g_{tt}} = \frac{a^2}{g_1} \frac{ \sqrt{C_4} B^2}{3 \sqrt{6}}.
\label{Eq:Gg1}
\eeq
This result is consistent with our earlier conclusion,
in order to keep the physical constant $G$ fixed,
when taking the continuum limit $a \to 0$ one has to tune the effective coupling constant $g_1 \propto a^{2}$. 
This means that in terms of the lattice volume $N(i)$ fluctuations should diverge,
and this happens when we approach a second or higher-order transition line.
Therefore it is important to determine the order of transition.
If the $B$-$C$ transition would be of the first-order
we may hope that the triple point is of higher order,
which is often true, and obtain the proper continuum limit while approaching this point.
Fig. \ref{Fig:G1Delta} shows the measured effective coupling constant $g_1$ for various values of $\Delta$.
Indeed when we approach the $B$-$C$ transition line $g_1$ tends to zero.
However, it is not so easy to find the exact path in $K_0-\Delta$ coupling constant plane,
as formula (\ref{Eq:Gg1}) depends also on other parameters, namely 
the effective four-volume shared by one spatial tetrahedron $C_4(\alpha)$
and the width $B$ of the distribution $\bar{n}(t)$,
and it will require further extensive computer-experiments.

{\bf The Universe size.}
Using relation (\ref{Eq:Gg1}) we can express the cut-off length $a$ in terms of the Planck length,
and thus estimate the size of the Universe generated in computer simulations.
Let us recall that in natural units $G = \ell_{Pl}^2$.
For the bare coupling constants $K_0 = 2.2, \Delta =0.6$ we measured the quantities: 
$K^{crit}_4 = 0.922, \xi = \frac{N_{32}}{N_{41}} = 1.30, \alpha = 0.5858, C_4 = 0.0317, g_1 = 0.038$,
which results in $a \approx 1.9 \ell_{Pl}$ and 
the linear size $\pi R$ of the universe built from $160000$ simplices is about $20 \ell_{Pl}$.
The quantum de Sitter universes studied here are therefore quite small, 
and quantum fluctuations around their average shape are large (cf. (\ref{Fig:AvNi})).
Surprisingly, the semiclassical minisuperspace
formulation gives an adequate description of the measured data,
at least for the volume profile.

To conclude, based on the numerical evidence
we reconstructed the discrete effective action
describing quantum fluctuations of the three-volume $N(i)$.
This allowed us to identify the effective action 
with the discretization of the minisuperspace action with a negative sign.
We also derived a relation between the physical gravitational constant
and the effective coupling constants appearing in the model,
giving a recipe of how to obtain a meaningful continuum limit.

\clearemptydoublepage

\chapter{Geometry of spatial slices}

\label{Chap:Slice}

{\noindent \it This chapter is based on the publication:
 J.~Ambj\o rn, A.~G\"orlich, J.~Jurkiewicz and R.~Loll,
  \emph{"Geometry of the quantum universe"},
  Phys. Lett. B {\bf 690}, 420 (2010).
}\\[4ex]

After investigating in Chapter \ref{Chap:Macroscopic} the four-dimensional geometry of the quantum Universe,
the next step is to look deeper into the geometry of spatial slices.
A spatial slice is a leaf of the imposed global proper-time foliation
and is labeled by a discrete time index $i$.
Each slice is built of equilateral spatial tetrahedra.
A hypersurface of this kind is a three-dimensional triangulation,
more precisely, a piecewise linear manifold of topology $S^3$.
However, it does not mean that the geometry of slices is close to
the geometry of a three-dimensional sphere.

\section{Hausdorff dimension}

Since in this Chapter we consider each slice as a separate object,
let us denote the number of tetrahedra building a slice $i$ 
by the discrete three-volume $n_3 \equiv N(i)$.
A basic observable defined on a slice,
is the number of tetrahedra $n(r, i_0)$ at a three-dimensional distance $r$ from some initial tetrahedron $i_0$.
A three-dimensional distance is based on the definition of tetrahedra neighborhood
within one slice,
while a four-dimensional distance is based on the definition of simplices neighborhood (cf. Section \ref{Sec:Geometry4D}).
This difference introduces a significant change in the shell
decomposition picture.
At distance $r = 0$ only the initial tetrahedron is counted and $n(0, i_0) = 1$.
Next, we move out by one step visiting all four neighbors of $i_0$, thus $n(1, i_0) = 4$.
Moving out by a further step, there will be $n(2, i_0)$ tetrahedra at a distance $r = 2$, 
and so on until all tetrahedra are visited, 
with the restriction that one tetrahedron is visited only once.
For such definition, $n(r, i_0)$ corresponds to an \emph{area} of the shell of radius $r$.
Summing up the area over all \emph{shells} gives the discrete volume of a slice $n_3$, 
\[ n_3 = \sum_{r = 0}^{r_{max}} n(r, i_0). \]
Further, $n(r)$ denotes the average of $n(r, i_0)$ over all $n_3$ initial tetrahedra $i_0$,
\[ n(r) = \frac{1}{n_3} \sum_{i_0 = 1}^{n_3} n(r, i_0). \]
In this Section we investigate scaling properties of $n(r)$ with respect to the slice volume $n_3$.
\begin{figure}[t]
\centering
\includegraphics[width=0.8\textwidth]{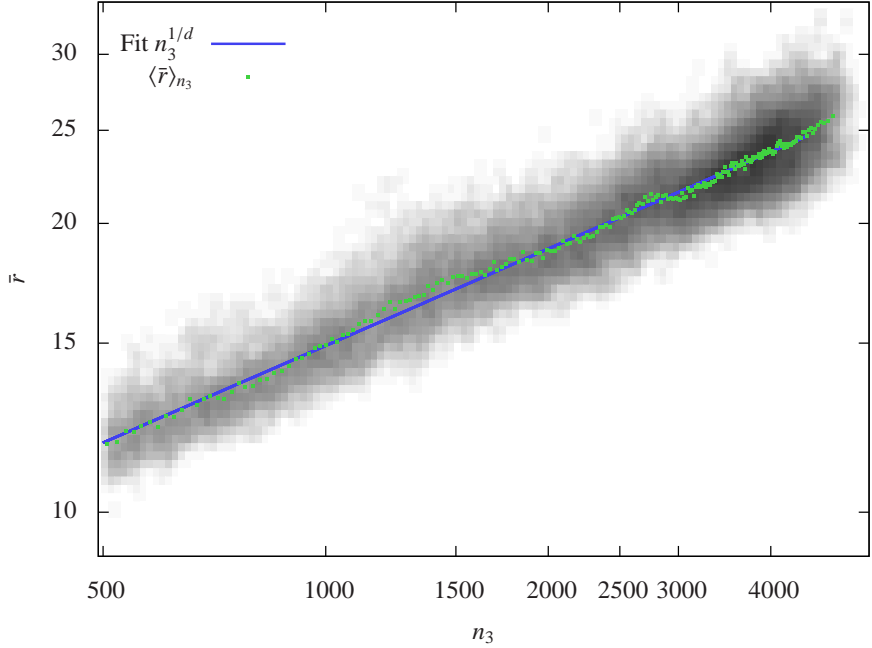}
\caption{
The \emph{cloud} visible on the plot presents the set of measured points $(n_3, \bar{r})$.
The darker pixel, the more events were observed in given range.
To clarify the picture, green points correspond to an average
of $100$ successive measurements sorted by $n_3$.
The blue line represents the fitted curve $\bar{r} \propto n_3^{1/d_h}$
using linear regression, $d_h = 2.984 \pm 0.024$.
The plot is in log-log scale.
}
\label{Fig:N3r}
\end{figure}

The Hausdorff dimension of spatial slices may be measured,
using the scaling properties, in a following way \cite{Reco}.
For a given slice of volume $n_3$, we define the average linear extent as
\[ \bar{r} \equiv \frac{1}{n_3} \sum_r r \cdot n(r) . \]
For each slice we obtain a data pair $\{n_3, \bar{r}\}$.
Such pairs are then gathered for all slices of a number of  Monte Carlo (MC) configurations.
A plot off all points belonging to some MC sample, shown in Fig. \ref{Fig:N3r} for $N_{41} = 160k$, looks like a cloud.
To make the average behavior more visible, the number of data points is reduced.
The results are sorted by $n_3$ and averaged over a sequence of $100$ consecutive points
(in contrary to \cite{Reco} where the sample is sorted by $\bar{r}$,
because fluctuations of $\bar{r}$ are larger than $n_3$,
the approach used here gives smaller finite size effects and better
linear behavior of the log-log plot.)
For Hausdorff dimension $d_h$, the expected relation between the slice volume
$n_3$ and the average linear extent $\bar{r}$ is following
\beq 
n_3 \propto \bar{r}^{d_h}.
\label{Eq:N3rScaling}
\eeq
The reduced data points (green dots) together
with the relation (\ref{Eq:N3rScaling}) fitted to the full data are presented 
in Fig. \ref{Fig:N3r}. 
The fitted value of the Hausdorff dimension is $d_h = 2.984 \pm 0.024$.
\begin{figure}[t]
\centering
\includegraphics[width=0.8\textwidth]{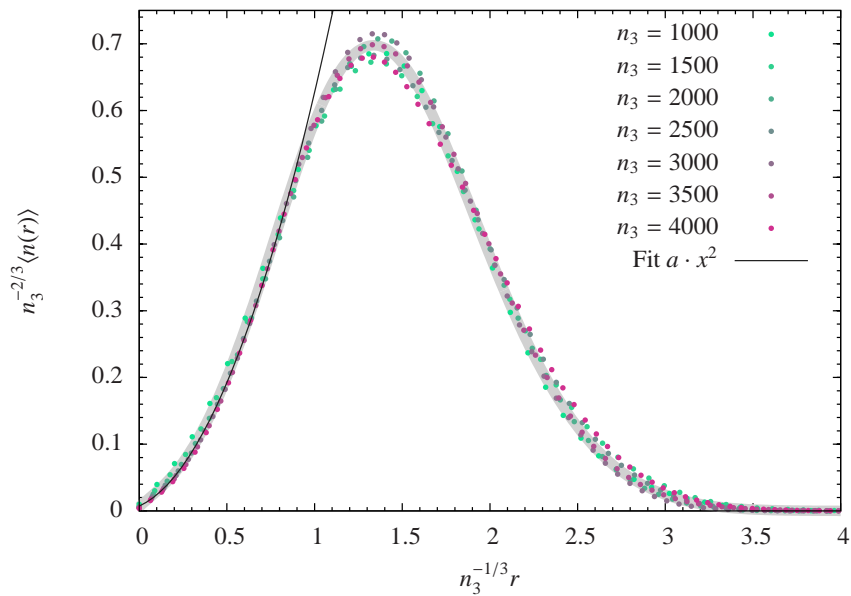}
\caption{
Scaled values of the radius $n_3^{-1/d_h} r$ and shell area $n_3^{-1 + 1/d_h} n(r)$ for $d_h = 3$.
Data points for various values of slice volume $n_3$ overlap.
The gray strip plots the scaled radial volume averaged over all data points.
The fitted curve $a r^2$ (black line) presents the small $r$ behavior and is consistent with Hausdorff dimension $d_h = 3$.
Measurements were performed at $K_0 = 2.2$ and $\Delta = 0.6$.}
\label{Fig:ScaledNr}
\end{figure}

Similar results can be obtained by a direct comparison of the scaled with volume $n_3$ values of the radial volume $n(r)$.
First, for a large number of Monte Carlo configurations,
slices with the same volume $n_3$  (more or less few tetrahedra) are collected into groups.
The average radial volume $n(r)$ within a group $n_3$ is denoted as $\langle n(r) \rangle_{n_3}$.
Next, the radius $r$ and the average volume $\langle n(r) \rangle_{n_3}$
are scaled and normalized in the following way
\beq
\left(r,\ \langle n(r) \rangle\right) \to \left( n_3^{-1/d_h} r,\ n_3^{-1 + 1/d_h} \langle n(r) \rangle \right) .
\label{Eq:ScalingNr}
\eeq
Analogically as in Section \ref{Sec:Geometry4D}, where we calculated the scaling dimension $d_H$
for $\bar{N}(i)$ (cf. Fig. \ref{Fig:ScaledAvNi} and \ref{Fig:ErrorDH}),
we define the error of the overlap of the scaled points and find such value of $d_h$
which minimizes the dispersion.
The best fit is obtained for $d_h = 2.94 \pm 0.05$.
Fig. \ref{Fig:ScaledNr} presents the 
measured values of $\langle n(r) \rangle$ scaled according to (\ref{Eq:ScalingNr}) with $d_h = 3$
and for various values of $n_3$ between $1000$ and $4000$ tetrahedra.

The Hausdorff dimension $d_h$ may be also estimated
from the small $r$ behavior of $\langle n(r) \rangle_{n_3}$.
The value $d_h$ determines the dependence of the volume of a ball on its radius $r$,
$\textrm{Vol}[B(r)] \propto r^{d_h}$
or equivalently for an area of a sphere 
\[ \textrm{Area}[S(r)] \propto r^{d_h - 1} ,\quad \textrm{when} \ r \to 0. \]
Because $n(r)$ is interpreted as the area of a sphere with a radius $r$,
the expected behavior of $\langle n(r) \rangle_{n_3}$ for small $r$ is $\langle n(r) \rangle_{n_3} \propto r^{d_h - 1}$.
Assuming that $\langle n(r) \rangle_{n_3} \propto r^{\alpha}$ for small $r$, $\alpha = d_h - 1$ is the only exponent
consistent with the scaling (\ref{Eq:ScalingNr}).
Fig. \ref{Fig:ScaledNr} shows that $d_h = 3$ is in agreement with the measured data.
This method of estimating the Hausdorff dimension, is however  more sensitive on the fit region
than studying the scaling properties.

To conclude, all methods of determining the Hausdorff dimension of spatial slices $d_h$ give consistent results.
The measured value is with a high accuracy equal to $d_h = 3$.
Moreover, the measured value of $d_h$ is independent of the coupling constants
$K_0$ and $\Delta$, as long as we stay well inside the phase $C$.
This results is true if we consider the ensemble average of the slice geometry.
As we shall see in the next Sections, this result does not mean
that individual spatial slices resemble a smooth three-dimensional geometry.

\section{Spectral dimension}

In this Section we determine spectral dimension $d_s$
by investigating diffusion process on triangulated slices
dominating in the CDT partition function in phase $C$.
We measure the spectral dimension in the same way as described in 
Section \ref{Sec:Geometry4D}.
The probability of finding a diffusing particle in tetrahedron 
$i$ after a diffusion time $\sigma$ and starting at tetrahedron $i_0$
is given by the probability density $\rho(i, i_0; \sigma)$.
The discrete diffusion equation, describing the evolution of 
the probability density, is an analogue of the equation (\ref{Eq:DiscreteDiffusion}), namely
\beq
\rho(i, i_0; \sigma + 1) = \frac{1}{4} \sum_{j \leftrightarrow i} \rho(j, i_0; \sigma),
\label{Eq:DiscreteDiffusion2}
\eeq
where the sum is over all tetrahedra $j$ adjacent to $i$.
For a starting tetrahedron $i_0$, chosen at random,
we set the initial condition $\rho(i, i_0; \sigma = 0) = \delta_{i i_0}$.
By iterating the diffusion equation, we calculate the return probability
$P(\sigma, i_0) \equiv \rho(i_0, i_0; \sigma)$ for successive discrete
diffusion steps $\sigma$.
Further, we compute the \emph{average return probability} $P(\sigma) \equiv \langle \langle P(\sigma, i_0) \rangle_{i_0} \rangle_{MC}$
by averaging over initial points and configurations. For each configuration we consider only the central slice.
Spectral dimension $d_s$ is obtained from the return probability using the definition (\ref{Def:Spectral})
\beq
d_s \equiv -2 \frac{\dd \log P(\sigma)}{\dd \log \sigma}.
\label{Def:Spectral2}
\eeq
In the case of Euclidean space $\R^d$, the spectral dimension and Hausdorff dimension are equal to the topological dimension $d_s = d_h = d$.
For a three-sphere $S^3$, spectral dimension $d_s$ is equal $3$ for short diffusion times.
Because of the finite volume, for longer times the zero mode of the Laplacian will dominate
and $d_s$ will tend to zero.
\begin{figure}
\centering
\includegraphics[width=0.8\textwidth]{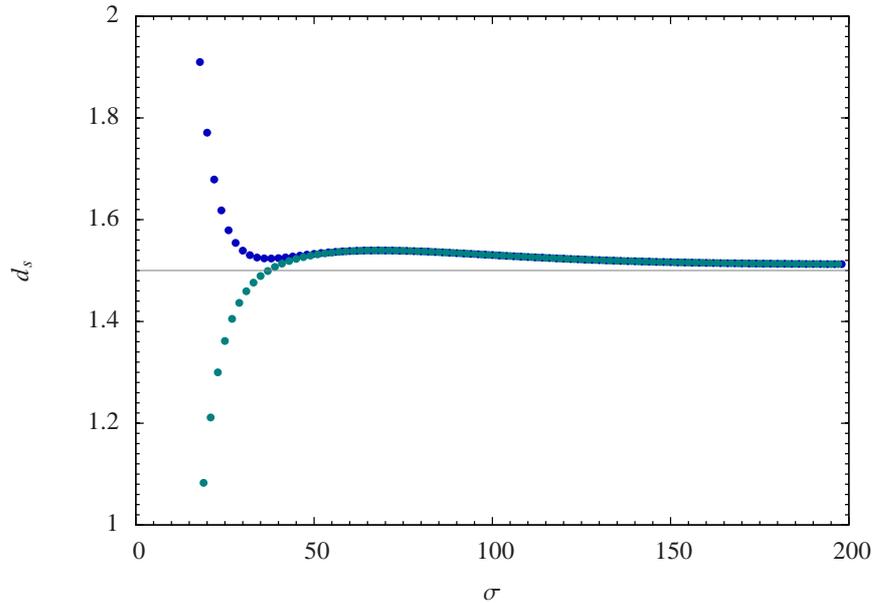}
\caption{
Spectral dimension $d_s$ of spatial slices as a function of diffusion time $\sigma$. 
For short diffusion times,
a split for even (blue) and odd (teal) values of $\sigma$ is observed
arising from the discrete structure.
The measured values of $d_s$ converge to thin line corresponding to $d_s = 1.5$. 
}
\label{Fig:SpectralDimension}
\end{figure}
Fig. \ref{Fig:SpectralDimension} shows values of the spectral dimension $d_s$
as a function of the diffusion time $\sigma$
determined by numerical simulation using the definition (\ref{Def:Spectral2})
for a randomly chosen typical configurations in phase C.
Due to the discrete lattice structure,
for small values of $\sigma$ a split for even and odd diffusion times is observed.
Because of finite volumes of slices, for very large $\sigma$, 
$d_s$ falls down to zero.
For the intermediate region,
there is a plateau of the spectral dimension at $d_s \approx 1.5$.

\section{The fractal structure of spatial slices}

The measured value of the spectral dimension of spatial slices $d_s \approx 1.5$
is significantly smaller than the Hausdorff dimension $d_h \approx 3$.
The difference between $d_h$ and $d_s$ is an indication of a fractal nature of slices.
We will now identify this fractal structure in a more direct way.
Three-dimensional spatial slices demonstrate a very similar structure
to triangulated universes present in three-dimensional
Euclidean Dynamical Triangulations (EDT).
It is known, that in three-dimensional EDT so-called {\it baby Universes} 
separated by {\it minimal necks} are observed \cite{Baby, Observing}.
Advocated by the presence of such objects in the Euclidean theory,
we look for their existence inside spatial slices.
\begin{figure}[!ht]
\centering
\includegraphics[width=0.4\textwidth]{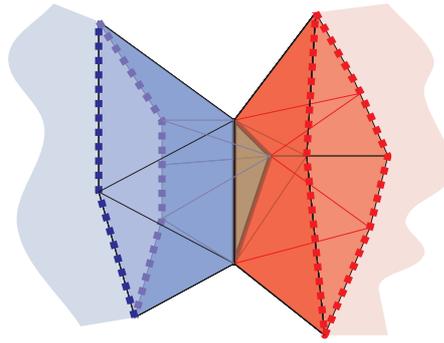}
\caption{
A visualization of a \emph{minimal neck} on
a two-dimensional triangulation.
The \emph{minimal neck} is marked with a thick black line 
consisting of three segments creating a triangle which itself is not a part 
of the triangulation. Cutting the triangulation along the \emph{minimal neck}
will separate it into two components: blue and red.
}
\label{Fig:Minbu}
\end{figure}
A \emph{minimal neck} denotes a set of exactly \emph{four} triangles 
forming a tetrahedron (i.e. each triangle is connected to each other),
which is not present in the triangulation.
It corresponds to the smallest nontrivial boundary of a three-dimensional
simplicial manifold built of tetrahedra.
The minimal necks provide the three-dimensional triangulation
with a tree-structure.
A simplicial manifold cut along a minimal neck,
by removing connections between tetrahedra which 
share the triangular faces belonging to the minimal neck,
is decomposed into two disconnected parts.
The two components form a simplicial manifold with a boundary
and both of them contain more than one tetrahedron (in fact at least four).
By closing the boundaries with tetrahedra we make the manifolds 
into two triangulation of a topology $S^3$.
Fig. \ref{Fig:Minbu} illustrates a \emph{minimal neck} in the case of a two-dimensional triangulation.
In two-dimensions a minimal neck is created of three links forming
a missing triangle and is drawn with a thick black line,
it separates the triangulation into a blue and red part.

It is natural to visualize this structure in the form of a tree graph.
The procedure of creating such tree is following:
\begin{itemize}
\itemsep = 0pt
\item We find all \emph{minimal necks}, 
i.e. sets of four triangles forming a tetrahedron 
which does not belong to the triangulation.
\item Each triangle is a common face of two tetrahedra.
\emph{Ungluing} the tetrahedra sharing elements of a \emph{minimal neck}
decomposes the triangulation into two parts.
\item We cut the triangulation along all \emph{minimal necks},
separating the slice into many disconnected pieces.
We represent each piece as a vertex.
To each vertex, we can assign the number of tetrahedra building a corresponding piece.
This number is called a \emph{vertex volume}.
\item We add an edge linking two vertices if they share a common \emph{minimal neck}.
Because the slice has a topology of a three-sphere,
by definition the set of vertices and edges forms a connected tree graph,
called a \emph{minbu tree}.
\end{itemize}
\begin{figure}[t]
\centering
\includegraphics[height=0.6\textwidth]{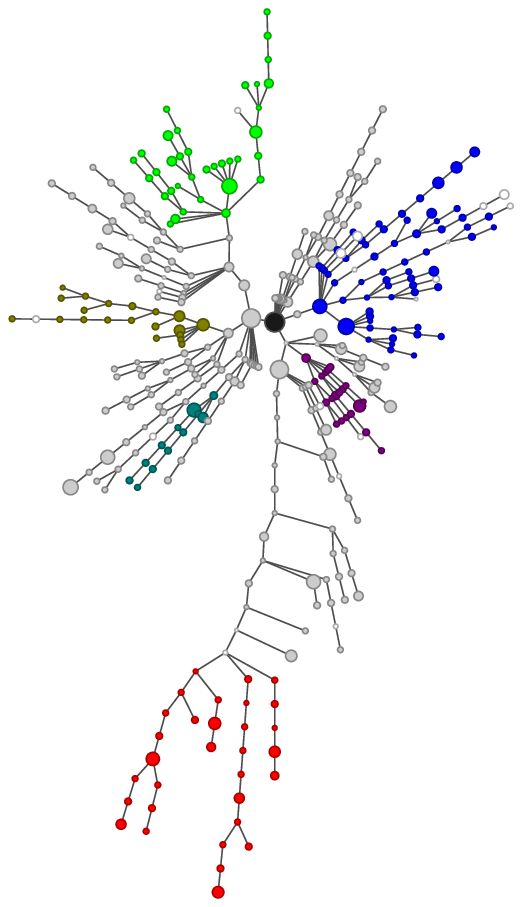}
\includegraphics[height=0.66\textwidth]{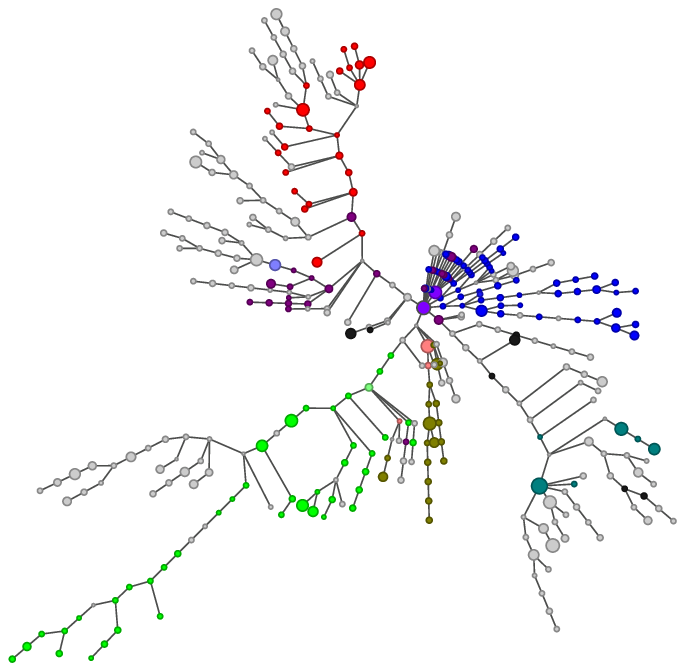}
\caption{
The tree structures of minimal necks, shown here for two consecutive spatial hypersurfaces,
reveal a fractal nature. 
The left tree corresponds to a demonstration slice $i$,
while the right to slice $i + 1$.
Some branches in tree $i$ are marked with a chosen color,
which also taints close vertices in adjacent tree.
The colors identify neighbor vertices in two hypersurfaces.
The diameter of vertices grow with their \emph{volume}.
}
\label{Fig:MinbuTree} 
\end{figure} 
Fig. \ref{Fig:MinbuTree} presents two \emph{minbu trees} created by the above procedure
and corresponding to two randomly chosen successive slices.
Considering one graph, we see that it is indeed a tree,
but what is most important it has a fractal structure (regarding for finiteness).
Very often, but at random places, a branch bifurcates into two or more branches.
Most probably, when the size of the slice grows to infinity,
one would observe a real fractal structure of an infinite \emph{branched polymer}.

The definition of a minbu tree gives very similar results
as the definition of diffusion trees used in case of spatial slices (cf. Section \ref{Sec:Geometry4D}).
The full four-dimensional geometries did not exhibit any fractality,
four-dimensional analogues of minimal necks appeared very seldom.
In contrast, if we had created a diffusion tree for a spatial slice,
we would observe the same fractal structure as presented in Fig. \ref{Fig:MinbuTree},
with even more bifurcation points.
The advantage of a minbu tree is that it does not introduce
any specified initial tetrahedra.
The graph in Fig. \ref{Fig:MinbuTree} is centered on the edge which divides the tree into two most equal parts
with respect to the three-volume.

The fractal structure suggest that leaves of the imposed time foliation have no semiclassical meaning.
One has to perform an average of observables over all triangulations as we did in previous Sections.
On the other hand,
we checked that three-dimensional distance hierarchy reflected 
by the minbu tree (every path connecting two tetrahedra in different vertices
must cross the appropriate minimal necks), is not an artifact
of the imposed foliation and persist in the full, four-dimensional embedding.
There are no drastic shortcuts through the four-dimensional simplicial complex.

This is corroborated by studying if the fractal structure survives between one time slice and the next.
Fig. \ref{Fig:MinbuTree} illustrates minbu trees for two consecutive slices $i$  and $i + 1$.
Using the full four-dimensional triangulation, we introduce the notion of neighbors on vertices of
two trees. 
A vertex of a tree $i$ is a neighbor of the vertex of a tree $i+1$,
if there exists
a pair of tetrahedra respectively shared by each vertex 
between which four-dimensional distance is $4$,
i.e. the shortest path between two slices passing through simplices 
$\{4, 1\}$, $\{3, 2\}$, $\{2, 3\}$ and finally $\{1, 4\}$.
Intuitively this means that two vertices are as close as possible 
in the four-dimensional simplicial manifold.
One vertex may have many neighbors and the relation is symmetric.
This notion allows us to study if the tree structure propagates over successive slices,
i.e. if it survives from one slice to the next or if there is no correlation.
For purposes of illustration, as shown in Fig. \ref{Fig:MinbuTree}, 
we have marked all vertices of a given tree branch at time $i$ in a chosen color.
For each vertex we then find its neighboring vertices in
time slice $i + 1$, and mark them with the same color, allowing also for mixing of
colors.
Clearly one can explicitly identify \emph{branches} in the two slices,
which proves that the fractal structure is correlated and not totally independent of each other.
Nevertheless, the branches may significantly change their shape.
The relation with further slices decays fast as the time step separation grows.
So far, we have only this qualitative notion at our disposal.
It is a challenge to define more quantitative methods measuring the correlation of successive trees,
i.e. an analog of scalar product acting in the space of trees
based on a four-dimensional definition of distance.

We conclude that the measurements of Hausdorff dimension 
evidently show that $d_h = 3$.
The spectral dimension $d_s$ is equal only half of this value, $d_s = 1.5$.
Such behavior indicates that the geometry of spatial slices is
completely different from a smooth geometry of a three-sphere.
Indeed, direct measurements on individual slices prove
that they are equipped with a fractal structure similar to branched polymers.
Quantum fluctuations can not be described by small Gaussian fluctuations
around some average geometry, suggesting that the specified hypersurfaces 
of constant time are not physical.
Still, the tree structure propagates over consecutive slices.

\clearemptydoublepage

\chapter{Implementation}
\label{Chap:Implementation}

The main tool used to obtain results within the framework of four-dimensional
Causal Dynamical Triangulations are numerical Monte Carlo simulations.
In CDT the partition function (\ref{Eq:ZCont}) is regularized by a discrete sum
(\ref{Eq:ZDisc}) over causal triangulations $\dT$.
Such approach allows to generate Monte Carlo triangulations and calculate
expectation values of observables according to (\ref{Eq:ExpVal}).
Therefore, we need to handle elements of the set of labeled causal triangulations $\tilde{\dT}$.
In this Chapter we introduce 
software solutions and data structures used to store information about objects
describing a triangulation.
Later, we describe details of the implementation of the Monte Carlo algorithm, 
especially the Monte Carlo moves.

Experience has shown that simulations are limited mainly
by the available CPU time, rather than by the amount of the used memory.
Therefore, the implementation of the algorithm uses
a maximal information philosophy.
The program stores in memory a full information 
about points, links, triangles and simplices.
This approach allows for quicker access to the needed information
at the expense of the increased memory demand.

All results presented in this work were obtained using 
a computer program written in C.
Therefore, we use the C language syntax to describe data structures.

\section{Parametrization of the manifold}

To perform the average (\ref{Eq:ExpVal}) the Monte Carlo algorithm
has to generate configurations according to the probability distribution
on an ensemble of causal triangulations $\tilde{\dT}$.
We have to define the way we code a simplicial complex
and store it in the computer memory.
To increase the acceptance of Monte Carlo moves,
and thus the efficiency of the algorithm,
we introduce many lists, enumerating the Lorentzian structures.

Once we have labeled vertices of a triangulation
and set its topology, the simplicial complex is given 
by the adjacency relations between four-simplices.
Four-simplices, which are also labeled, share five vertices.
The connections between simplices induce the adjacency relations on all sub-simplices
to which we also attach labels.
Each label is an integer number between $1$ and some maximal value. 
A label uniquely identifies sub-simplices appearing in the triangulation,
and instead of writing \emph{"label of an object"} we shall use a shortcut \emph{"object"}.
The topology $S^1 \times S^3$ is ensured by construction.
As we shall see, we start with a minimal configuration
with the designated topology and introduce
only local Monte Carlo moves which do not spoil the topology.
Therefore, to perform simulations the program has to code such 
objects as \emph{points}, \emph{edges}, \emph{triangles} and four-simplices.
It is not necessary to store information about tetrahedra ($3$-simplices)
which if needed may be easily extracted.
In this Section we describe most of the data structures used in this context.

\subsection*{Points}

The points are labeled by a dynamical sequence of labels.
The number of all accessible point labels is denoted as \verb#nn0#.
Therefore possible vertex labels are \verb#1 ... nn0#.
The number of used point labels \verb#num0# is equal to the number of vertices
in the triangulation $N_0$.
Some Monte Carlo moves are possible to be realized only on a local structure
containing a point with a coordination number equal $8$.
Coordination number of a sub-simplex is denoted as the number of four-simplices
sharing a given sub-simplex.
The list of used labels is stored in the array \verb#list0# which has a following structure:
\bi
\itemsep=-1ex
\topsep=-1ex
\partopsep=-1ex
\parsep=-1ex
\item Positions \verb#1 ... num0# are the used labels:\\[-4ex]
\bi
\topsep=-1ex
\itemsep=-0.5ex
\partopsep=-1ex
\parsep=-1ex
\item Positions \verb#1 ... num80# are the labels of points with coordination 8 (used in move $\bar{\textrm{4}}$).
\item Positions \verb#num80 + 1 ... num0# are the rest of used (labels of) points.
\ei
\item Positions \verb#num0 + 1 ... nn0# are the unused labels.
\ei
The properties of all points are stored in the array \verb#p# indexed by integer \emph{point labels}:
\begin{verbatim}
point p[nn0];
\end{verbatim}
The structure \verb#point# (declared in the file \verb#types.h#) holds information about a vertex:
\begin{verbatim}
typedef struct
{
    int i;      /* Index in list0                             */
    int n;      /* Coordination number                        */
    int l;      /* a random link which shares this vertex     */
    int t;      /* a random triangle which shares this vertex */
    int s;      /* a random simplex which shares this vertex  */
    int time;   /* time coordinate, number of the time slice  */
} point;
\end{verbatim}
The fields of the structure \verb#point# have a following interpretation:
\bi
\item Field \verb#i# contains index of this vertex in the array \verb#list0#,\\
 i.e. \verb#list0[p[num].i] = num#.
\item The coordination number \verb#n# denotes the number of simplices, which share this vertex.
\verb#n# is not less than 8.
\item Field \verb#l# is a label of a (random, among many) link sharing this vertex.
\item Field \verb#t# is a label of a (random, among many) triangle sharing this vertex.
\item Field \verb#s# is a label of a (random, among many) simplex that shares this vertex (used in move $\bar{\textrm{4}}$).
\item Field \verb#time# stores the integer time coordinate of a point,
i.e. the  number of a slice to which the point belongs.
Two points may be connected by a link only if their time difference is not larger than one. 
\ei

\subsection*{Links}
 
Links are labeled by a dynamical sequence of labels.
The list of used labels is in the array \verb#list1# with a following structure,
analogical to array \verb#list0#:
\begin{itemize}
\item Positions \verb#1 ... num1# are the used labels:
\begin{itemize}
\item Positions \verb#1 ... num61s# are spatial links with a coordination 6 (used in move $\bar{\textrm{5}}$).
\item Positions \verb#num61s + 1 ... num61s + num41t# (next \verb#num41t#) are time links with coordination 4 (used in move $\bar{\textrm{2}}$).
\item Positions \verb#num61s + num41t + 1 ... num1# are the rest of used links.
\end{itemize}
\item Positions \verb#num1 + 1 ... nn1# are the unused labels.
\end{itemize}
The data about all links are stored in array \verb#l# indexed by links labels:
\begin{verbatim}
digon l[nn1];
\end{verbatim}
The structure \verb#digon# (defined in \verb#types.h#) holds information about a link:
\begin{verbatim}
typedef struct
{
    int p[2]; /* Labels of endpoints            */
    int n;    /* Coordination number            */
    int s;    /* Simplex which shares this link */
    int i;    /* Position in list1              */
    int h;    /* Hash code of this link         */
    int n2;   /* Next link in a hash table list   */
} digon;
\end{verbatim}
The meaning of the fields is as follows:
\bi
\item The labels \verb#p# of the endpoints. The labels are sorted: \verb#p[0] < p[1]#.
\item The coordination number \verb#n# denotes the number of simplices, that share this link.
\verb#n# is at least $4$ for time links, and at least $6$ for spatial links.
\item Field \verb#s# is a label of one (random, among many) four-simplex containing a link with given label.
\item Field \verb#i# contains the position of the link in the array \verb#list1#:\\ \verb#list1[l[num].i] = num#.
\item Field \verb#h# contains the hash code for the link:\\ \verb#l[num].h = hlc(l[num].p[0], l[num].p[0])#.
\item Field \verb#n2# contains the label of the next link in the singly-linked list of links with the same hash-code \verb#h#.
\ei

\subsubsection*{Hash map}
In move 2 a time link with a coordination 4 is created,
while in move 5 a spatial link with coordination 5 is created.
It might happen that a link with the same endpoints already exists.
In this case we have to abort the move.

In order to perform a fast search of a link given the labels of its endpoints,
the \emph{hash table} is used.
Knowing the labels of vertices \verb#p[0]# $<$ \verb#p[1]#
we calculate the \emph{hash code} \verb#h# using the hash function \verb#hlc#
\begin{verbatim}
h = hlc(p[0], p[1]) = ((p[0] << 10) xor p[1]) 
\end{verbatim}
defined in file \verb#core/hash.h#.
The first element of a singly-linked list of links with the same \emph{hash code} \verb#h#
is given by \verb#hlt[h]#.
The next element in the list is accessed via the field \verb#n2# in the structure \verb#digon#.
To find if a given link exists, we have to look through this list - very short on average.

\subsection*{Triangles}
The triangles are labeled by a dynamical sequence of labels stored in the array \verb#list2#:
\bi
\item Positions \verb#1 ... num32t# are \emph{time} triangles with a coordination 3 (used in move 3).
There are no time triangles with coordination less than 3.
\item Positions \verb#num32t + 1 ... num32t + num42s# are \emph{spatial} triangles with a coordination 4 (used in move 5). 
There are no spatial triangles with a coordination less than 4.
\item Positions \verb#num32t + num42s + 1 ... num2# are the rest of used links.
\item Positions \verb#num2 + 1 ... nn2# are the unused labels.
\ei
The properties of all triangles are stored in the array \verb#t# indexed by a triangle \emph{label}:
\begin{verbatim}
triangle t[nn2];
\end{verbatim}
The structure \verb#triangle# (declared in \verb#types.h#) holds information about a triangle:
\begin{verbatim}
typedef struct
{
    int p[3];  /* Labels of vertices                 */
    int n;     /* Coordination number                */
    int s;     /* simplex which shares this triangle */
    int i;     /* Position in list2                  */
    int h;     /* Hash code of this triangle         */
    int n2;    /* Next triangle in hash table list   */
} triangle;
\end{verbatim}
\bi
\item Field \verb#p# stores sorted labels of vertices: \verb#p[0] < p[1] < p[2]#.
\item Field \verb#n# is a coordination number of a triangle, i.e. athe number of simplices, that share this triangle.
It is not less than 3.
\item Field \verb#s# is a label of a simplex which shares this triangle (used in move 3).
\item Field \verb#i# contains index of this triangle in the array \verb#list2#.
\item Field \verb#h# contains the hash code of a triangle:\\ \verb#t[num].h = htc(t[num].p[0], t[num].p[0], t[num].p[2])#.
\item Field \verb#n2# contains the label of the next triangle in the singly-linked list of triangles with the same hash-code \verb#h#.
\ei

\subsubsection*{Hash map}
In move 3 a time triangle with a coordination 3 is created,
while in move $\bar{\textrm{5}}$ a spatial triangle with a coordination 4 is created.
It might happen that this triangle already exists (a triangle with the same vertices).
In this case we have to abort the move.

To perform a fast check if a triangle with given vertices exists,
a \emph{hash table} is used.
Given the labels of vertices \verb#p[0]# $<$ \verb#p[1]# $<$ \verb#p[2]#
we calculate the \emph{hash code} \verb#h# using the hash function \verb#htc#
\begin{verbatim}
h = htc(p[0], p[1], p[2]) = (p[0] * 256) xor (p[1] * 16) xor p[2] 
\end{verbatim}
defined in a file \verb#core/hash.h#.
The first element of a singly-linked list of triangles with the same \emph{hash code} \verb#h#
is given by \verb#htt[h]#.
The next element in the list is accessed via the field \verb#n2# in the structure \verb#triangle#.
Similarly as for links, to find if a given triangle exists, we have to look through this list - very short on average.

\subsection*{Simplices}
In CDT, there are two types of simplices:
\bi
\item type $\{4, 1\}$ and $\{1, 4\}$ with four vertices in one time layer, 
\item type $\{3, 2\}$ and $\{2, 3\}$ with three vertices in one time layer. 
\ei
The list of simplex labels is stored in the array \verb#list4# with a following order:
\bi
\item Positions \verb#1 ... num444# are simplices of type $\{4, 1\}$ and $\{1, 4\}$.
\item Positions \verb#num444 + 1 ... num4# are simplices of type $\{3, 2\}$ and $\{2, 3\}$.
\item Positions \verb#num4 + 1 ... nn4# are the unused labels.
\ei
The properties of all simplices are stored in the array \verb#s# indexed by a simplex \emph{label}:
\begin{verbatim}
pentachoron s[nn4];
\end{verbatim}
The structure \verb#pentachoron# (declared in \verb#types.h#) holds information about a simplex:
\begin{verbatim}
typedef struct
{
    int p[5];   /* Labels of vertices */
    int s[5];   /* Labels of adjacent 4-simplices */
    int l[10];  /* Labels of links shared by simplex */
    int t[10];  /* Labels of triangles shared by simplex */
    int i;      /* Position in list4 */
    int rez[5]; /* Reserved. Aligned to 32 */
} pentachoron;
\end{verbatim}
The fields of the structure \verb#pentachoron# store a following information:
\bi
\item Field \verb#p# stores labels of vertices of a simplex.
\item Field \verb#s# stores labels of the neighboring simplices. 
Simplex \verb#s[i]# is the neighbor opposite to \verb#p[i]#.
\item Field \verb#l# stores labels of links shared by a simplex.
The links in the list \verb#l# are sorted w.r.t. the first endpoint, and then w.r.t. the second one,
where the ordering of the endpoints is given by \verb#p[i]#.
\item Field \verb#t# stores labels of triangles shared by a simplex.
The triangles sorted w.r.t. the first, second than third endpoint
where the ordering of endpoints is given by \verb#p[i]#.
\item Field \verb#i# is the index of a simplex in the array \verb#list4#.
\ei
On simplices, we finish the description of the data structures used to store full information about the triangulation.
The general idea of the Monte Carlo program
boils down to initialize above structures 
and sequential calling procedures that modify data in accordance with Monte Carlo moves. 

\section{Monte Carlo Simulations}

The idea behind the Monte Carlo simulations
is to approximate the sum appearing in the partition function (\ref{Eq:ZDisc})
by a sum over a finite number of Monte Carlo configurations.
More precisely, given an observable $\mathcal{O}[\cT]$, e.g. a distribution of a three-volume $N(i)$ of spatial slices,
we would like to calculate its expectation value $\langle \mathcal{O}[\cT] \rangle$.
In the CDT framework the geometries $[g]$ are already restricted to a discrete set of simplicial manifolds $\mathcal{T}$ and
\begin{equation}
\langle \mathcal{O}[\cT] \rangle = \frac{1}{Z} \sum_{\tilde{\mathcal{T}} \in \tilde {\mathbb{T}}} \frac{1}{N_{0}[\tilde{\mathcal{T}}] !} \mathcal{O}[\tilde{\mathcal{T}}] e^{- S[\tilde{\mathcal{T}}]},
\label{eq:ExDisc}
\end{equation}
where the partition function $Z$ is given by eq. (\ref{Eq:ZDisc}) with the action (\ref{Eq:SRegge}) and 
the $\tilde{\cT}$ corresponds to the labeling of $\cT$.

Monte Carlo simulations generate a finite set of configurations $\{\cT^{(1)}, \dots, \cT^{(K)}\}$
and allow to approximate the average (\ref{eq:ExDisc}) by a summation over it,
\begin{equation}
\langle \mathcal{O}[\cT] \rangle \approx \frac{1}{K} \sum_{i=1}^{K} \mathcal{O}[\cT^{(i)}] .
\label{eq:ExApp}	
\end{equation}

Let us notice, that no factor $\frac{1}{N_{0}[\tilde{\mathcal{T}}] !} e^{- S[\tilde{\mathcal{T}]}}$ is needed,
since configurations are generated according to the probability distribution $P[\tilde{\cT}] = \frac{1}{Z}\frac{1}{N_{0}[\tilde{\mathcal{T}}] !} e^{- S[\tilde{\mathcal{T}]}}$,
which means that more probable geometries will more likely appear in the set $\{\cT^{(1)}, \dots, \cT^{(K)}\}$. 
We implement the Metropolis-Hastings algorithm to generate 
a Markov Chain using Monte Carlo methods \cite{Metropolis, Hastings}. 
We define and implement local transformations, called Monte Carlo moves,
which modify a configuration $\cA$.
The Monte Carlo move applied to a triangulation $\cA$ transforms it into
a triangulation $\cB$.
A series of such moves defines a Markov chain,
the next configuration depends only on the last and on the applied move.
After one step configurations $\cA$ and $\cB$ are almost identical,
but after sufficiently large number of moves separating them,
the initial and final configurations are independent. 
If the moves are chosen randomly,
the algorithm performs a random walk in the phase-space of configurations, 
i.e. causal triangulations $\dT$. 
The moves are however not chosen completely at random.
The moves and their weights, i.e. probability of being realized,
must fulfill some constraints in order to properly probe 
the ensemble of causal triangulation $\tilde{\cT}$ according to (\ref{eq:ExDisc}).
The algorithm, after satisfying  the conditions described below, 
ensures that it samples the configuration space according to a probability
distribution
\[ P[\tilde{\cT}] \equiv \frac{1}{Z} \frac{1}{N_{0}[\tilde{\mathcal{T}}] !} e^{- S[\tilde{\mathcal{T}}]}. \]
The Monte Carlo moves must fulfill following conditions:\\
\point{Ergodicity.} 
The moves generating the Markov chain have to be ergodic.
It means that it is possible to reach any topologically equivalent 
causal triangulation by a sequence of such moves.
Our scheme of assembling an appropriate set of elementary moves
is to first select moves that are ergodic within the spatial slices.
This is clearly a necessary condition for ergodicity in the full triangulations.
Namely, those are moves $4$ and $5$ together with their inverse moves
\footnote{The numeration of moves corresponds to the convention used in the computer program.}.
These moves act on spatial slices like 
three-dimensional Pachner moves
which are equivalent to the primal Alexander moves
and thus proved to be ergodic and preserving the $S^3$ topology of the space \cite{Pachner,Alexander,GrossVarsted}.

Then we supplement them by moves $2$ and $3$ acting within two spatial slices.
They do not affect connections between tetrahedra building the spatial slices
and can be understood as a Lorentzian variant, i.e. combination, 
of the four-dimensional Pachner moves,
in the sense that they are compatible with the discrete time slicing of our causal
geometries \cite{Dyna, Background}.
The moves are described in detail in the next Section.

\point{Topology and causality.}
Because move $3$ is self-dual,
the CDT program is using a set of seven local moves,
being a combination of Pachner moves.
By construction, these moves do not destroy the topology $S^1 \times S^3$ of the spacetime,
and preserve the global proper-time foliation.
Moves affecting the spatial slices preserve the $S^3$ topology of the space.
The two features define the entropy factor.
In fact, we know neither what class of geometries we should consider in quantum gravity
nor the proper measure. The CDT model is a trial to define these objects.

\point{Detailed balance condition.}
The previous two points are independent of the probability distribution of configurations
specified by the action $P[\cT] = \frac{1}{Z} e^{-S[\cT]}$.
Given a move, it transforms a configuration $\mathcal{A}$ into configuration $\mathcal{B}$.
In order to ensure that the Monte Carlo algorithm 
probes the configuration space with probability $P[\cT]$,
the move has to be accepted with some probability weight $W(\mathcal{A} \rightarrow \mathcal{B})$
and satisfy the detailed balance condition:
\[ P(\mathcal{A}) W(\mathcal{A} \rightarrow \mathcal{B}) = P(\mathcal{B}) W(\mathcal{B} \rightarrow \mathcal{A}) . \]
This condition determines the weights, however not uniquely. 
The probability $P(\mathcal{A})$ depends on the considered model.
The transition probability $W(\cA \rightarrow \cB)$ is a product of two factors:
\[ W(\cA \rightarrow \cB) = W^{(1)}(\cA \rightarrow \cB) \cdot W^{(2)}(\cA \rightarrow \cB).\]
The first $W^{(1)}(\cA \rightarrow \cB)$ is determined by the way 
we choose a position in $\cA$ where the move is performed
and by the move type.
 The second $W^{(2)}(\cA \rightarrow \cB)$ is chosen such that
the detailed balance condition is satisfied.
Once we have decided what Monte Carlo moves, and how we implement them,
the weight $W^{(1)}(\cA \rightarrow \cB)$ is fixed and depends on the internal probability
of choosing a move.
A good Monte Carlo algorithm has the weights $W^{(2)}(\cA \rightarrow \cB)$ and $W^{(2)}(\cB \rightarrow \cA)$
close to $1$.
This is equivalent to a high acceptance of moves.

\point{Autocorrelation and thermalization time.}
An accurate approximation of the average (\ref{eq:ExApp}) requires 
a suitably large sample of independent configurations $\{\cT^{(1)}, \dots, \cT^{(K)}\}$. 
Configurations are collected in intervals, 
measured in a number of Monte Carlo steps,
whose length determine correlations between configurations.
Intuitively, the larger is the number of Monte Carlo moves which separate triangulation $\cT^{(k)}$ and $\cT^{(k+1)}$
the more uncorrelated they are.
The separation should be chosen properly based on the autocorrelation time,
which can be estimated using some slowly changing variables,
like a width of the volume distribution $N(i)$ or its mass center position.
We will refrain from discussing the details of this numerical procedure.
Before we can start measurements, the configurations have to undergo a thermalisation procedure,
which consists on sufficiently long execution of the algorithm without performing measurements.

\point{Tuning of $K_4$.}
The Regge action (\ref{Eq:SRegge}) depends on three bare coupling constants $K_0, \Delta$ and $K_4$.
For technical reasons of simulations it is preferable to enforce that the total number 
of four-simplices $N_{41}$ fluctuates around some prescribed value \verb#nvolume#
during the Monte Carlo simulation. 
The coupling constant $K_4$ is related to the bare cosmological
coupling constant.
If the value of $K_4$ is smaller than its critical value $K_4^c$,
the total volume $N_4$ tends to explode.
In the opposite case $K_4 > K_4^c$, the configuration collapses.
Thus $K_4$ should be tuned to its critical value.
In practice we use a modified Regge action by adding a term $+\epsilon |N_{41} - \verb#nvolume#|$
which, for $K_4 \approx K_4^c$, ensures that $N_{41}$ fluctuates around \verb#nvolume#.
Alternatively, we can add a quadratic term $+\epsilon (N_{41} - \verb#nvolume#)^2$
which is also studied. Presented results are consistent in both cases.

\begin{figure}
\begin{center}
\includegraphics[width=0.9\textwidth]{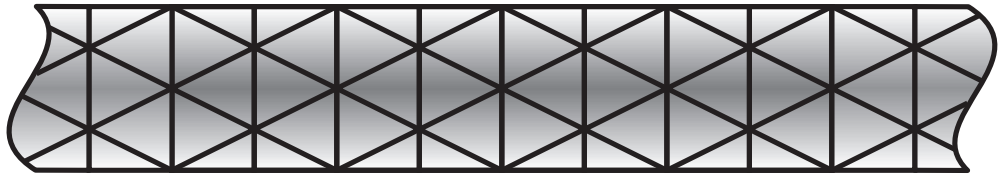}
\caption{A two-dimensional visualization of a starting configurations. 
It presents a minimal triangulation of a thin tube with glued ends,
thus having a topology of a torus $S^1 \times S^1$.
In four dimensions the topology of our interests is $S^1 \times S^3$.}
\label{fig:starting}
\end{center}
\end{figure}

\noindent\textbf{Starting configurations.}
Before we can apply the moves and generate a Markov chain
we must create the initial configuration.
We construct a minimal configuration with a topology $S^1 \times S^3$.
Fig. \ref{fig:starting} presents a two-dimensional analogue of the starting configuration.
First, we assemble the simplest three-dimensional simplicial complex with a topology $S^3$,
corresponding to a spatial slice at a fixed time.
It is a boundary of a 4-simplex and contains five tetrahedra \emph{glued} to each other. 
Let us label five vertices of this complex by $0, 1, 2, 3, 4$.
Then the five tetrahedral interfaces are denoted as
\[[0, 1, 2, 3], [0, 1, 2, 4], [0, 1, 3, 4], [0, 2, 3, 4]\ \textrm{and}\ [1, 2, 3, 4]. \]
Next, we take two such structures, 
first at time $i$ with vertices $0, 1, 2, 3, 4$,
second at time $i+1$ with vertices $\overline{5}, \overline{6}, \overline{7}, \overline{8}, \overline{9}$.
And join them with a minimal number of 4-simplices.
Each of five interfaces at time $i$ becomes a base of a simplex
with a top vertex at time $i+1$, thus creating five $\{4, 1\}$ four-simplices.
Let us denote these by
\[ \{0, 1, 2, 3, \overline {5}\}, 
   \{0, 1, 2, 4, \overline {6}\}, 
   \{0, 1, 3, 4, \overline {7}\}, 
   \{0, 2, 3, 4, \overline {8}\}, 
   \{1, 2, 3, 4, \overline {9}\}.\]
These are supplemented by 10 simplices of the type $\{3, 2\}$:
\[
 \{0, 1, 2, \overline {5, 6}\}, 
 \{0, 1, 3, \overline {5, 7}\}, 
 \{0, 2, 3, \overline {5, 8}\}, 
 \{1, 2, 3, \overline {5, 9}\}, 
 \{0, 1, 4, \overline {6, 7}\}, 
\]
\[
 \{0, 2, 4, \overline {6, 8}\}, 
 \{1, 2, 4, \overline {6, 9}\}, 
 \{0, 3, 4, \overline {7, 8}\}, 
 \{1, 3, 4, \overline {7, 9}\}, 
 \{2, 3, 4, \overline {8, 9}\}. 
\]
Likewise, mirror reflected simplices are created,
10 simplices of the type $\{2, 3\}$:
\[
 \{0, 1, \overline {5, 6, 7}\}, 
 \{0, 2, \overline {5, 6, 8}\}, 
 \{0, 3, \overline {5, 7, 8}\}, 
 \{0, 4, \overline {6, 7, 8}\}, 
 \{1, 2, \overline {5, 6, 9}\}, 
\]
\[
 \{1, 3, \overline {5, 7, 9}\}, 
 \{1, 4, \overline {6, 7, 9}\}, 
 \{2, 3, \overline {5, 8, 9}\}, 
 \{2, 4, \overline {6, 8, 9}\}, 
 \{3, 4, \overline {7, 8, 9}\}, 
\]
and finally 5 simplices of the type $\{1, 4\}$:
\[
  \{0, \overline {5, 6, 7, 8}\}, 
  \{1, \overline {5, 6, 7, 9}\}, 
  \{2, \overline {5, 6, 8, 9}\}, 
  \{3, \overline {5, 7, 8, 9}\}, 
  \{4, \overline {6, 7, 8, 9}\}.
\]
The result of these operations is a four-dimensional segment with a topology $[0, 1] \times S^3$.
In total, one segment is composed of 
five simplices of type $\{4, 1\}$, ten simplices of type $\{3, 2\}$,
ten simplices of type $\{2, 3\}$ and five simplices of type $\{1, 4\}$
summing up to 30 simplices.
The boundary of a segment consists of two identical components, built of five interfaces.
A number of segments can be iteratively joined one after another 
forming a simplicial manifold of a desired time length.
In order to obtain a starting configuration with a topology $S^1 \times S^3$,
i.e. periodic boundary conditions,
the interfaces at the beginning and at the end are glued together.

\subsection*{Algorithm}
The scheme of the algorithm is as follows:
\begin{enumerate}
\item We select the type of a move at random.
The probabilities are chosen in such a way that the numbers of accepted moves of each type are approximately the same.
\item Each move type is related to a characteristic structure of  simplices,
which must be met in order to be accepted.
We choose at random a point, link, triangle or simplex that satisfy these conditions. 
The design of the program provides a quick access to such elements.
\item We check if the new simplicial complex  corresponds
 to a triangulation of a manifold.
One must avoid creating  simplices or subsimplices already present in the triangulation.
Also, the move must not destroy the foliation.
\item We compute the Metropolis weights $W^{(2)}$ and accept or reject
 the move with proper probabilities.
\item We update the lattice structures.
\end{enumerate}

In the next Section we describe all types of moves,
in particular we describe the conditions which must be satisfied in order to perform a move.

\section{Monte Carlo Moves}

We start the discussion of Monte Carlo moves with recalling
the Regge action which determines the probability distribution (\ref{eq:ExDisc})
on the ensemble of causal triangulations.
The implementation of moves may be found in files \verb#moves/move??.c# (\verb#??# denotes a particular move name).
The action used in the  program is parametrized as:
\[ S[\mathcal{T}] = - (K_0 + 6 \Delta) N_0 + K_4 N_4 + \Delta N_{41}, \]
where $K_0$, $K_4$ and $\Delta$ are bare coupling constants and
$N_0$ is the number of vertices,
$N_4$ is the number of simplices,
$N_{41}$ is the number of simplices of the type $\{4, 1\}$.
In the remaining part of this Section we focus on particular types of moves.

\subsection*{Move 2}
Move $2$ replaces the interface \av{0, 1, 2, 3} between the two four-simplices by
a link \av{4, 5} dual to the interface.
Here the numbers denote labels of vertices involved in the move,
and the bracket \av{\dots} denotes a tetrahedron built on  given vertices.
The new link has a coordination number $4$.
We can only create a new time (and not space) link with a coordination 4.
This can easily be checked before the move is attempted. The first step in the
move is to pick a
$\{3,2\}$ or $\{2,3\}$ four-simplex from the list and to chose the interface.
The neighboring four-simplex can be
either again a  $\{3,2\}$ or $\{2,3\}$ four-simplex or a $\{4,1\}$ or 
$\{1,4\}$ four-simplex. The move is accepted with a different weight in
these two cases, since
in the first case there are two ways it can be realized and in the second only one way,
this affects $W^{(1)}$.
The operation of the move $2$ can by schematically written as
\bea
\{0, 1, 2, 3, 4\} &+& \{0, 1, 2, 3, 5\} \\
&\Downarrow&\\
\{0,  1,  2,  4,  5\} +
\{0,  1,  3,  4,  5\} &+&
\{0,  2,  3,  4,  5\} +
\{1,  2,  3,  4,  5\}
\eea
The effect of this move is:
\begin{itemize}
\item The number of points does not change. $\Delta N_0 = 0$.
\item New time link \av{4, 5} with  coordination number $4$ is created.
\item Four new time triangles  with a coordination number $3$ are created.
\item Two simplices of type $\{2, 3\}$ are created. $\Delta N_{14} = 0$, $\Delta N_{23} = 2$.
\end{itemize}
The weight for the move depends on the Regge action and is given by
\[ w_2 \equiv \frac{P(\cB)}{P(\cA)} =  e^{-2 K_4} ,\]
where $\cA$ denotes a triangulation before performing the move
and $\cB$ after performing the move.
This relation determines the value of $W^{(2)}(\cA \rightarrow \cB)$.

\subsection*{Move $\bar{\textbf{2}}$}

It is an inverse of move 2. 
It requires a (time) link with a coordination number 4.
Its action can be summarized by
\bea
\{ 0, 1, 2, 4, 5\} + \{ 0, 1, 3, 4, 5\} &+&
\{ 0, 2, 3, 4, 5\} + \{ 1, 2, 3, 4, 5\}\\
&\Downarrow&\\
\{0, 1, 2, 3, 4\} &+& \{0, 1, 2, 3, 5\} \\
\eea
\begin{itemize}
\item The number of points does not change. $\Delta N_0 = 0$.
\item Time link $[4, 5]$ with a coordination number $4$ is removed.
\item Four time triangles with a coordination number $3$ are removed.
\item Two simplices of type $\{2, 3\}$ are removed. $\Delta N_{14} = 0$, $\Delta N_{23} = -2$.
\end{itemize}
Analogically as in previous move, the weight for the move is $w_{\bar{2}} = e^{2 K_4}$.

\subsection*{Move 3}
A self-dual move. 
It requires a time triangle \av{0, 1, 2} with a coordination number 3,
and replaces it by a dual triangle \av{3, 4, 5} with the same coordination number.
A triangle with a coordination 3 can only be a time 
triangle. The spatial triangles necessarily have coordination 4 or more.
It can be easily seen by a following argument: a spatial triangle is an 
interface between two spatial tetrahedra. Therefore it will belong to at least
4 four-simplices, for which these tetrahedra are bases and the tips point up
or down. In other words creating spatial triangles with a coordination 3, 
although superficially possible, creates a topological defect.

\arraycolsep 0em
\bea
\{0,  1,  2,  3,  4\} + \{0,  1,  &2&, 3,  5\} + \{0,  1,  2,  4,  5\} \\
&\Downarrow&\\
\{0,  1,  3,  4,  5\} + \{0,  2,  &3&, 4,  5\} + \{1,  2,  3,  4,  5\} 
\eea
\bi
\item	No new points nor links are created. $\Delta N_0 = 0$.
\item	Triangle \av{0, 1, 2} is removed.\\
		Triangle \av{3, 4, 5} is created.
\item	The number of simplices does not change. $\Delta N_{14} = \Delta N_{4} = 0$.
\ei
All changes are zero, the weight is $w_3 = 1$.

\subsection*{Move 4}
This move adds a new vertex at a given time slice $i$.
It acts on a configuration of $\{1, 4\}$ and $\{4 ,1\}$ simplices. 
Let the two simplices be
$\{0, 1, 2, 3, \underline{5}\}$ and $\{ 0, 1, 2, 3,\overline{6}\}$, where bars
indicate the time position of points.
We produce a new point $\{4\}$ between the two simplices.
The move adds 6 new simplices of type $\{1, 4\}$ to our system, and the vertex $\{4\}$ has a coordination number equal $8$.
\arraycolsep 0em
\bea
\{ 0,  1,  2,  3,  \underline{5} \} &+& \{ 0,  1,  2,  3,  \overline{6} \} \\
&\Downarrow&\\
\{ 0,  1,  2,  4,  \underline{5} \} + \{ 0,  1,  3,  4,  \underline{5} \} &+& \{ 0,  2,  3,  4,  \underline{5} \} + \{ 1,  2,  3,  4,  \underline{5} \} + \\
\{ 0,  1,  2,  4,  \overline{6} \} + \{ 0,  1,  3,  4,  \overline{6} \} &+& \{ 0,  2,  3,  4,  \overline{6} \} + \{ 1,  2,  3,  4,  \overline{6} \} 
\eea
\bi
\item	One new point $\{4\}$ with a coordination number $8$ is created.\\ 
		This property is used as a signal for the inverse move. $\Delta N_0 = 1$.
\item	Four new spatial links with a coordination number $6$ are created.\\
		Two new time links with a coordination number $4$ are created.
\item	Six new spatial triangles with  coordination number $4$ are created.\\
		Eight new time triangles with a coordination number $3$ are created.
\item	Six new simplices of the type $\{1, 4\}$ are created $\Delta N_{14} = 6$, $\Delta N_{23} = 0$.
\ei
The weight for this move is $w_{4} = e^{- 6 K_4 + K_0}$.

\subsection*{Move $\overline{\bf 4}$}
This move is an inverse of move 4.
We pick point $\{4\}$ with a coordination 8 from the list of vertex labels.
Such a point may be shared only by simplices of type $\{1, 4\}$
and the neighborhood satisfies the necessary constraints. 
In this move the spatial interface \av{0, 1, 2,3} is created and we have to check
if such interface does not already exist. If \emph{yes}, we abort the move.
\arraycolsep 0em
\bea
\{ 0,  1,  2,  4,  \underline{5} \} + \{ 0,  1,  3,  4,  \underline{5} \} &+& \{ 0,  2,  3,  4,  \underline{5} \} + \{ 1,  2,  3,  4,  \underline{5} \} + \\
\{ 0,  1,  2,  4,  \overline{6} \} + \{ 0,  1,  3,  4,  \overline{6} \} &+& \{ 0,  2,  3,  4,  \overline{6} \} + \{ 1,  2,  3,  4,  \overline{6} \} \\
&\Downarrow&\\ 
\{ 0,  1,  2,  3,  \underline{5} \} &+& \{ 0,  1,  2,  3,  \overline{6} \} 
\eea
\bi
\item	Point $4$ with a coordination $8$ is removed. $\Delta N_0 = -1$
\item	Four spatial links with a coordination $6$ are removed.\\
		Two time links with a coordination $4$ are removed.
\item	Six spatial triangles with a coordination $4$ are removed.\\
		Eight time triangles with a coordination $3$ are removed .
\item	Six simplices of the type $\{1, 4\}$ are removed. $\Delta N_{14} = -6$, $\Delta N_{23} = 0$.
\ei
The weight for this move is
\[ w_{\bar{4}} = e^{6 K_4 - K_0} .\]

\subsection*{Move 5}
This move consists on a restructuring of the triangle between four simplices.
This move requires a spatial triangle \av{0, 1, 2} with  coordination 4.
By construction it is shared by four simplices of the type $\{1, 4\}$.
Vertices $0, 1, 2, 3, 4$ have the same time position.
Vertex $5$ is in the previous layer and vertex $6$ is in the next layer (or reversed).
A starting configuration must have 4 four-simplices in the following setup:
\bea
\{ 0,  1,  2,  3,  \underline{5}\} +
\{ 0,  1,  2,  4,  \underline{5}\} &+& 
\{ 0,  1,  2,  3,  \overline{6}\} + 
\{ 0,  1,  2,  4,  \overline{6}\} \\
&\Downarrow&\\
\{ 0,  1,  3,  4,  \underline{5} \} +
\{ 0,  2,  3,  4,  \underline{5} \} +
\{ 1,  2,  3,  4,  \underline{5} \} &+& 
\{ 0,  1,  3,  4,  \overline{6} \} +
\{ 0,  2,  3,  4,  \overline{6} \} +
\{ 1,  2,  3,  4,  \overline{6} \}
\eea
\bi
\item	The number of points doesn't change. $\Delta N_0 = 0$.
\item	One new spatial link \av{3, 4} with  coordination $6$ is created. \\
		This property is used as a signal for the inverse move.
\item	Two new time triangles with a coordination $3$ are created.\\
		Three new spatial triangles with a coordination $4$ are created.\\
		Spatial triangle \av{0, 1, 2} with a coordination $4$ is removed.
\item	Two new simplices of the type $\{1, 4\}$ are created. $\Delta N_{14} = 2$, $\Delta N_{23} = 0$.
\ei 
The weight for the move is $w_{5} = e^{- 2 K_4 - 2 \Delta}$.

\subsection*{Move $\overline{5}$}
This move is an inverse of move 5. 
We pick a spatial link \av{3, 4} with a coordination 6.
Such a link is always shared by $6$ simplices of the type $\{1,4\}$.
\bea
\{ 0,  1,  3,  4,  \underline{5} \} +
\{ 0,  2,  3,  4,  \underline{5} \} +
\{ 1,  2,  3,  4,  \underline{5} \} &+& 
\{ 0,  1,  3,  4,  \overline{6} \} +
\{ 0,  2,  3,  4,  \overline{6} \} +
\{ 1,  2,  3,  4,  \overline{6} \}\\
&\Downarrow&\\
\{ 0,  1,  2,  3,  \underline{5}\} +
\{ 0,  1,  2,  4,  \underline{5}\} &+& 
\{ 0,  1,  2,  3,  \overline{6}\} + 
\{ 0,  1,  2,  4,  \overline{6}\}
\eea
\bi
\item	The number of points doesn't change. $\Delta N_0 = 0$.
\item	Spatial link \av{3, 4} with a coordination $6$ is removed.
\item	Move creates a spatial triangle \av{0, 1, 2} with coordination $4$, unless it already exists.
      If this is the case, the move is aborted.
		This property is used as a signal for the move 5.\\
		Two time triangles with a coordination $3$ are removed.\\
		Three spatial triangles with a coordination $4$ are removed.\\
\item	Two simplices of the type $\{1, 4\}$ are removed. $\Delta N_{14} = -2$, $\Delta N_{23} = 0$.
\ei
The weight for the move is
\[ w_{\bar{5}} = e^{2 K_4 + 2 \Delta} .\]

At this point we finish the description of the implementation of the Monte Carlo algorithm.
We presented the basics necessary to create a program generating
Monte Carlo configurations in the framework of 
four-dimensional Causal Dynamical Triangulations.
A more detailed and technical description is beyond the scope of this thesis.
We shall omit e.g. issues of the random number generator,
preliminary acceptance, features specific to causal triangulations
which allow to reject a move at the initial stage,
and other solutions affecting the efficiency of the implementation.

Although we have devoted only one Chapter to the problem of the causal triangulations generator,
it should be noted that the software development was the major part of the work described in this dissertation.

A copy of the \emph{source code} together with the \emph{user guide} and the \emph{documentation}
is attached to this dissertation.
The author distributes this package on request through the  e-mail: \verb#atg@th.if.uj.edu.pl#

\clearemptydoublepage

\chapter*{Conclusions}
  \addcontentsline{toc}{chapter}{Conclusions}
\markboth{Conclusions}{Conclusions}

The aim of this thesis is to present a comprehensive review of recent results obtained
within the framework of the four-dimensional model of Causal Dynamical Triangulations (CDT).
We provided  answers to questions like:
\emph{how} does a background geometry emerge dynamically,
\emph{what} does it correspond to and \emph{how} to describe quantum fluctuations around the average geometry.

The model of Causal Dynamical Triangulations is a non-perturbative and background independent 
approach to quantum gravity.
The foundations of this model are very simple.
It is a mundane lattice field theory 
with a piecewise linear manifold serving as a regularization of general relativity.
The introduction of Wick rotation allows to use very powerful Monte Carlo techniques
and calculate quantum expectation values of observables.

Based on the Monte Carlo measurements the CDT model predicts the existence of three phases.
Introduction of a qualitative order parameter describing typical geometries appearing 
in a given phase, allows to make a one-to-one correspondence with 
three phases of an anisotropic theory of a Lifshitz scalar,
being an inspiration for Ho\v{r}ava-Lifshitz gravity model.
This two models, in addition to formal similarities 
involving the imposition of a global time-foliation,
exhibit the same scale dependence of the spectral dimension.
From a physical viewpoint, 
particularly interesting is the phase in which the scale factor 
as a function of time behaves as a bell-shaped distribution
spontaneously breaking the time-translational symmetry.
This phase is called de Sitter phase.

In order to calculate a meaningful average of the three-volume
we took into account the nonuniform volume distribution and 
introduced the procedure which breaks the time-translational freedom.
This step was crucial to evaluate the average scale factor distribution
and confirmed that a background geometry emerges dynamically.
The average geometry behaves like a well-defined four-dimensional manifold, 
both the Hausdorff dimension and spectral dimension equal $4$ at large scales.
We showed that the background geometry corresponds to a Euclidean de Sitter space,
i.e. a four-sphere, an isotropic and homogeneous solution of the vacuum Einstein field equations
with a positive cosmological constant.
Therefore CDT produces a picture of the Universe with 
superimposed finite quantum fluctuations around the classical trajectory.
In terms of the lattice spacing, configurations resemble 
a four-dimension spheroid
elongated in the time direction.

The elimination of a translational mode enables 
to observe the emergence of the background geometry,
but as a consequence also allows to investigate quantum fluctuations
around the average geometry.
Using Monte Carlo methods we measured the covariance matrix 
of scale factor fluctuations. 
Applying the semiclassical expansion we reconstructed 
the effective discrete action describing the three-volume.
In the CDT model no reduction of degrees of freedom is introduced.
The three-volume distribution is obtained by integrating out all degrees of freedom except 
the scale factor.
The semiclassical expansion takes into
account both the path integral measure and the bare action,
making the analysis truly non-perturbative.
The extracted effective action may be identified 
with the discretization of the minisuperspace action with an opposite sign.
The effect of the reversed sign is a consequence of the entropy factor.
Due to the identification, the effective coupling constant can be
related to the physical gravitational constant,
giving a recipe of how to obtain a meaningful continuum limit
and expressing the lattice constant in terms of physical units.

Understanding of the geometric nature of three-dimensional spatial slices,
interpreted as leaves of the imposed global proper-time foliation,
may be of crucial importance also for other attempts to quantize gravity.
Although the measurements show that the Hausdorff dimension is equal $3$,
the measured spectral dimension is only half of this value,
suggesting a fractal structure of spatial slices.
Indeed, the fractality is confirmed by a direct analysis of tree structures 
defined in terms of so-called \emph{minimal necks}.

{\it Outlook.}
There are many unresolved questions.
The author would like to list few points he finds most interesting:
\vspace*{-6pt}
\begin{itemize}
\itemsep=-1ex
\item What is the role of simplices of type 
$\{3, 2\}$ and $\{4, 1\}$ in better understanding of the critical behavior when approaching the phase transition?
\item What is the order of $B$-$C$ phase transition and how to obtain a proper continuum limit?
\item Whether the Ho\v{r}ava-Lifshitz scenario is realized
in our model, i.e. if the anisotropic space-time scaling in the UV-limit is observed?
\item How to improve the Monte Carlo algorithm to increase its efficiency
near the phase transition?
\end{itemize}
\vspace*{-6pt}
A physically relevant theory must describe matter fields.
The work on incorporating multiple massive scalar fields and massive point particle is already in progress.
We developed an efficient method of calculating field propagator on a random lattice 
based on a diffusion process. An important question is 
\vspace*{-6pt}
\begin{itemize}
\parskip=0pt
\topsep=-1ex
\itemsep=-1ex
\item
How to observe gravitational interaction between matter in our model
and how to establish the relation between the effective coupling constant 
and the conventional Newton's gravitational constant?
\end{itemize}

\clearemptydoublepage

\appendix
\chapter{Derivation of the Regge action}

\begin{figure}
\begin{center}
\includegraphics[width=0.8\textwidth]{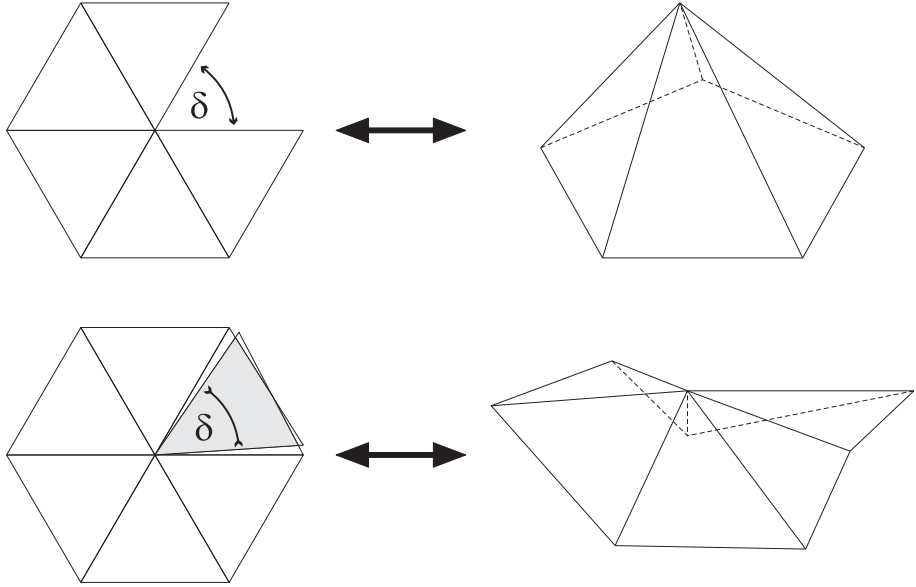}
\end{center}
\caption{An example of a positive (top) and negative (bottom) deficit angle.
The first triangulation corresponds to a cone with a positive singular curvature
at the peak, while the second to a saddle.}
\label{Fig:Deficyt}
\end{figure}

In this Appendix we derive the Euclidean version of the Regge action
which, together with the introduced measure, defines the partition function (\ref{Eq:ZDisc})
of the model of Causal Dynamical Triangulations.
For simplicity, we start directly with
the Einstein-Hilbert action (\ref{Eq:SEH}) 
after performing a Wick rotation
of the spacetime to the Euclidean region with the metric signature $\{+, +, +, +\}$.
The simplices building piecewise linear manifolds
are embedded in the Euclidean space $\R^4$.
Simplices are illustrated in Fig. \ref{Fig:Sympleksy}.
At the end we show that the derived action is indeed obtained
by a Wick rotation of the Lorentzian Regge action.
The Euclidean Einstein-Hilbert action is given by \cite{Hawking, Gibbons}
\beq
S_{EH}^{Euc} [g_{\mu \nu}]\equiv - \frac{1}{16 \pi G} \int_\mathcal{M} \dd^4 x \sqrt{g} (R - 2 \Lambda) = - \int_\mathcal{M} \dd^4 x \sqrt{g} \left( \frac{1}{2} k R - \lambda\right).
\eeq
where $k = \frac{1}{8 \pi G}$, $\lambda = \frac{\Lambda}{8 \pi G}$.
The Regge action is obtained by a direct evaluation of the Einstein-Hilbert action 
at a piecewise linear triangulation.
It is straightforward to calculate
that part of the action which depends on the cosmological constant, 
as it is proportional to the total four-volume.
It is given by a sum of contributions coming from all individual simplices.
We distinguish two types of simplices, namely $\{4, 1\}$ and $\{3, 2\}$.
Properties of all simplices within one type are identical
and their geometry is completely determined by the lengths $a_s$ and $a_t$.
The total volume is therefore given by
\beq
\int_\mathcal{M} \dd^4 x \sqrt{g} \lambda = \lambda \left( N_{41} \textrm{Vol}^{\{4, 1\}} + N_{32} \textrm{Vol}^{\{3, 2\}} \right),
\label{Eq:ReggeVol}
\eeq
where $N_{41}$ is the number of simplices of the type $\{4, 1\}$, which have a volume $\textrm{Vol}^{\{4, 1\}}$,
and $N_{32}$ is the number of simplices of the type $\{3, 2\}$, which have a volume $\textrm{Vol}^{\{3, 2\}}$.
Integration of the scalar curvature requires some attention.
The curvature is singular and localized at \emph{hinges} (triangles).
It occurs that, in the case of triangulations,
it can be simply expressed in terms of deficit angles
around those \emph{hinges}.
Let us illustrate this on a two-dimensional example.
In this case, 
the scalar curvature is a distribution with a support on vertices,
and the angle deficit is calculated around them.
A triangulation consisting of
$6$ equilateral triangles sharing one vertex 
and periodically connected along sides may be \emph{drawn}
on a plane $\dR^2$. Because the plane is flat, 
so is the vicinity of the common point.
The sum of angles adjoint to this point equals $\theta = 2 \pi$,
the  angle deficit $\delta = 2  \pi - \theta = 0$ and scalar curvature vanishes.
If such triangulation is built of smaller number of triangles,
as shown an the top of Fig. \ref{Fig:Deficyt},
it forms a cone and the angle deficit is positive, as is the curvature.
Likewise, a larger number of triangles gathered around one vertex
produces a saddle, illustrated an the bottom of Fig. \ref{Fig:Deficyt}, with a negative curvature.

A generalization to higher dimensions was proposed by Regge in \cite{Regge}.
In this case besides the deficit angles the integral of the curvature 
depends on a volume of homogeneous \emph{hinges}.
It was shown that the integral over a neighborhood $\cO$ containing one \emph{hinge} $\triangle$
is given by
\[ - \frac{1}{2} k \int_\mathcal{O} \dd^4 x \sqrt{g}\  R = - k (2 \pi - \theta_\triangle) \mathrm{Vol}_\triangle  ,\]
where $\theta_\triangle$ denotes the sum of dihedral angles around the triangle $\triangle$
and  $\mathrm{Vol}_\triangle$ denotes its area.
The integral over the whole manifold $\cM$ is obtained by summing contributions from all \emph{hinges}.
By a $(p, 3 - p)$-triangle we will denote a triangle with $p$ vertices in one slice and $3 - p$ in next slice.
A $\{4, 1\}$-simplex contains two types of triangles: $4$ $(3, 0)$-triangles and $6$ $(2, 1)$-triangles.
A $\{3, 2\}$-simplex contains $1$ $(3, 0)$-triangle, $6$ $(2, 1)$-triangles and $3$ $(1, 2)$-triangles.
Each simplex shares $10$ triangles and contributes to the total dihedral angle $\theta_\triangle$.
Using these properties we can write the global integral as
\begin{align}
 - \frac{1}{2} k \int_\mathcal{M} \dd^4 x \sqrt{g}\ R & = - 2 \pi k \sum_\triangle \mathrm{Vol}_\triangle + \nonumber\\
& + k\, N_{41} \left(4 \Theta^{\{4, 1\}}_{(3, 0)} \textrm{Vol}_{(3, 0)} + 6 \Theta^{\{4, 1\}}_{(2, 1)} \textrm{Vol}_{(2, 1)}\right) + \nonumber\\
& + k\, N_{32} \left(\Theta^{\{3, 2\}}_{(3, 0)} \textrm{Vol}_{(3, 0)} + 6 \Theta^{\{3, 2\}}_{(2, 1)} \textrm{Vol}_{(2, 1)} + 3 \Theta^{\{3, 2\}}_{(1, 2)} \textrm{Vol}_{(1, 2)}\right)
\label{Eq:ReggeCurv}
\end{align}
where $\Theta^{\{q, 5 - q\}}_{(p, 3 - p)}$ denotes the dihedral angle between two \emph{faces} of $\{q, 5 - q\}$-simplex
sharing a $(p, 3 - p)$-triangle, $\textrm{Vol}_{(p, 3 - p)}$ is the triangle volume.
As before all of these quantities are functions of $a_s$ and $a_t$.
The sum over all triangles $\triangle$ equals $\sum_\triangle \mathrm{Vol}_\triangle = N_{(3, 0)} \textrm{Vol}_{(3, 0)} + (N_{(2, 1)} + N_{(1, 2)}) \textrm{Vol}_{(2, 1)}$
and $N_{(p, 3- p)}$ denotes the number of triangles of appropriate type.
In order to evaluate the Euclidean Einstein-Hilbert action for a specific triangulation,
one must know volumes and angles associated with simplices.
Let us explicitly write down the positions of vertices
of the two types of simplices embedded in the Euclidean space $\R^4$.
We incorporate the length of spatial links $a_s$ into the coupling constant
and rescale the edge lengths so that $a_s = 1, a_t = \sqrt{\alpha}$.
The fourth coordinate defines the original time direction.
The vertex coordinates of a $\{4, 1\}$-type simplex are:
\begin{displaymath}
\begin{array}{rrrrr}
    v_1 = \Big\{&- \frac{1}{2},& - \frac{1}{2 \sqrt{3}},& - \frac{1}{2 \sqrt{6}},& 0 \Big\}, \\ 
    v_2 = \Big\{&  \frac{1}{2},& - \frac{1}{2 \sqrt{3}},& - \frac{1}{2 \sqrt{6}},& 0 \Big\}, \\ 
    v_3 = \Big\{& 				 0,&     \frac{1}{\sqrt{3}},& - \frac{1}{2 \sqrt{6}},& 0 \Big\}, \\ 
    v_4 = \Big\{& 				 0,& 							  0,&     \sqrt{\frac{3}{8}},& 0 \Big\}, \\ 
    v_5 = \Big\{& 				 0,& 							  0,&     0,& \sqrt{\frac{8 \alpha - 3}{8}} \Big\}. \\ 
\end{array}
\end{displaymath}
The distances between vertices $\{1,\dots,4\}$ are equal $a_s = 1$,
and the lengths of links ending in vertex $5$ are equal $a_t = \sqrt{\alpha}$.
Using these values we can calculate the simplex volume
and the dihedral angle between two three-dimensional hyperplanes passing 
through a given triangle and respectively the two remaining vertices of the simplex
\begin{align*}
 \textrm{Vol}^{\{4, 1\}} &= \frac{\sqrt{8 \alpha - 3}}{96}, \\
 \cos [\Theta^{\{4, 1\}}_{(3, 0)}]  &= \frac{1}{2 \sqrt{6 \alpha - 2}}, \\
 \cos [\Theta^{\{4, 1\}}_{(2, 1)}]  &= \frac{2 \alpha - 1}{6 \alpha - 2}.
\end{align*}
For a non-degenerate simplex we get the condition $\alpha > \frac{3}{8}$.
Similarly, we find the vertex coordinates of a $\{3, 2\}$-type simplex.
Now two vertices lie in the \emph{upper} plane
\begin{displaymath}
\begin{array}{rrrrr}
    v_1 = \Big\{&- \frac{1}{2},& - \frac{1}{2 \sqrt{3}},&  0,& 0 \Big\}, \\ 
    v_2 = \Big\{&  \frac{1}{2},& - \frac{1}{2 \sqrt{3}},&  0,& 0 \Big\}, \\ 
    v_3 = \Big\{& 				 0,&     \frac{1}{\sqrt{3}},&  0,& 0 \Big\}, \\ 
    v_4 = \Big\{& 				 0,& 							  0,&  -\frac{1}{2},&   \sqrt{\frac{12 \alpha - 7}{12}} \Big\}, \\ 
    v_5 = \Big\{& 				 0,& 							  0,&   \frac{1}{2},&  \sqrt{\frac{12 \alpha - 7}{12}} \Big\}. \\ 
\end{array}
\end{displaymath}
Again the links lying in the same \emph{horizontal} plane have the length $a_s = 1$, 
and the rest have $a_t = \sqrt{\alpha}$.
The volume of a  $\{3, 2\}$-simplex and corresponding dihedral angles,
are now given by
\begin{align*}
 \textrm{Vol}^{\{3, 2\}} &= \frac{\sqrt{12 \alpha - 7}}{96}, \\
 \cos [\Theta^{\{3, 2\}}_{(3, 0)}]  &= \frac{6 \alpha - 5}{6 \alpha - 2}, & \textrm{Vol}_{(3, 0)} &= \frac{\sqrt{3}}{4}, &\\
 \cos [\Theta^{\{3, 2\}}_{(2, 1)}]  &= \frac{1}{2 \sqrt{3 \alpha -1} \sqrt{4 \alpha - 2}}, & \textrm{Vol}_{(2, 1)} &= \frac{\sqrt{4 \alpha - 1}}{4}, &\\
 \cos [\Theta^{\{3, 2\}}_{(1, 2)}]  &= \frac{4 \alpha - 3}{8 \alpha - 4}, & \textrm{Vol}_{(1, 2)} &= \frac{\sqrt{4 \alpha - 1}}{4}. &
\end{align*}
For a non-degenerate simplex we get the condition $\alpha > \frac{7}{12}$.
The area of triangles is the same for both types of simplices and depends only on $a_s$ and $a_t$.

To simplify equations (\ref{Eq:ReggeVol}) and (\ref{Eq:ReggeCurv}) we use relations, 
specific for Causal Dynamical Triangulations, between numbers of appropriate sub-simplices.
Despite the fact that the simplices are a subset of Euclidean space,
the triangulation is equipped with the causal structure
which is a consequence of the original Lorentzian spacetime
and each slice is a triangulation of a three-sphere.
Because each spatial tetrahedron is shared by exactly two $\{4, 1\}$-simplices,
and those tetrahedra itself build a simplicial manifold one gets the equality
between the number of spatial triangles and $\{4, 1\}$-simplices
\[ N_{(3, 0)} = N^{\{4, 1\}}.\]
Using the above equality, the Euler identity and the Dehn-Sommerville conditions for a general four-dimensional triangulation
\cite{Dyna, FourDim} we get
\[ N_{(2, 1)} + N_{(1, 2)} = 2 N_0 + N_4 + N_{32}. \]
By inserting this relation into the sum of (\ref{Eq:ReggeVol}) and (\ref{Eq:ReggeCurv}), 
we get the Regge action for a triangulation.
It is given by a very simple linear function of three independent global parameters,
the number of vertices $N_0$, the number of simplices $N_4$ and the number $N_{41}$ of $\{4, 1\}$-simplices,
\beq
S = - (K_0 + 6 \Delta) N_0 + K_4 N_4 + \Delta N_{41} ,
\label{Eq:SReggeEuc}
\eeq
where the coefficients, called \emph{bare} coupling constants, are nontrivial functions
of the coupling constants present in the continuum Einstein-Hilbert action 
\begin{align}
K_0 + 6 \Delta &= \sqrt{4 \alpha - 1} \pi k, \\
K_4 &= \frac{\sqrt{12 \alpha - 7}}{96} \lambda + \frac{\sqrt{3}}{4} k \arccos \frac{6 \alpha - 5}{6 \alpha - 2} + \nonumber \\
    &+ \sqrt{4 \alpha - 1} k \left( \frac{3}{2} \arccos \frac{1}{2 \sqrt{3 \alpha -1} \sqrt{4 \alpha - 2}} + \frac{3}{4} \arccos \frac{4 \alpha - 3}{8 \alpha - 4} - \pi \right),  \\
K_4 + \Delta &= \frac{\sqrt{8 \alpha - 3}}{96} \lambda + \sqrt{3} k \left( \arccos \frac{1}{2 \sqrt{6 \alpha - 2}} - \frac{\pi}{2} \right) + \nonumber \\
             &+ \sqrt{4 \alpha - 1} k \left( \frac{3}{2} \arccos \frac{2 \alpha - 1}{6 \alpha - 2} - \frac{\pi}{2} \right). 
\label{Eq:AppCoupling}
\end{align}

Equivalently, one can start with the  Einstein-Hilbert action (\ref{Eq:SEH})
and evaluate it at a Lorentzian piecewise linear manifold made up of 
simplices being a subset of Minkowski spacetime, as described in \cite{Dyna}
(here we use opposite sign convention $\ a_t^2 = \alpha \cdot a_s^2$).
In the Lorentzian case $\alpha$ is negative and the simplex volumes are given by 
\[ \textrm{Vol}_{Lor} (\alpha)= \sqrt{-1} \textrm{Vol}_{Euc} (\alpha) \]
The same relation holds for volumes of time-like triangles.
The equations for angles are identical in both cases,
in the Lorentzian case angles may take complex values. 
Inserting the Lorentzian counterparts of volumes and angles into the Einstein-Hilbert action
and taking advantage of equations (\ref{Eq:ReggeVol}) and (\ref{Eq:ReggeVol}),
we get the Lorentzian Regge action.
The further step is to analytically continue $\alpha$ from negative values
to positive values through the lower-half complex plane ($\sqrt{-1} = -i$).
For $\alpha > \frac{7}{12}$ the Lorentzian action becomes purely imaginary.
The Euclidean action defined by relation $-S^{Euc} = i S^{Lor}$
coincides with equation (\ref{Eq:SReggeEuc}).

\clearemptydoublepage

\chapter{Constrained propagator}

Let us consider a multivariate normal distribution with a $T$ component
random vector $\bx$ and zero mean.
We denote the full propagator inverse by $\tilde \bP$.
Further we impose a constraint on the random vector,
i.e. we measure $\bx$ only when it is orthogonal to $\bq$
\[ P[\bx] = \frac{1}{Z[0]} e^{- \frac{1}{2} \bx^\tau \tilde \bP \bx}\ \delta(\mathbf{q}^\tau \bx), \quad\mathbf{q}^\tau \mathbf{q} = 1. \] 
For definiteness we normalized the eigenvector $\bq$ corresponding to the zero mode 
The partition function of this simple statistical model with a source term $\bk^{\tau} \bx$ is given by
\[ Z[\mathbf{k}] = \int \dd^T \ \bx e^{- \frac{1}{2} \bx^\tau \tilde \bP \bx + \mathbf{k}^\tau \bx}\ \delta(\mathbf{q}^\tau \bx). \]
The propagator of the constrained model is given by a derivative of the constrained partition function,
\[ \bC_{i j} = \langle x_i x_j \rangle_c = \left. \frac{\partial^2 \ln Z[k]}{\partial k_i \partial k_j} \right|_{\mathbf{k} = 0}.\]
In order to derive a relation between the constrained propagator $\bC$ and full propagator $\tilde \bC \equiv \tilde \bP^{-1}$,
we use the integral representation of the delta function.
The function $Z[\bk]$ is given by
\begin{align*}
Z[\mathbf{k}] &= \int \dd^T \bx \, \dd \lambda \ e^{- \frac{1}{2} \bx^\tau \tilde \bP \bx + \mathbf{k}^\tau \bx + i \lambda \mathbf{q}^\tau \bx} \\
              &= \sqrt{\frac{(2 \pi)^T}{\det \tilde P}} \int \dd \lambda \ e^{\frac{1}{2} (\mathbf{k} + i\lambda \mathbf{q})^\tau  \tilde \bC  (\mathbf{k} + i\lambda \mathbf{q})} \\
              &= \sqrt{\frac{(2 \pi)^T}{\det \tilde P}} \sqrt{\frac{2 \pi}{\mathbf{q}^\tau \tilde \bC \mathbf{q} }} \exp \left[ \frac{1}{2} \mathbf{k}^\tau \tilde \bP^{-1} \mathbf{k} - \frac{1}{2 \mathbf{q}^\tau \tilde \bC \mathbf{q} } \mathbf{k}^{\tau} \tilde \bC \, \mathbf{q} \mathbf{q}^{\tau} \tilde \bC \, \mathbf{k}  \right].
\end{align*}
and the constrained propagator $\bC$ as a function of the full propagator $\tilde \bC$ is given by
\beq
\bC =  \tilde \bC - \frac{1}{\Tr \ \mathbf{A} \tilde \bC } \tilde \bC \mathbf{A} \tilde \bC, \quad \mathbf{A}  \equiv \mathbf{q} \mathbf{q}^\tau .
\label{Eq:BC}
\eeq
The matrix $\bP$ is defined as an inverse of $\bC$ on a subspace orthogonal to $\mathbf{q}$,
\beq
\bP \bC = \bI - \mathbf{A} .
\label{Eq:PC}
\eeq
It is easy to check that 
\beq
\bP  = (\bI - \mathbf{A}) \, \tilde \bP \, (\bI - \mathbf{A}),
\label{Eq:BP}
\eeq
solves equation (\ref{Eq:PC}), where we used relation (\ref{Eq:BC}).
The constraint inverse propagator $\bP$ can be expressed by the constrained propagator 
in the following way
\beq 
\bP = (\bC + \bA)^{-1} - \bA.
\label{Eq:PCA}
\eeq
From (\ref{Eq:BC}) and (\ref{Eq:BP})
it follows that $\bP$ and $\bC$ act in a space orthogonal to $\mathbf{q}$
\[ \bP \mathbf{q} = \bC \mathbf{q} = 0 .\]

\clearemptydoublepage

\clearemptydoublepage

\chapter*{The author's list of publications}
\addcontentsline{toc}{chapter}{\bf The author's list of publications}

\markboth{The author's list of publications}{The author's list of publications}

\begin{longtable}{r@{ \,} l}

\multicolumn{2}{l}{\bf Peer-reviewed publications}\\[2ex]

[1] &  J.~Ambj\o rn, A.~G\"orlich, J.~Jurkiewicz, R.~Loll, \\
    & \textit{Geometry of the quantum universe},\\
    & Phys. Lett. B {\bf 690}, 420 (2010) [arXiv:1001.4581].\\[1.5ex]
[2] & J.~Ambj\o rn, A.~G\"orlich, S.~Jordan, J.~Jurkiewicz, R.~Loll, \\
    & \textit{CDT meets Ho\v rava-Lifshitz gravity},\\
    & Phys. Lett. B {\bf 690}, 413 (2010) [arXiv:1002.3298].\\[1.5ex]
[3] & J. Ambj\o rn, A. G\"orlich, J. Jurkiewicz, R. Loll, \\
    & {\it Nonperturbative quantum de Sitter universe}, \\
    & Phys. Rev. D \textbf{78}, 063544 (2008) [arXiv:0807.4481].\\[1.5ex]
[4] & J. Ambj\o rn, A. G\"orlich, J. Jurkiewicz, R. Loll,\\
    & {\it Planckian Birth of the Quantum de Sitter Universe},\\
    & Phys. Rev. Lett. \textbf{100}, 091304 (2008) [hep-th/0712.2485].\\[1.5ex]
[5] & V. Corato, A. G\"orlich, P. Korcyl, P. Silvestrini, L. Stodolsky, J. Wosiek,\\
    & {\it Simulations of quantum gates with decoherence},\\
    & Phys. Rev. B \textbf{75}, 184507 (2007) [cond-mat/0611445].\\[1.5ex]
[6] & Z. Burda, A. G\"orlich, B. Wac\l{}aw, \\
    & {\it Spectral properties of empirical covariance matrices for data with power-law tails},\\
    & Phys. Rev. E \textbf{74}, 041129 (2006) [physics/0603186]. \\[1.5ex]
[7] & Z. Burda, A. G\"orlich, J. Jurkiewicz, B. Wac\l{}aw,\\
    & {\it Correlated Wishart Matrices and Critical Horizons},\\
    & Eur. Phys. J. B \textbf{49}, 319 (2006)  [cond-mat/0508341].\\ [1.5ex]
[8] & Z. Burda, A. G\"orlich, A. Jarosz, J. Jurkiewicz,\\
    & {\it Signal and noise in correlation matrix},\\
    & Physica A \textbf{343} 295 (2004) [cond-mat/0305627]. \\

\newpage

\multicolumn{2}{l}{\bf Conference proceedings}\\[2ex]

[9] & A. G\"orlich, \\
    & {\it Background Geometry in 4D Causal Dynamical Triangulations}, \\
    & Acta Phys. Pol. B \textbf{39}, 3343 (2008).\\[1.5ex]
[10] & J. Ambj\o rn, A. G\"orlich, J. Jurkiewicz, R. Loll,\\
     & {\it The Quantum Universe},\\
     & Acta Phys. Pol. B \textbf{39}, 3309 (2008).\\[1.5ex]
[11] & J.~Ambj\o rn, A.~G\"orlich, J.~Jurkiewicz, R.~Loll, \\
     & {\it The emergence of Euclidean de Sitter space-time}\\
     & PATH INTEGRALS: NEW TRENDS AND PERSPECTIVES, \\
     & Proceedings of the 9th International Conference, \\
     & ed. W. Janke, A.Pelster, \\
     & World Scientific Publishing.\\[1.5ex]

\multicolumn{2}{l}{\bf Other publications}\\[2ex]

[12] & A. G\"orlich, A. Jarosz, \\
     & {\it Addition of Free Unitary Random Matrices},\\
     & [math-ph/0408019].\\[1.5ex]
[13] & J.~Ambj\o rn, A.~G\"orlich, J.~Jurkiewicz, R.~Loll, \\
     & {\it CDT-—an Entropic Theory of Quantum Gravity},\\
     & [arXiv:1007.2560].
\end{longtable}

\clearemptydoublepage

\chapter*{Acknowledgments}
\addcontentsline{toc}{chapter}{\bf Acknowledgments}

I am very grateful to my supervisor Professor Jerzy Jurkiewicz
for introducing me into the fascinating topic of quantum gravity.
For scientific inspiration, constant help and support which made this thesis possible.

I am specially thankful to Professor Jan Ambj\o{}rn and Professor Renate Loll
for a fruitful collaboration in a nice atmosphere and hospitality during my visits to Utrecht.

I acknowledge the collaboration and rousing discussions with Samo Jordan, Jakub Gizbert-Studnicki and Tomasz Trze\'s{}niewski.

Last but not least, I wish to thank my family whose support allowed me to finish the thesis.

I wish to acknowledge financial support by the Polish Ministry of Science grant N~N202 034236 (2009-2010).

\clearemptydoublepage

\end{document}